\documentclass[12pt]{amsart}
\usepackage{amssymb}
\usepackage{amsmath}
\usepackage{amsfonts}
\usepackage{mathtools}
\usepackage{mathrsfs}
\usepackage{graphicx}
\usepackage{placeins}
\usepackage{color}
\usepackage[onehalfspacing]{setspace}
\usepackage{caption}
\usepackage{subcaption}
\usepackage{natbib}
\usepackage{enumerate}
\usepackage[utf8]{inputenc}
\usepackage{tikz-cd}
\usepackage[charter,cal=cmcal]{mathdesign}
\usepackage{epigraph}
\usepackage{dcolumn}
\usepackage{placeins}
\usepackage[all,2cell]{xy}
\UseAllTwocells
\xyoption{arc}
\xyoption{rotate}
\usepackage{bibunits}
\usepackage[colorlinks=true,citecolor=blue,urlcolor=blue,pdfpagemode=UseNone,pdfstartview=FitH,hypertexnames=false]{hyperref}

\DeclareTextFontCommand{\bi}{%
	\fontseries\bfdefault 
	\itshape
}

\DeclareMathOperator*{\argmin}{\arg\!\min}

\makeatletter
\def\section{\@startsection{section}{1}
	\z@{0.8\linespacing\@plus\linespacing}{.7\linespacing}{\Large}}

\def\subsection{\@startsection{subsection}{2}
	\z@{.5\linespacing\@plus.7\linespacing}{.7\linespacing}{\large}}

\def\subsubsection{\@startsection{subsubsection}{3}
	\z@{.5\linespacing\@plus.7\linespacing}{-.5em}{\normalfont\bfseries}}
\makeatother

\newtheorem{theorem}{Theorem}[section]

\newtheorem{lemma}{Lemma}[section]

\theoremstyle{definition}

\theoremstyle{definition}
\newtheorem{assumption}{Assumption}[section]

\theoremstyle{definition}

\calclayout

	\title{}
	
\begin{document}
			\vspace*{5ex minus 1ex}
		\begin{center}
			\Large \textsc{Synthetic Decomposition for Counterfactual Predictions}
			\bigskip

			\normalsize

		\end{center}
		
		\date{%
			\today%
		}
		
		\vspace*{3ex minus 1ex}
		\begin{center}
		Nathan Canen and Kyungchul Song\\
		\textit{University of Warwick \& CEPR and University of British Columbia}\\
		
		\bigskip
		\bigskip

		\end{center}
		
		\thanks{We thank Victor Aguirregabiria, Tim Armstrong, Xu Cheng, EunYi Chung, Wayne Gao, Bruce Hansen, Yu-Chin Hsu, Hide Ichimura, Hiro Kasahara, Ying-Ying Lee, Vadim Marmer, Ismael Mourifie, Jack Porter, Vitor Possebom, Frank Schorfheide, Paul Schrimpf, Xiaoxia Shi and participants in the seminars in CIREQ Montreal Econometrics Conference, Conference on Econometrics for Modern Data Structures, Seoul National University, SETA 2022, University of Calgary, University of California, Irvine, University of Illinois Urbana-Champaign, University of Norte Dame, University of Pennsylvania, University of Toronto, University of Victoria, University of Wisconsin-Madison for valuable comments. All errors are ours. We also thank Ratzanyel Rinc\'{o}n for his excellent research assistance. Song acknowledges that this research was supported by Social Sciences and Humanities Research Council of Canada. Corresponding Address: Kyungchul Song, Vancouver School of Economics, University of British Columbia, 6000 Iona Drive, Vancouver, BC, V6T 1L4, Canada, kysong@mail.ubc.ca}
		\address{Department of Economics, University of Warwick, Coventry, CV4 7AL, United Kingdom}
		\email{nathan.canen@warwick.ac.uk}
		\address{Vancouver School of Economics, University of British Columbia, 6000 Iona Drive, Vancouver, BC, V6T 1L4, Canada}
		\email{kysong@mail.ubc.ca}

		\fontsize{12}{14} \selectfont 
		\begin{bibunit}[econometrica]		

\begin{abstract}
    It is challenging to conduct counterfactual predictions when the policy variable goes beyond its pre-policy support. However, in many cases, information about the policy of interest is available from other (``source'') regions where a similar policy has already been implemented. In this paper, we propose a novel method of cross-population extrapolation that uses data from such source regions to predict a new policy in a target region. Using data from the source regions, we construct a synthetic outcome-policy relationship that is most similar to the relationship in the target region based on pre-policy data. This synthetic relationship identifies the counterfactual prediction under a transferability condition which is weaker than those often used in decomposition and transfer methods. We apply our proposal to predict average teenage employment in Texas following a counterfactual increase in the minimum wage.
\medskip

{\noindent \textsc{Key words:} Counterfactual Predictions, Decomposition Analysis, Ex Ante Policy Evaluation, Cross-Population Extrapolation, Synthetic Decomposition}
\medskip

{\noindent \textsc{JEL Classification: C30, C54}}
\end{abstract}
\maketitle

\bigskip
\bigskip
\bigskip
\bigskip

\section{Introduction}

Policymakers' questions are often centered around the prediction of a new policy's outcome, such as predicting the effect of a new job training program, the welfare implication of a proposed merger of firms, or the employment effect of a minimum wage increase. Such questions are hard to answer because the new policy's outcome is unobserved. For example, when a state in the U.S. considers increasing its minimum wage to a level never seen before within that state, the policy is beyond its historical variations. As pointed out by \cite{Wolpin:13:Book}, the predominant approach in such a situation is to employ a parametric specification of the policy-outcome relationship and extrapolate it to a post-policy environment. However, when the new policy extends beyond the support of its historical variations, the counterfactual prediction loses its nonparametric identification, making the prediction potentially sensitive to the selected parametrization.

\begin{figure}[t]
	\begin{center}
		\includegraphics[scale=0.52]{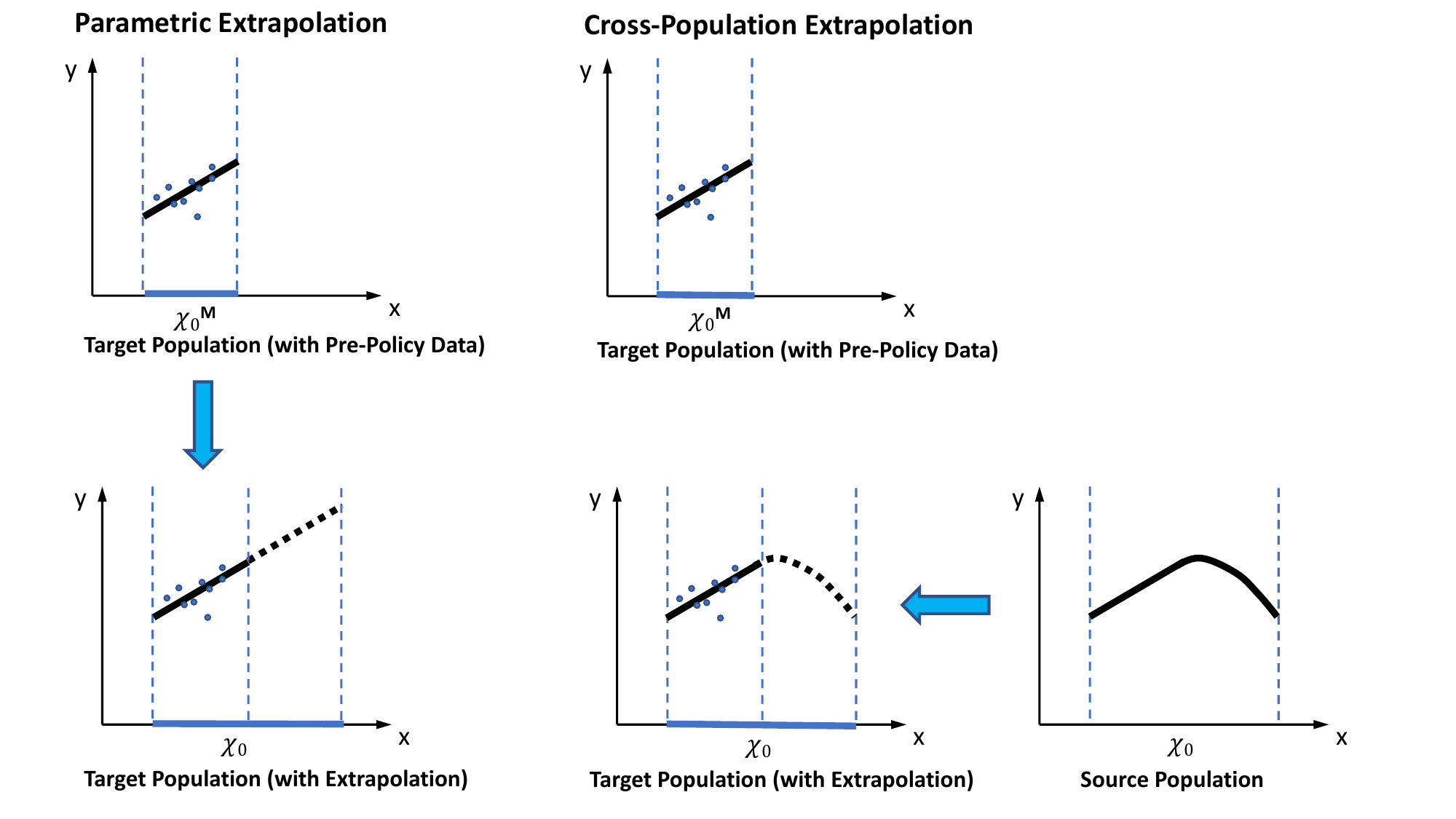}
        \captionsetup{width=1\textwidth}
		\caption{Parametric Extrapolation and Cross-Population Extrapolation}
    
		\label{fig:Extrapolation}

        \medskip

        \parbox{\textwidth}{\footnotesize Note: The figure illustrates the difference between parametric extrapolation and cross-population extrapolation. In parametric extrapolation, the researcher uses a parametric specification of the policy-outcome relationship in the target population and extrapolates it outside the support of the pre-policy variations of the policy variable. In cross-population extrapolation, the researcher uses data from other populations that have richer variations and transfer the relationship to the target population.}
	\end{center}
\end{figure}

In this paper, we consider cross-population extrapolation as an alternative to parametric extrapolation (see Figure \ref{fig:Extrapolation} for a schematic comparison.) In many scenarios of \textit{ex ante} program evaluations, researchers have data from other populations that have experienced a ``similar'' policy. For instance, when predicting the average teenage employment following a minimum wage increase in Texas, one might look at the policy's effects in California, Oregon, and Washington. In this setting, the researcher may consider using data from other states and ``transfer'' policy predictions to the population of Texas instead of relying solely on a parametric specification of the policy-outcome relationship in Texas. Yet, it is unclear how one should determine which source populations are most relevant for the prediction in the target population.

The transfer of empirical features from one context to another has been a longstanding practice in economics. Examples include calibration methods in macroeconomics (see \cite{Gregory/Smith:93:Elsevier} for a review), decomposition methods in labor economics (see, e.g. \cite{Fortin/Lemieux/Firpo:11:Handbook} for a review), and program evaluations using experimental data to evaluate a structural model (see, e.g., \cite{Todd/Wolpin:06:AER}, \cite{Attanasio/Meghir/Santiago:11:ReStud}) or to assess the generalizability of an experiment result to other populations (see \cite{Hotz/Imbens/Mortimer:05:JOE}). To the best of our knowledge, little attention has been paid to the transfer problem from multiple heterogeneous source populations in predicting the effect of new policies in a target population. 

In this paper, we present a novel approach of cross-population extrapolation from multiple source populations to generate counterfactual predictions. In this approach, we consider structural equations that represent causal relationships between the policy variable and the outcome variable in the source regions. We use those equations as a fixed set of basis functions and construct their weighted average to capture the outcome-policy relationship in the target region. To identify those weights, we draw insights in \cite{Todd/Wolpin:08:AES} and \cite{Wolpin:13:Book}, recognizing that structural equation models often involve the policy variable in an index (called the \textit{policy component} here) which exhibits variations at the individual level. Using the policy component, we can divide the target population into two groups: a \textit{matched group} and an \textit{unmatched group}. The matched group comprises individuals whose post-policy value of the policy component can be matched with the pre-policy value of another individual within the group. We derive weights that align the weighted average of outcome-policy relationships from source populations to closely match the outcome-policy relationship of the target population within the matched group, utilizing pre-policy data from the matched group to guide this alignment. With these weights identified, we can generate counterfactual predictions for new policies that have not been implemented previously. Since our proposal is closely related to the Kitagawa-Oxaca-Blinder decomposition method, we call our method a \textit{synthetic decomposition method}.

Every prediction that uses cross-population extrapolation from a single source population, either explicitly or implicitly, relies on a transferability condition requiring the same structural relationship to hold in the target population as in the source population. When there are multiple source populations, this transferability condition can be weakened to what we call the \textit{synthetic transferability condition}. The condition stipulates that the true counterfactual relationship in the target population (say, between minimum wages and employment in Texas) is represented by a weighted average of the outcome-policy relationships derived from the source populations (say, California, Oregon, Washington).

The synthetic transferability condition is tantamount to modeling the counterfactual relationship as a point in the convex hull of a fixed, finite set of basis functions. However, unlike nonparametric sieves such as trigonometric or polynomial series, these basis functions are derived from the structural relationships identified in the source populations. Limiting the model to the convex hull ensures that the qualitative properties of the structural relationships, such as monotonicity or concavity, are preserved. This preservation maintains the interpretability of the resulting synthetic relationship in practical applications.\footnote{While extending our proposal to the linear span instead of the convex hull is straightforward, the linear span does not necessarily preserve such qualitative properties.}

Foremost, the synthetic transferability condition emerges as notably less stringent when contrasted with earlier transferability conditions focusing on a single source population, e.g., the standard decomposition method in Labor Economics, or extrapolation of experimental results, where it is commonly invoked, often implicitly. Unlike these conditions, it does not require the source populations to have the same structural relationship as the target population. Furthermore, the synthetic transferability condition has testable implications which a researcher can use to gauge its plausibility in data. We explain this in detail later in the paper.

The synthetic decomposition method applies to a wide range of counterfactual prediction settings. The method is built on semiparametric outcome-policy relationships that are nonseparable in the (potentially multi-dimensional) unobserved heterogeneity. This flexibility allows the researcher to derive a semiparametric outcome-policy relationship from a structural model that specifies peoples' incentives and choices differently across the populations. Furthermore, the method accommodates various forms of policies, including policies that transform a certain individual-level exogenous variable (e.g., demographic-dependent tax subsidies) or an aggregate-level exogenous variable (e.g., minimum wages). The policy can be one that changes a structural parameter or a coefficient of a certain variable. We can further allow spillovers to be present and for weights to depend on covariates. We present fully worked out examples below. 

We illustrate our procedure with an empirical application studying the effect of a counterfactual increase in minimum wages in Texas to US\$9. (The prevailing minimum wage is the federal level of US\$7.25, set in 2009.) Such an increase is subject to extensive policy and academic interest, as shown by it being a central policy proposal in the 2022 Texas gubernatorial elections.\footnote{See \url{tinyurl.com/2cmv7fhz} for its presence and analysis within one of the candidate's policy platforms.} However, the extensive minimum wage literature in labor economics predominantly focuses on \textit{ex post} analyses of minimum wage increases (\cite{Neumark:19:GER}). We implement our proposed method using Current Population Survey (CPS) data and estimate that an increase in minimum wages would decrease average (teenage) employment by 6.5-9 percentage points on a baseline of approximately 29\% if minimum wages in Texas were US\$9. In doing so, our synthetic comparison for Texas (i) accounts for the heterogeneous skill distributions and demand conditions across states (\cite{Flinn:11:Book}), (ii) does not require the researcher to choose the comparison unit (e.g., whether to focus on geographically close or distant states - see \cite{Dube/Lester/Reich:10:ReStat} and \cite{Neumark:19:GER} for a discussion), (iii) accounts for the difference in causal relationships between minimum wages and employments across states (\cite{Flinn:02:ReStud}), (iv) does not rely on parametric extrapolation, which is a concern in this literature - see \cite{Flinn:06:Ecma, Gorry/Jackson:17:CEP, Neumark:19:GER}, for example.\medskip

\noindent \textbf{Related Literature} The importance of \textit{ex ante} policy evaluation in economics is widely acknowledged (e.g. \cite{Heckman/Vytlacil:05:Eca}, \cite{Heckman:10:JEL}, \cite{Wolpin:07:AER} and \cite{Wolpin:13:Book}). Many applications hinge on a parametric structural model for counterfactual policy analysis. However, predictions can be sensitive to the chosen parametric specification, especially when nonparametric identification fails due to data constraints. Our method presents an approach that utilizes data sets from other populations.

The practice of combining datasets from diverse populations is widespread in economics. Since the seminal paper by \cite{LaLonde:86:AER}, experimental data have been used to assess various non-experimental methods (\cite{Heckman/Ichimura/Todd:97:ReStud}, \cite{Heckman/Ichimura/Smith/Todd:98:Eca}, \cite{Dehejia/Wahba:99:JASA} and \cite{Smith/Todd:05:JOE}), as well as to construct nonexperimental comparison groups for policy impact assessment (\cite{Friedlander/Robins:95:AER} and \cite{Hotz/Imbens/Mortimer:05:JOE}). The literature reveals heterogeneity in treatment effects across experiment sites and limitations in generalization to scaled-up settings (\cite{Heckman/Smith:00:YEAC}, \cite{Allcott:15:QJE}, \cite{Bold_et_al:18:JPubE} and \cite{Wang/Yang:21:NBER}), prompting interest in assessing treatment effect generalizability beyond experimental domains (\cite{Dehejia:03:JBES}, \cite{Stuart/Cole/Bradshaw/Leaf:11:JRSS}, \cite{Meager:19:AEJ}, \cite{Vivalt:20:JEEA}, \cite{Bandiera_et_al:21:AER}, \cite{Gechter/Samii/Dehejia/Pop-Eleches:19:arXiv}, and \cite{Ishihara/Kitagawa:21:arXiv}). On the other hand, \cite{Todd/Wolpin:06:AER} and \cite{Attanasio/Meghir/Santiago:11:ReStud} proposed using experimental data to estimate and evaluate structural models. Recently, \cite{Menzel:WP:23} developed a method of generating a linear predictor of conditional average treatment effects for a target site using the baseline pre-intervention outcome data from multiple sites, and \cite{Andrews_et_al:WP:22} a method of building forecast intervals for measures of transfer error of economic models across different domains.

Closer to this paper is the literature of decomposition methods in labor economics. Since the seminal papers by \cite{Oaxaca:73:IER} and \cite{Blinder:73:JHR}, the literature has adopted a decomposition method to generate counterfactual predictions using data sets from other populations (\cite{Juhn/Murphy/Pierce:93:JPE}, \cite{DiNardo/Fortin/Lemieux:96:Eca}). A growing interest has also been paid to the role of the decomposition methods in generating causal inference and counterfactual predictions, with the required transferability condition made more explicit (\cite{Fortin/Lemieux/Firpo:10:NBER}, \cite{Kline:11:AERPP}, \cite{Chernozhukov/Fernandez/Melly:13:Eca},  \cite{Rothe:10:JoE}, \cite{Ao/Calonico/Lee:21:JBES} and \cite{Hsu/Lai/Lieli:22:JBES}). Interestingly, as we show later, the transferability condition in the literature of decomposition methods is closely connected with the conditions used for extrapolation of causal inference in the literature (\cite{Hotz/Imbens/Mortimer:05:JOE}, \cite{Hartman_et_al:15:JRSS}, \cite{Athey_et_al:20:WP} and \cite{Gechter/Meager:22:WP}.)

Our synthetic decomposition method draws inspiration from the increasingly popular synthetic control method in econometrics and applied research (see \cite{Abadie:21:JEL} for a review of the method and literature). Both methods construct a synthetic counterfactual quantity using data from multiple populations. A distinctive feature of our method is the use of the invariance of the structural relationship between the outcome and the policy component. This invariance allows us to form a matched group and identify the synthetic relationship, rather than the synthetic outcome. To the best of our knowledge, this feature does not have a direct analogue in the synthetic control method. While our method shares similarities with the literature on model averaging and combining forecasts (see \cite{Timmermann:06:Handbook} and \cite{Steel:20:JEL} for a review), it is specifically designed for generating counterfactual predictions using data from multiple source populations. Moreover, our approach emphasizes the importance of explicit transferability conditions necessary for identifying relevant causal relationships.

The rest of the paper proceeds as follows. In Section 2, we present our main proposal of the synthetic decomposition method and discuss conditions for the method to work and related literature. In Section 3, we provide procedures of estimation and construction of confidence intervals, assuming that we observe a random sample of data from each population. In Section 4, we present an empirical application that studies the prediction problem of average employment when the minimum wage increases in Texas. In Section 5, we conclude. In the Supplemental Note, we present general conditions for the proposed confidence intervals to be uniformly asymptotically valid, as well as formal results and proofs. The Supplemental Note also contains some more details on the empirical application.

\section{Synthetic Decomposition for Counterfactual Predictions}

\subsection{Counterfactual Predictions with Multiple Populations}

\subsubsection{The Target Population and Counterfactual Predictions}

Suppose that there is a region planning to implement a new policy and we aim to predict its outcome. For each sample unit (individual or a firm) $i$ in the region, the outcome variable denoted by $Y_i$ is generated as follows:
\begin{align}
	\label{RF 0}
	Y_i &= g_0\left(\mu_0(X_i,v_0), X_i, U_i\right), \text{ before the policy, and},\\ \notag
    Y_i &= g_0\left(\mu_0(X_i,v_0^*), X_i, U_i\right), \text{ after the policy}, 
\end{align}
where $U_i$ is an unobserved random vector, $X_i$ observed random vector, $v_0$ and $v_0^*$ aggregate policy variables, and $\mu_0$ and $g_0$ are structural functions invariant to the policy. Here the policy simply refers to a transform of $v_0$ into $v_0^*$. The subscript $0$ in $g_0$, $\mu_0$ and $v_0$ expresses that they belong to the region 0. We call $\mu_0(X_i,v_0)$ and $\mu_0(X_i,v_0^*)$ the \bi{policy components} which we assume to be a parametric function as follows:
\begin{align}
	\label{parametrization}
	\mu_0(X_i,v_0) = \mu(X_i,v_0;\beta_0) \text{ and } \mu_0(X_i,v_0^*) = \mu(X_i,v_0^*;\beta_0).
\end{align}
We require the policy components to exhibit individual-level variations. This requirement is met in many settings, as we show through examples below.\medskip

\noindent \textbf{Example 1 (Transforming an Individual Covariate): } The policy changes $X_i$ into $f(X_i)$ for some map $f$ for each individual $i$. We can accommodate this setting by taking $v_0 = 0$, $v_0^* = 1$, $\mu_{0}(x,0) = x$ and $\mu_{0}(x,1) = f(x)$. For example, suppose that $X_i = (X_{1,i}, X_{2,i})$ where $X_{1,i}$ represents an individual's income and $X_{2,i}$ other demographic characteristics. The policy of interest is an income subsidy of an amount $\delta>0$ for each individual $i$ with $X_i$ in a set $A$. Then, we can take 
\begin{align*}
	f(x) = (x_1 + \delta,x_2) 1\{x \in A\} + (x_1, x_2) 1\{x \notin A\}.
\end{align*}
While the amount $\delta$ is the same across individuals, the policy components $\mu_0(X_i,v_0)$ and $\mu_0(X_i,v_0^*)$ generally exhibit variations at the individual level.\footnote{It is important to note that this simple setting of counterfactual prediction is already different from the standard program evaluation setting. Here, the potential outcomes are given as follows: 
\begin{align*}
	Y_i(0) = g_0\left(\mu_0(X_i,v_0),U_i\right) \text{  and  } Y_i(1) = g_0\left(\mu_0(X_i,v_0^*),U_i\right).
\end{align*}
Unlike the standard program evaluation setting, \textit{everybody is treated} here. Furthermore, we focus on an \textit{ex ante} policy evaluation where we do not observe the outcome of the policy for the target region $0$. (See \cite{Heckman/Vytlacil:07:Handbook} and \cite{Wolpin:13:Book} for the problem of policy analysis in such a setting.)} $\blacksquare$\medskip

\noindent \textbf{Example 2 (Changing a Structural Parameter): } The policy component $\mu_{0}(X_i,v_0)$ can be a parametric function where $v_0$ and $v_0^*$ represent the parameter values before and after the policy. For example, 
\begin{align*}
    \mu_0(x,v_0) = x^{\prime} v_0 \text{  and  } \mu_0(x,v_0^*) = x^{\prime} v_0^*.
\end{align*}
A special case is shutting down the impact of the $j$-th covariate in $X_i$. This can be expressed by taking $v_0 = v_0^*$ except for the $j$-th entry $v_{0,j}^*$ of $v_0^*$ being set to be zero.  $\blacksquare$\medskip

\noindent \textbf{Example 3 (Changing an Aggregate Variable): } The variable $v_0$ in the policy component can be an aggregate variable such as a tax rate or a minimum wage. For example, $\mu_0(X_i,v_0)$ can be a component in the labor supply decision, and $v_0$ denotes the minimum wage in region $0$. The counterfactual policy could be an increase in the minimum wage $v_0$ by $\Delta$, so that the new minimum wage becomes $v_0^* = v_0 + \Delta$. While the aggregate policy variable $v_0$ does not vary at the individual level, the policy component $\mu_0(X_i,v_0)$ does in many settings. $\blacksquare$\medskip

We are interested in predicting the average outcome after the policy, defined as
\begin{align}
\label{theta_00}
\theta_0 \equiv \mathbf{E}_0\left[ g_0\left(\mu_0(X_i,v_0^*), X_i, U_i \right) \right] = \int m_0\left(\mu_0(x,v_0^*),x\right) dP_0(x),
\end{align}
where $\mathbf{E}_0$ denotes the expectation with respect to the population $P_0$ in region $0$, and the average response function $m_0$ is defined as follows:
\begin{align}
	\label{m_0}
	m_0\left(\overline \mu,x \right) &=  \int g_0(\overline \mu,x,u)dP_{0,U | X}(u \mid x),
\end{align}
where $P_{0,U | X}$ denotes the conditional distribution of $U_i$ given $X_i$ in the target population (before the policy). The map $m_0$ summarizes the structural relationship between the outcome and the policy component in the model. When $X_i$ and $U_i$ are independent, it reduces to the average structural function (ASF) introduced by \cite{Blundell/Powell:03:Adv}. Unlike the ASF, the map $m_0\left(\overline \mu,x\right)$ is \textit{causal} in the first argument $\overline \mu$ but not in the second argument $x$, because in this model the causal relation between $U_i$ and $X_i$ is left ambiguous.\footnote{Alternatively, we might be interested in the distribution of the outcomes:
\begin{align*}
	p_0(A; \overline \mu,x) \equiv \int 1\left\{g_0(\overline \mu,x,u) \in A \right\}dP_{0,U | X}(u \mid x), \text{ for each set } A.
\end{align*}
The quantity represents the conditional probability of $Y_i$ taking values from a set $A$ given $X_i = x$, when $\mu_0(X_i,v_0)$ is \textit{fixed to be} $\overline \mu$. The proposal of this paper carries over to this alternative with straightforward modifications.} 

\begin{figure}[t]
		\begin{center}
			\includegraphics[scale=0.50]{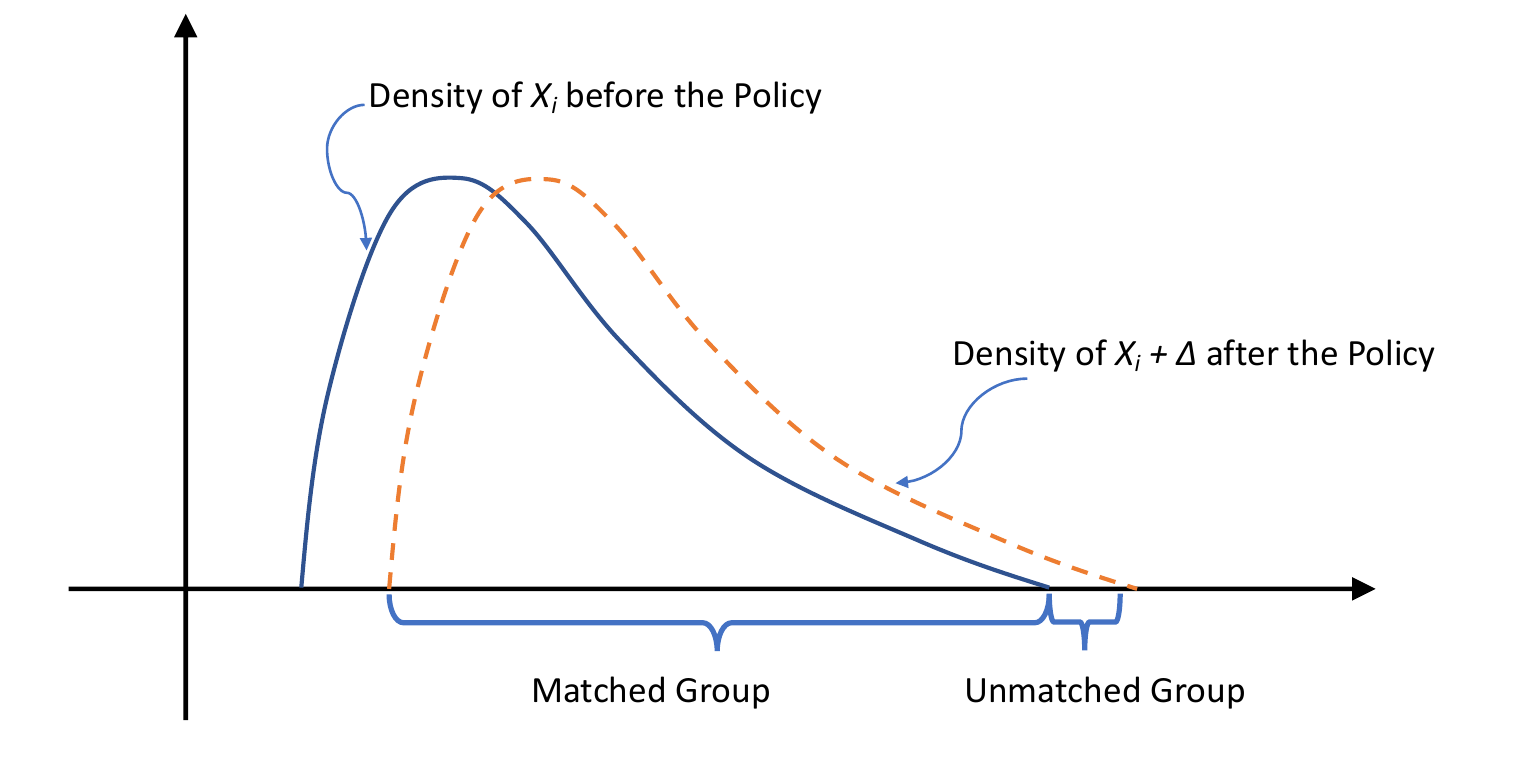}
			\caption{The Pre-Policy and Post-Policy Distributions of $X_i$ in the Target Population: \footnotesize Consider a setting where $Y_i = \mu_0(X_i) + U_i$, $U_i$ and $X_i$ are independent, and the policy shifts $X_i$ to $X_i + \Delta$ for some known vector $\Delta$. Then, the pre-policy outcomes of people in the matched group are compared with the post-policy outcomes of other people in the same population. The matched group is given by $\{x \in \mathcal{X}_0: \mu_0(x + \Delta) = \mu_0(\tilde x) \text{ for some } \tilde x \in \mathcal{X}_0$\}.}
			\label{fig:MatchedGroup}
		\end{center}
	\end{figure}

We assume that the population $P_0$ has not experienced the policy yet. Hence, we observe $(Y_i,X_i)$ before the policy but observe only $X_i$ after the policy. In this paper, we follow the basic ideas of \cite{Todd/Wolpin:06:AER} and \cite{Wolpin:13:Book} and use the outcome-policy relationship on a matched group for counterfactual policy predictions. We let $\mathcal{X}_0$ be the support of $X_i$ in the target population and partition it as follows: 
\begin{align*}
	\mathcal{X}_0 = \mathcal{X}_0^\mathsf{M} \cup \mathcal{X}_0^\mathsf{U},
\end{align*}
where the set $\mathcal{X}_0^\mathsf{M}$, called the \bi{matched group}, is defined as
\begin{align}
	\label{matched group}
	\mathcal{X}_0^\mathsf{M} = \left\{x \in \mathcal{X}_0: \mu_0(x,v_0^*) = \mu_0(\tilde x,v_0) \text{ for some } \tilde x \in \mathcal{X}_0 \right\},
\end{align}
and the set $\mathcal{X}_0^\mathsf{U} = \mathcal{X}_0 \setminus \mathcal{X}_0^\mathsf{M}$ represents the \bi{unmatched group}. Roughly speaking, the set $\mathcal{X}_0^\mathsf{M}$ is the set of values $x$ such that the post-policy value $\mu_0(x,v_0^*)$ matches up with the pre-policy value  $\mu_0(\tilde x,v_0)$ for some $\tilde x \in \mathcal{X}_0$. Figure \ref{fig:MatchedGroup} provides an illustrative example.
We require the identification of the post-policy conditional average outcome only for the matched group.
\begin{assumption}
	\label{assump: support target}
	(i) The matched group $\mathcal{X}_0^\mathsf{M} \subset \mathcal{X}_0$ is nonempty and identified.
    
    (ii) $m_0(\mu_0(x,v_0^*),x)$ is identified for all $x \in \mathcal{X}_0^\mathsf{M}$.
\end{assumption}

Assumption \ref{assump: support target} is a mild condition that is often satisfied in practice. For example, suppose that $\mu_0(x,v_0^*)$ and $\mu_0(x,v_0)$ are identified for all $x$ in the support of $X_i$, due to the use of parametrization in (\ref{parametrization}), and $m_0(\overline \mu,x)$ is identified for all $(\overline \mu,x)$ in the support of $(\mu_0(X_i,v_0),X_i)$. Then, Assumption \ref{assump: support target} is satisfied if the supports of the policy components $\mu_0(X_i,v_0)$ and $\mu_0(X_i,v_0^*)$ overlap.

Note that Assumption \ref{assump: support target} requires $m_0(\mu_0(x,v_0^*),x)$ to be identified only for $x$ in the matched group $\mathcal{X}_0^\mathsf{M}$, not for all $x \in \mathcal{X}_0$. Hence, the assumption is too weak to identify $\theta_0$ when the policy sends the policy component outside of its pre-policy support. To address this challenge, we propose using information from other populations which have already implemented a similar policy.

\subsubsection{Multiple Source Populations}

We assume that there are $K$ regions (e.g., countries, states, markets, etc.). For each region $k=1,...,K$, the outcome $Y_i$ is generated as follows:
\begin{align}
	\label{RF k}
	Y_i = g_k\left(\mu_k(X_i,v_k), X_i, U_i\right),
\end{align} 
where the unit $i$ represents the sample unit (i.e., an individual or a firm) in the region $k$ and the map $g_k$ governs the causal relationship between the outcome variable $Y_i$ and the exogenous variables $(X_i,U_i)$ in region $k$. Note that the structural function, $g_k$, has a subscript $k$ as it varies across regions, reflecting different causal relationships (e.g., due to different institutional details in the population). We call the distribution $P_k$ of $(Y_i,X_i,U_i)$ in region $k$, $k=1,...,K$, a \bi{source population}, and call the previous population $P_0$ the \bi{target population}.  Similarly as before, we define the average response function for the source population $k$ as
\begin{align}
	\label{m_k}
	m_k \left(\overline \mu,x\right) =\int g_k(\overline \mu, x, u)dP_{k,U | X}(u \mid x),
\end{align}
where $P_{k,U | X}$ denotes the conditional distribution of $U_i$ given $X_i$ in population $k$. 

Unlike the target population, each source population has experienced a similar policy so that the conditional average outcome $m_k(\mu_k(x,v_0^*),x)$ is identified for all $x \in \mathcal{X}_0$. 

\begin{assumption}
	\label{assump: support source}
	For all $k=1,...,K$ and all $x \in \mathcal{X}_0$, $m_k\left(\mu_k(x,v_0^*),x\right)$ is identified.
\end{assumption}

We can view Assumption \ref{assump: support source} as an ``eligibility condition'' for any population to serve as a source population for the prediction problem. In many settings, it is not hard to find source populations that satisfy this assumption, especially when the observations are made over multiple periods. To see this, consider a setting where the post-policy components $\mu_k(x,v_0^*)$ are identified, and $m_k(\overline \mu,x)$ is identified for each $(\overline \mu,x)$ on the support of $(\mu_k(X_i,v_0^*),X_i)$ with $X_i$ drawn from the population $k$. Then, Assumption \ref{assump: support source} is satisfied if for all $k=1,2,...,K$,
\begin{align}
	\label{support comp}
	\left\{\mu_k(x,v_0^*): x \in \mathcal{X}_0 \right\} \subset \left\{\mu_k(x,v_0^*): x \in \mathcal{X}_k \right\},
\end{align}
where $\mathcal{X}_k$ denote the support of $X_i$ in the source population $P_k$. In a multi-period setting, we often choose one target period for prediction, while observing each source population over multiple periods. This tends to result in wider variations of $\mu_k(x,v_0^*)$ with $x \in \mathcal{X}_k$ than those of $\mu_k(x,v_0^*)$ with $x \in \mathcal{X}_0$. In a later section of empirical application, we give supporting evidence that this latter condition is satisfied.

\subsection{Synthetic Decomposition}

Our approach uses the maps, $m_k(\mu_k(\cdot,v_0^*),\cdot)$, $k=1,...,K$, as a fixed set of basis functions to construct a synthetic map for the target region. From here on, we call the map $m_k(\mu_k(\cdot,v_0^*),\cdot)$ the \bi{conditional average outcome}. 

Let $\Delta_{K-1}$ denote the $(K-1)$-simplex, i.e., $\Delta_{K-1} = \{\boldsymbol{w} \in \mathbf{R}^K: \sum_k w_k = 1, w_k \geq 0, k=1,...,K\}$. For each $\boldsymbol{w} = (w_1,...,w_K) \in \Delta_{K-1}$, define
 \begin{align}
 	\label{theta w}
 	\theta(\boldsymbol{w}) &= \int_{\mathcal{X}_0^\mathsf{M}} m_0\left(\mu_0(x,v_0^*),x\right)  dP_0(x) + \sum_{k=1}^K \int_{\mathcal{X}_0^\mathsf{U}} m_k\left(\mu_k(x,v_0^*),x\right) w_k dP_0(x).
 \end{align}
The relevance of $\theta(\boldsymbol{w})$ to the prediction in the target population depends on the choice of $\boldsymbol{w}$. In choosing an appropriate $\boldsymbol{w}$, we take an approach inspired by the synthetic control method. The synthetic control method chooses a weight $\boldsymbol{w}$ that exhibits the best pre-treatment fit of the outcomes between the target and source regions. In our setting, the individual outcomes on the matched group plays the role of the pre-treatment outcomes. Just as the synthetic control method requires that the weighted post-treatment outcomes in the donor pool match the counterfactual post-treatment untreated outcomes of the units in the treated region, we introduce a transferability condition that requires the target conditional average outcomes to lie in the convex hull of those in the source regions on the unmatched group. 

First we define the set
\begin{align}
	\label{id w}
   	\mathbb{W}_0 = \argmin_{\boldsymbol{w} \in \Delta_{K-1}} \rho^2(\boldsymbol{w}),
\end{align}
and
\begin{align}
	\label{obj}
	 \rho^2(\boldsymbol{w}) = \int_{\mathcal{X}_0^\mathsf{M}} \left( m_0\left(\mu_0(x,v_0^*),x\right) - \sum_{k=1}^K m_k \left(\mu_k(x,v_0^*),x \right)w_k \right)^2 dP_0(x).
\end{align}
A weight in the set $\mathbb{W}_0$ in (\ref{id w}) brings the synthetic conditional average outcome as close as possible to the target conditional average outcome. The integral in the pseudo-distance $\rho$ is taken only on the matched group. Therefore, the map $\rho$ is identified from data, as both the synthetic and target conditional average outcomes are identified on the matched group. 

\begin{assumption}[Synthetic Transferability] 
	\label{assump: synthetic transferability}
	There exists $\boldsymbol{w} = (w_1,...,w_K) \in \mathbb{W}_0$ such that
\begin{align}
	\label{synth hypothesis}
	m_0\left(\mu_0(x,v_0^*),x\right) =  \sum_{k=1}^K m_k \left(\mu_k(x,v_0^*),x \right)w_{k}, \text{ for all } x \in \mathcal{X}_0.
\end{align}
\end{assumption}
 
The synthetic transferability condition posits that the counterfactual conditional average outcome in the target population can be represented on the unmatched group using a set of conditional average outcomes from source populations as basis functions. Unlike traditional basis functions used in series estimation methods, such as trigonometric or polynomial functions, this approach employs the conditional average outcomes from source populations as basis functions, which are taken to encode transferable causal information between the policy variable and the outcome variable.\footnote{Although it is possible to relax this assumption to an approximate transferability condition — where the approximation error diminishes at a certain rate as the number of source populations increases — we do not explore this relaxation here, as it does not contribute significantly to the innovative aspects of our paper.} The condition generalizes transferability conditions used in the literature in a setting with a single source population. Later, we further relax the assumption so that the weight is allowed to depend on the covariates. In the next section, we provide a detailed discussion on the assumption and its testable implications.

We can reformulate the optimization problem in (\ref{id w}) as a quadratic programming problem. First, define 
\begin{align}
	\label{m}
	\boldsymbol{m}(x) = \left[m_1\left(\mu_1(x,v_0^*),x\right),...,m_K\left(\mu_K(x,v_0^*),x\right)\right]^{\prime}.
\end{align}
Then, we can rewrite
\begin{align}
	\label{characterization}
	\mathbb{W}_0 = \argmin_{\boldsymbol{w} \in \Delta_{K-1}} \enspace \boldsymbol{w}' H \boldsymbol{w} - 2 \boldsymbol{w}'\boldsymbol{h},
\end{align}
where, with $\boldsymbol{m}(x)$ defined in (\ref{m}), $H$ and $\boldsymbol{h}$ are given by
\begin{align*}
	H &= \int_{\mathcal{X}_0^{\mathsf{M}}} \boldsymbol{m}(x) \boldsymbol{m}(x)' dP_0(x) \text{ and }\\
	\boldsymbol{h} &=  \int_{\mathcal{X}_{0}^\mathsf{M}} \boldsymbol{m}(x) m_0\left(\mu_0(x,v_0^*),x\right) d P_0(x).
\end{align*}
The theorem below shows that the parameter $\theta_0$ is partially identified without the invertibility of $H$, but with invertibility, it is point-identified.

\begin{theorem}
    \label{thm: identification}
	Suppose that Assumptions \ref{assump: support target}-\ref{assump: synthetic transferability} hold. Then, the identified set for $\theta_0$ is given by 
    \begin{align*}
        \left\{\theta(\boldsymbol{w}): \boldsymbol{w} \in \mathbb{W}_0 \right\}.
    \end{align*} 
    In particular, when $H$ is invertible, $\mathbb{W}_0 = \{\boldsymbol{w}_0\}$ for some $\boldsymbol{w}_0 \in \Delta_{K-1}$, and $\theta_0 = \theta(\boldsymbol{w}_0)$.
\end{theorem}

\noindent \textbf{Proof: } By Assumption \ref{assump: synthetic transferability}, we have $\theta_0 = \theta(\boldsymbol{w}_0)$ for some $\boldsymbol{w}_0 \in \mathbb{W}_0$. By Assumptions \ref{assump: support target}-\ref{assump: support source}, the map $\theta: \Delta_{K-1} \rightarrow \mathbf{R}$ is identified. Therefore, the identified set for $\theta_0$ is $\{\theta(\boldsymbol{w}): \boldsymbol{w} \in \mathbb{W}_0\}$. As for the second statement, since $\Delta_{K-1}$ is compact and $\rho^2(\boldsymbol{w})$ is continuous and strictly convex in $\boldsymbol{w}$, $\rho^2$ has a unique minimizer in $\Delta_{K-1}$. However, $\rho^2(\boldsymbol{w}_0) = 0$. Hence $\mathbb{W}_0 = \{\boldsymbol{w}_0\}$. By Assumption \ref{assump: support target}, the map $\rho$ is identified, so are $\boldsymbol{w}_0$ and $\theta_0 = \theta(\boldsymbol{w}_0)$. $\blacksquare$\medskip

When the synthetic transferability condition fails, the prediction $\theta(\boldsymbol{w}_0)$ is still derived from the weighted average of the outcome-policy relationships which is chosen to be \textit{as close as possible to meeting the synthetic transferability condition}, based on their predictive performance on the pre-policy support of $X_i$ in the target population. Naturally, those source populations with the outcome-policy relationships more similar to that of the target population on the matched group receive a higher weight by design.

\subsection{Discussion on the Synthetic Transferability Condition}

Let us now discuss how Assumption \ref{assump: synthetic transferability} compares to other often-invoked transferability conditions in the literature. In fact, most transferability conditions in the literature can be viewed as the special case of the synthetic transferability condition with a single source region (i.e., $K=1$). More specifically, when $K=1$ (i.e., a single source region), Assumption \ref{assump: synthetic transferability} collapses to:
\begin{align}
	\label{transferability}
	m_1\left(\mu_1(x,v_0^*),x\right) = m_0\left(\mu_0(x,v_0^*),x\right), \quad x \in \mathcal{X}_0.
\end{align}
The condition (\ref{transferability}) says that the conditional average outcome in the source population is identical to that in the target population. This condition is stronger than the synthetic transferability condition, as the latter condition allows the conditional average outcome in the target population to be different from any of those in the source populations. The condition (\ref{transferability}) is explicitly used in various settings of extrapolating results from a single source region to a target region as we discuss below.

\subsubsection{Comparison with Decomposition Methods}

Decomposition methods are often used in the labor economics literature to compare outcomes before and after the policy - e.g., \cite{Oaxaca:73:IER} and \cite{Blinder:73:JHR}\footnote{Vitor Possebom kindly let us know that there was an early appearance of a similar idea in \cite{Kitagawa:55:JASA}.}, and extensions such as \cite{Juhn/Murphy/Pierce:93:JPE} and \cite{DiNardo/Fortin/Lemieux:96:Eca}.  The causal interpretation of the decomposition method was studied by \cite{Kline:11:AERPP}. See also \cite{Fortin/Lemieux/Firpo:11:Handbook} for an extensive review of this literature. However, the transfer of results from a source population to a target population requires various forms of transferability conditions (see \cite{Chernozhukov/Fernandez/Melly:13:Eca}, \cite{Rothe:10:JoE}, \cite{Ao/Calonico/Lee:21:JBES}, and \cite{Hsu/Lai/Lieli:22:JBES}.)

To see the connection between Assumption \ref{assump: synthetic transferability} and the decomposition method, consider a setting where at time $t_0$, the policy is not implemented and $Y_i$ is generated as follows: 
\begin{align*}
	Y_i = g_0\left(\mu_0(X_i,v_0),X_i,U_i\right), \text{ with } X_i \sim P_{0,X},
\end{align*}
and at time $t_1 > t_0$, the policy is implemented and $Y_i$ is generated as follows: 
\begin{align*}
	Y_i = g_1\left(\mu_1(X_i,v_0^*),X_i,U_i\right), \text{ with } X_i \sim P_{1,X}.
\end{align*}
Then, the difference between the expected outcomes at times $t_1$ and $t_0$ is given by
\begin{align}
	\label{decomposition}
	\mathbf{E}_1\left[Y_i\right] - \mathbf{E}_0 \left[Y_i\right] &= \int m_1\left(\mu_1(x,v_0^*),x \right) \left(dP_{1,X}(x) - dP_{0,X}(x)\right) \\ \notag
	&\quad \quad + \int \left(m_1\left(\mu_1(x,v_0^*),x\right) - m_0(\mu_0(x,v_0),x)\right) dP_{0,X}(x).
\end{align}
Thus, the change in the mean of $Y_i$ before and after the policy is decomposed into the component due to the change in the distribution of $X_i$ and the component due to the change in the conditional average outcome. If the change of the conditional average outcome given $X_i$ between times $t_0$ and $t_1$ is only due to the policy, i.e., the condition (\ref{transferability}) holds, the second term on the right hand side of (\ref{decomposition}) can be interpreted as an average causal effect of the policy in the target population: 
\begin{align*}
	\int \left(m_0\left(\mu_0(x,v_0^*),x\right) - m_0\left(\mu_0(x,v_0),x\right)\right) dP_{0,X}(x).
\end{align*}
Thus, the causal interpretation of the decomposition method follows from the transferability condition (\ref{transferability}).

\subsubsection{Comparison with Transferability Conditions in Extrapolating Experimental Results}

A growing attention has been paid to the issue of external validity in field experiments, such as when the results obtained from experiments are not replicated in their scaled-up implementation. (See \cite{Allcott:15:QJE}, \cite{Bold_et_al:18:JPubE}, and \cite{Wang/Yang:21:NBER} and references therein. See also \cite{Duflo:04AWBC} and \cite{Muralidharan/Niehaus:17:JEP} for the review of these issues and the literature.)

The transferability condition with a single source region (\ref{transferability}) is related to the conditional external validity conditions used in this literature (see \cite{Hotz/Imbens/Mortimer:05:JOE}, \cite{Hartman_et_al:15:JRSS}, \cite{Athey_et_al:20:WP}, and \cite{Gui:22:WP}). To see the connection, let us write the outcomes before and after the policy as the potential outcomes 
\begin{align*}
   Y_i(0) = g_0\left(\mu_0(X_i,v_0),X_i,U_i\right) \text{ and } Y_i(1) = g_0\left(\mu_0(X_i,v_0^*),X_i,U_i\right).
\end{align*}
As for the source population, we assume that the experiment is already conducted, and the outcomes are generated as follows:
\begin{align*}
	Y_i(1) = g_1\left(\mu_1(X_i,v_0^*),X_i,U_i\right).
\end{align*}
Following \cite{Hotz/Imbens/Mortimer:05:JOE}, we define $D_i = 1$ if the individual unit belongs to the source population $k=1$ and $0$ if it belongs to the target population $k=0$. Then, part of the mean version of the location unconfoundedness condition in \cite{Hotz/Imbens/Mortimer:05:JOE} is written as 
\begin{align}
	\label{location unconfoundedness}
	\mathbf{E}[Y_i(1) \mid D_i = 1, X_i] = \mathbf{E}[Y_i(1) \mid D_i = 0, X_i],
\end{align}
(see e.g. \cite{Hartman_et_al:15:JRSS}, \cite{Athey_et_al:20:WP} and \cite{Gechter/Meager:22:WP} for variants of this condition). Since we can write 
\begin{align*}
	\mathbf{E}[Y_i(1) \mid D_i = 1, X_i] &= m_1(\mu_1(X_i,v_0^*),X_i) \text{ and }\\
	\mathbf{E}[Y_i(1) \mid D_i = 0, X_i] &= m_0(\mu_0(X_i,v_0^*),X_i),
\end{align*}
the condition (\ref{location unconfoundedness}) corresponds to the transferability condition (\ref{transferability}) with a single source region.

\subsubsection{A Placebo Test for Synthetic Transferability}
\label{subsubsec: testable implications} We may consider using a placebo test. For this, we take an alternative policy $\overline v_{0}$ that is different from the target policy $v_0^*$, such that the policy component $\mu_0(x,\overline v_0)$ stays within the support of the pre-policy data in the target population. Let us define our new target parameter as follows: 
\begin{align*}
	\overline \theta_0 = \mathbf{E}_0\left[ g_0\left(\mu_0(X_i,\overline v_0), X_i, U_i \right) \right].
\end{align*}
 In this case, we have two different ways to identify $\overline \theta_0$. The first way is to identify $\overline \theta_0$ using observations only from the target population without invoking the synthetic transferability condition: 
 \begin{align}
	\label{expression1}
	\overline \theta_0 = \int m_0\left(\mu_0(x,\overline v_0),x\right) dP_0(x).
 \end{align}
 The second way is to identify $\overline \theta_0$ using the synthetic transferability condition and data from the source populations. In this case, we identify $\overline \theta_0$ as follows:
 \begin{align}
	\label{expression2}
	\overline \theta_0 = \int_{\mathcal{\overline X}_0^\mathsf{M}} m_0\left(\mu_0(x,\overline v_0),x\right)  dP_0(x) + \sum_{k=1}^K \int_{\mathcal{\overline X}_0^\mathsf{U}} m_k\left(\mu_k(x,\overline v_0),x\right) \overline w_{0,k} dP_0(x),
\end{align}
where $(\overline w_{0,1},...,\overline w_{0,K})$ and $\mathcal{\overline X}_0^\mathsf{M}$ are defined in the same way as $(w_{0,1},...,w_{0,K})$ and $\mathcal{X}_0^\mathsf{M}$ except that $v_0^*$ is replaced by $\overline v_0$. Under the synthetic transferability condition at the ``new'' policy $\overline v_0$, both identifications of $\overline \theta_0$ should work. Hence, we can test whether the two estimators of $\overline \theta_0$ (according to (\ref{expression1}) and (\ref{expression2})) are different with statistical significance. If they are different, this suggests evidence against the synthetic transferability condition. While this is not a direct test of synthetic transferability for the original policy $v_0^*$, it provides a useful diagnostic tool for the plausibility of the method.

\subsection{Examples}

\subsubsection{Minimum Wages and Labor Supply}\label{sec: example_minwage} Minimum wages are among the most widely debated labor market policies. When studying the effects of a counterfactual raise in a minimum wage on employment, rather than past raises, the literature often uses search-and-bargaining models (e.g., \cite{Flinn:06:Ecma, Ahn/Arcidiacono/Wessels:11:JBES, Flinn/Mullins:15:IER}). In Section \ref{sec: empirical app}, we carefully rewrite the model of \cite{Ahn/Arcidiacono/Wessels:11:JBES} to fit our framework. We give a brief overview here. 

Let $Y_{i,j} \in \{0,1\}$ denote the employment status of worker $i$ after a match with firm $j$, $X_{i}$ as worker $i$'s observable characteristics (age, race, etc.), $\underline{W}_k$ as the prevailing minimum wage in region $k$ that worker $i$ is subject to, and $U_{i,j}$ as a match-specific unobservable (shocks, unobserved types) drawn from a CDF $F_k$. As we explain in Section \ref{sec: empirical app}, the wage generation in \cite{Ahn/Arcidiacono/Wessels:11:JBES} can be written as:
\begin{align}
	\label{wage_ex}
	W_{i,j} = \max\{\delta_k M_{i,j}, \underline W_k\},
\end{align}
where $\delta_k \in (0,1)$ is a parameter that represents the worker $i$'s bargaining strength in region $k$, $M_{i,j}$ is the match productivity drawn for worker $i$ with firm $j$. Following \cite{Ahn/Arcidiacono/Wessels:11:JBES}, we parameterize the generation of $M_{i,j}$ as follows: 
\begin{align*}
    \log M_{i,j} = X_i^{\prime} \gamma_k + U_{i,j},
\end{align*}
where $U_{i,j}$ is independent of $X_i$. The employment indicator, $Y_{i,j}$, equals one if the match surplus is higher than the wage:
\begin{align}
	Y_{i,j} = 1\left\{M_{i,j} \ge W_{i,j}\right\} = 1\left\{M_{i,j} \ge \underline W_k\right\} = 1\left\{X_i^{\prime} \gamma_k + U_{i,j} \ge \log(\underline W_k)\right\}.	\label{employment_ex}
\end{align} 

Now, suppose that the minimum wage in Texas increases from $\underline W_0$ to $\underline W_0^*$ and we want to predict its effects on employment. Then, Texas is taken to be the target region $0$. In order to apply the synthetic decomposition method, we first set the policy component for each source region $k$ as 
\begin{align*}
    \mu_k(X_i,\underline W_k) = X_i^{\prime} \gamma_k - \log(\underline W_k).
\end{align*}
Hence, it follows from (\ref{employment_ex}) that:
\begin{align*}
g_k(\overline \mu,x,u) = 1\{\overline \mu +u \ge 0\}.
\end{align*}
The average response function $m_k$ for state $k$ is identified as $m_k(\overline \mu,x) = 1 - F_k(-\overline \mu)$. The function $m_k$ is identified as the share of workers in state $k$ whose productivity is higher than the minimum wage in state $k$. As we explain in Section \ref{sec: empirical app}, the policy component parameter $\gamma_k$ is identified from a semiparametric censored regression of wages: 
\begin{align*}
    W_{i,j} = \max\left\{\log \delta_k + X_i^{\prime} \gamma_k + U_{i,j}, \underline W_k\right\}.
\end{align*}
Then, we can identify the conditional average outcome as 
\begin{align*}
	m_k(\mu_k(x,\underline W_0^*),x) = \mathbf{E}\left[Y_{i,j(i)} \mid \mu_k(X_i,\underline W_0^*) = \mu_k(x,\underline W_0^*)\right],
\end{align*}
where $j(i)$ denotes the firm matched with worker $i$. Note that we do not need to parametrize the distribution of $U_{i,j}$.

Now, let us check the plausibility of Assumptions \ref{assump: support target} and \ref{assump: support source}. First, the matched group $\mathcal{X}_0^\mathsf{M}$ takes the following form: 
\begin{align}\label{support_emp}
	\mathcal{X}_0^\mathsf{M} = \left\{x \in \mathcal{X}_0: x^{\prime} \gamma_0 - \log \underline W_0^* = \tilde x^{\prime} \gamma_0 -\log \underline W_0, \text{ for some } \tilde x \in \mathcal{X}_0 \right\},
\end{align}
where we denote the support of $X_i$ in the target population by $\mathcal{X}_0$. Assumption \ref{assump: support target} requires that this set is not empty. This condition is often satisfied in practice as long as the new minimum wage is not too far away from the old one.

For source populations $k=1,2,...,K$, we observe the post-policy outcomes, and hence the map $\mathbf{E}[Y_{i,j(i)} \mid \mu_k(X_i,\underline W_0^*) = \cdot]$, for individual $i$ in region $k$, is identified on the support of $\mu_k(X_i,\underline W_0^*)$, $X_i \in \mathcal{X}_0$. Therefore, Assumption \ref{assump: support source} is satisfied if, for all $k=1,...,K$, 
\begin{align*}
	\left\{ x'\gamma_k: x \in \mathcal{X}_0 \right\} \subset \left\{ x'\gamma_k: x \in \mathcal{X}_k \right\}.
\end{align*}
In practice, it is not hard to find source populations that satisfy this condition especially when the observations are made over multiple periods.

\subsubsection{Tax Policy and Immigration} Changes to income tax rates may immediately affect tax revenue, but they may also change the composition of the population. For instance, high earners may choose to emigrate when facing higher taxes. This matters for welfare, as such high earners are highly mobile and pay a large share of taxes.\footnote{In 2016, the top 1\% of households in the U.S. earned 16\% of the total income while paying 25\% of all federal taxes. However, their income, accumulated wealth and favorable immigration policies permit straightforward changes to residence status, making them very responsive to tax policy. See \url{https://doc-research.org/2019/01/global-mobility-wealthy-push-pull-factors/} for a policy overview.} (See \cite{Scheuer/Werning:QJE:17} for a theoretical investigation and \cite{Moretti/Wilson:17:AER}, \cite{Akcigit/Baslandze/Stantcheva:16:AER}; \cite{Kleven/Landais/Saez:13:AER} and \cite{Kleven/Landais/Saez/Schultz:14:QJE} for evidence on the effects of \textit{past} changes to tax policies, including Danish and Spanish reforms.)

To evaluate the effects of a decrease in tax rates in country $0$ (e.g., U.K.) on high earners' immigration, we could follow \cite{Kleven/Landais/Saez:13:AER} and model this as a discrete choice problem. A high earner $i$'s preference, $V_{i,k}$, for living in country $k$ depends on the average tax rate $\tau_k$ on the wage $W_i$ the individual would face, and is specified as follows:
\begin{eqnarray}
V_{i,k} = \alpha \log(1-\tau_{k}) + \alpha \log(W_i) + Z_i^{\prime}\beta_k + U_{i,k}.
\end{eqnarray} 
The first two terms represent the (concave) preferences over net-of-tax wages, $Z_i^{\prime}\beta_k$ captures heterogeneity of worker preferences for each country (which may depend on age, nationality, etc.), with $Z_i$ denoting the observed characteristics of the individual $i$, and $U_{i,k}$ represents unobserved heterogeneity that is independent of $(W_i,Z_i)$ and i.i.d.\ across individuals and regions. (Note that for high earners, the average tax rate is approximately equal to the marginal tax rate which is the same across all the high earners.) Then, individual $i$'s decision to live in region $k$ is represented by a binary indicator $Y_{i}$ as follows:
\begin{eqnarray}
Y_{i} = 1\left\{V_{i,k} > \max_{j \neq k} V_{i,j}\right\}.
\end{eqnarray}

To apply the synthetic decomposition method, we take the policy component for the individual as: for each $k=0,1,...,K,$ and for each individual $i$ in region $k$ with $X_i = (W_i,Z_i)$,
\begin{align*}
    \mu_k(X_i,\tau_k) = \alpha \log(1-\tau_{k}) + \alpha \log(W_i) + Z_i^{\prime}\beta_k.
\end{align*}
The target country is the U.K., and the policy of interest is lowering tax rates in the U.K. so that 
\begin{align*}
    \mu_{0}(X_i,\tau_0^*) = \alpha \log(1-\tau_{0}^{*}) + \alpha \log(W_i) + Z_i^{\prime}\beta_0,
\end{align*}
for $\tau_0^* < \tau_0$. Again, we find that the conditional average outcome for region $k$ is identified as
\begin{align*}
	m_k\left(\mu_k(x,\tau_0^*),x\right) = h_k\left(\mu_k(x,\tau_0^*),\mu_{-k}(x,\tau_{-k})\right),
\end{align*}
where 
\begin{align*}
	h_k\left(\mu_k(x,\tau_0^*),\mu_{-k}(x,\tau_{-k})\right) = \mathbf{E}_k\left[Y_i \mid \mu_k(X_i,\tau_0^*) = \mu_k(x,\tau_0^*), \mu_{-k}(X_i,\tau_{-k}) = \mu_{-k}(x,\tau_{-k})\right],
\end{align*}
and $\mathbf{E}_k$ denotes the conditional expectation in population $k$ and $\mu_{-k}(x,\tau_{-k}) = (\mu_j(x,\tau_{j}))_{j \ne k}$. The plausibility of Assumptions \ref{assump: support target} and \ref{assump: support source} can be evaluated in a manner similar to the previous example.

\subsection{Extensions}
\subsubsection{Covariate-Dependent Weights}

The synthetic transferability condition assumes that the weights are the same across different demographic groups. This may be restrictive in some applications with the same population. For example, suppose that we have two source regions $1$ and $2$, where a high education group in region 1 is matched better with a high education group in the target region than region 2, whereas a low education group in region 2 is matched better with a low education group in the target region than in region 1. Our approach can accommodate this situation flexibly by allowing the weight to depend on the education indicator.

Suppose that the outcome $Y_i$ is generated as in (\ref{RF 0}) and (\ref{RF k}) for each individual $i$. As before, the policy alters $v_0$ to $v_0^*$ in the target population, and the parameter of interest $\theta_0$ is defined as in (\ref{theta_00}). The average response functions, $m_k$, $k=1,...,K$, are also given as in (\ref{m_0}) and (\ref{m_k}). Suppose that $X_i = (X_{i,1},X_{i,2})$, where for each population $k=0,1,...,K$, the random vectors $X_{i,1}$ and $X_{i,2}$ have $\mathcal{X}_{k,1}$ and $\mathcal{X}_{k,2}$ as their supports respectively. We let $\mathcal{X}_0$ denote the support of $X_i$ in the target population as before. 

For the identification of $\theta_0$, we require the identification of a matched group and the conditional average outcomes. For each $x_2 \in \mathcal{X}_{0,2}$, define 
\begin{align*}
	\mathcal{X}_{0,1}(x_2) = \left\{x_1 \in \mathcal{X}_{0,1}: (x_1,x_2) \in \mathcal{X}_0\right\},
\end{align*}
which is the $x_2$-section of the support $\mathcal{X}_0$. We define the conditional matched group as follows: for each $x_2 \in \mathcal{X}_{0,2}$,
\begin{align*}
	\mathcal{X}_{0,1}^\mathsf{M}(x_2) &= \left\{x_1 \in \mathcal{X}_{0,1}(x_2): \mu_0(x_1,x_2,v_0^*) = \mu_0(\tilde x_1,x_2,v_0), \text{ for some } \tilde x_1 \in \mathcal{X}_{0,1}(x_2) \right\} \text{ and }\\
	\mathcal{\tilde X}_0^\mathsf{M} &= \left\{(x_1,x_2) \in \mathcal{X}_0: x_2 \in \mathcal{X}_{0,2} \text{ and } x_1 \in \mathcal{X}_{0,1}^\mathsf{M}(x_2) \right\}.
\end{align*}
Now, we maintain Assumption \ref{assump: support source} but modify Assumption \ref{assump: support target} as follows. 
\begin{assumption}
	\label{assump: support target cond}
	(i) $\mathcal{\tilde X}_0^\mathsf{M}$ is nonempty and identified. 
	
	(ii) For each $x \in \mathcal{\tilde X}_0^\mathsf{M}$, $m_0(\mu_0(x,v_0^*),x)$ is identified.
\end{assumption}
In general, we have $\mathcal{\tilde X}_{0}^\mathsf{M} \subset \mathcal{X}_0^\mathsf{M}$, where the set $\mathcal{X}_0^\mathsf{M}$ is defined in (\ref{matched group}). Hence, if $\mu(x,v_0^*)$ and $\mu(x,v_0)$ are identified for each $x \in \mathcal{X}_0$, Assumption \ref{assump: support target cond}(i) is stronger than Assumption \ref{assump: support target}(i), while Assumption \ref{assump: support target cond}(ii) is weaker than Assumption \ref{assump: support target}(ii). 

For example, suppose that $\mu_0(x,v_0) = x_1 \beta_1 + x_2 \beta_2 - v_0$. Then, we have 
\begin{align*}
	\mathcal{X}_0^\mathsf{M} &= \left\{x \in \mathcal{X}_0: x_1 \beta_1 + x_2 \beta_2 - v_0^* = \tilde x_1 \beta_1 + \tilde x_2 \beta_2 - v_0, \text{ for some } (\tilde x_1,\tilde x_2) \in \mathcal{X}_{0} \right\}, \text{ and }\\
	\mathcal{\tilde X}_0^\mathsf{M} &= \left\{x \in \mathcal{X}_0: x_1 \beta_1 - v_0^* = \tilde x_1 \beta_1 - v_0, \text{ for some } \tilde x_1 \in \mathcal{X}_{0,1}(x_2) \right\}.
\end{align*}
\begin{figure}[t]
	\begin{center}
		\includegraphics[scale=0.45]{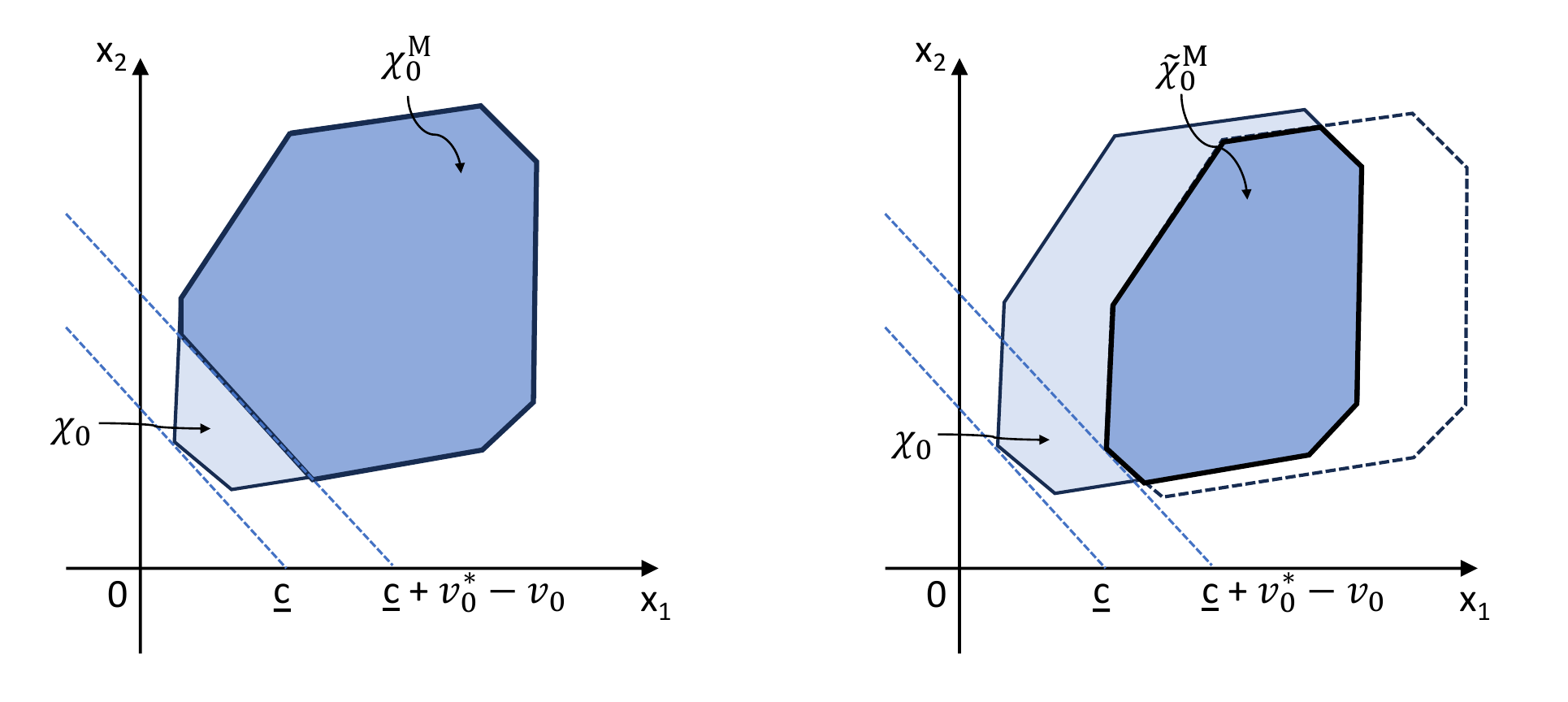}
		\caption{The Illustration of the Matched Group in a Setting with Covariate-Dependent Weights: \footnotesize We consider a setting with the policy component taking the form: $\mu_0(x_1,x_2,v_0) = x_1 \beta_1 + x_2 \beta_2 - v_0$ with $\beta_1 = \beta_2 = 1$, and $v_0^* > v_0$. The support of $X_i = (X_{i,1},X_{i,2})$ is depicted as the area with a light shade on both panels. The matched group $\mathcal{X}_0^\mathsf{M}$ in the original setting is depicted as a darker polygon on the left panel, whereas the matched group $\mathcal{\tilde X}_0^\mathsf{M}$ for the setting with the weights allowed to depend on $X_{i,2}$ is depicted as a darker polygon on the right panel. It is clearly seen that $\mathcal{X}_0^\mathsf{M}$ contains $\mathcal{\tilde X}_0^\mathsf{M}$. Both matched groups can be empty if $v_0^*$ is too far away from $v_0$.}
		\label{fig:OverlapCond}
	\end{center}
\end{figure}
(See Figure \ref{fig:OverlapCond} for an illustration of the matched group in this example.)

Similarly as above, the synthetic prediction takes the following form: for each map $\boldsymbol{w}:\mathcal{X}_{0,2} \rightarrow \Delta_{K-1}$, 
\begin{align*}
	\tilde \theta(\boldsymbol{w}) &= \int_{\mathcal{\tilde X}_0^\mathsf{M}} m_0\left(\mu_0(x,v_0^*),x\right)  dP_0(x) + \sum_{k=1}^K \int_{\mathcal{\tilde X}_0^\mathsf{U}} m_k\left(\mu_k(x,v_0^*),x\right) w_{k}(x_2) dP_0(x),
\end{align*}
where $\mathcal{\tilde X}_0^\mathsf{U} = \mathcal{\tilde X}_0 \setminus \mathcal{\tilde X}_0^\mathsf{M}$ denotes the unmatched group.

To motivate the choice of the weight, we define the goodness-of-fit on the matched group:
\begin{align}
	\label{obj2}
	\tilde \rho^2(\boldsymbol{w}) = \int_{\mathcal{\tilde X}_0^\mathsf{M}} \left( m_0\left(\mu_0(x,v_0^*),x\right) - \sum_{k=1}^K m_k\left(\mu_k(x,v_0^*),x\right) w_k(x_2) \right)^2 dP_0(x),
\end{align} 
and let 
\begin{align}
	\label{obj3}
	\mathbb{W}_0 = \argmin_{\boldsymbol{w}:\mathcal{X}_{0,2} \rightarrow \Delta_{K-1}} \tilde \rho^2(\boldsymbol{w}).
\end{align} 

We consider the following form of synthetic transferability: for all $x = (x_1,x_2) \in \mathcal{X}_{0}$, 
\begin{align*}
	m_0(\mu_0(x,v_0^*),x) = \sum_{k=1}^K m_k(\mu_k(x,v_0^*),x)w_k(x_2),
\end{align*}
for some $\boldsymbol{w} \in \mathbb{W}_0$. This condition is weaker than the previous synthetic transferability condition because the weight given to each source region $k$ can vary across different people in the target region depending on the value of their covariate $x_2$.

Then, we obtain the set of counterfactual predictions for the target region $0$ as 
\begin{align*}
	\left\{ \tilde \theta(\boldsymbol{w}): \boldsymbol{w} \in \mathbb{W}_0 \right\}.
\end{align*}
To compare this with one without the covariate-dependency of weights, suppose that $\mathcal{\tilde X}_0^\mathsf{M} = \mathcal{X}_0^\mathsf{M}$. The quantity $\tilde \rho^2(\mathbf{\tilde w}_0)$ is still smaller than that in (\ref{obj}), because the domain of the minimizers is larger. This means that the covariate-dependent weight will exhibit a better fit than the previous weights. 

As before, we can reformulate the optimization problem (\ref{obj3}) as follows: for each $x_2 \in \mathcal{X}_{0,2}$,
\begin{align*}
	\mathbb{W}_0(x_2) = \argmin_{\boldsymbol{w} \in \Delta_{K-1}}  \enspace \boldsymbol{w}' H(x_2) \boldsymbol{w} -  2 \boldsymbol{w}'\boldsymbol{h}(x_2),
\end{align*}
where 
\begin{align*}
	H(x_2) &= \int_{\mathcal{\tilde X}_{0}^\mathsf{M}} \boldsymbol{m}(x) \boldsymbol{m}(x)' dP_{0}(x \mid x_2)\\
	\boldsymbol{h}(x_2) &= \int_{\mathcal{\tilde X}_{0}^\mathsf{M}} \boldsymbol{m}(x) m_0(\mu_0(x,v_0^*),x) dP_{0}(x \mid x_2),
\end{align*} 
and $P_0( \cdot \mid x_2)$ denotes the conditional distribution of $X_i$ given $X_{i,2} = x$. Thus, when $\mathcal{X}_{0,2}$ is a small, finite set, the finite dimensional quadratic programming gives a fast solution.

\subsubsection{Spillover of Policy Effects Across Regions}

A policy in one region can often have a spillover effect on other regions. For example, the immigration of high-skilled workers in response to a change in tax policy in a target region (as in \cite{Kleven/Landais/Saez:13:AER}) would affect the number of immigrants in source regions.  We show that such a situation with spillover effects can be accommodated in our proposal. 

Consider two types of spillover effects. The first is the spillover effect of past policies from the source regions on the target region. The spillover effect is already reflected in the data when the policymaker considers implementing a new policy on the target population. For instance, the source countries with lower taxes have already received high-skilled immigrants from the target region. If the spillover effect is entirely mediated through the variations in $X_i$, its presence does not alter anything in our proposal because it is among the many sources of exogenous variations in $X_i$. If the spillover effect creates correlation between $X_i$ and $U_i$, we need to carefully search for an identification method using instrumental variables or resorting to a control function approach (e.g., \cite{Blundell/Matzkin:14:QE}) to satisfy Assumption \ref{assump: support source}.

The second spillover effect is that of the new policy in the target population on other regions. This is a spillover effect that is not yet reflected in the data and is part of the policy's effect on the target population. For example, a decrease in tax rates in the target region (say, the U.K.) would induce immigration away from source regions (e.g., Spain). Our definition of the pre-policy population will then be the population that consists of people before the migration induced by the policy, and likewise the post-policy population will be the population that consists of people after the migration. Therefore, the policy effect, according to our definition, includes both the effect on the people who do not migrate as a consequence and the composition effect that arises due to the migration.\footnote{For example, suppose that the policy not only changes $v_0$ into $v_0^*$, but also alters the distribution $P_0$ into $P_0 \circ f^{-1}$ for some map $f$. The latter change corresponds to changing $X_i$ into $f(X_i)$. Now, the post-policy prediction includes both the effects, so that we can take
\begin{align*}
	\theta_0 &= \int_{f(\mathcal{X}_0)}  m_0\left(\mu_0(x,v_0^*),x\right) d(P_0 \circ f^{-1})(x) = \int_{\mathcal{X}_0}  m_0\left(\mu_0(f(x),v_0^*),f(x)\right) dP_0(x).
\end{align*}
Hence, by redefining the policy, we can study the effect of a policy that has a spillover effect through migration. However, in contrast to the previous situations, we may need to estimate the ``policy'' as it includes its composition effect through migration from the target region.}

\subsubsection{Synthetic Decomposition with Multi-Period Observations}
\label{subsec: aggregate shocks}

In many empirical applications, we observe individuals over multiple time periods (either in panel data or in repeated cross-sectional data). Here we overview how our method applies to this multi-period setting. 

For each $k =0,1,...,K$ and $t=1,...,T$, we let $N_{k,t}$ denote the set of the sample units in the region $k$ in period $t$. Let $n_{k,t} = |N_{k,t}|$ be the size of the sample from region $k$ in period $t$. Consider the generation of outcomes for populations $k=0,1,...,K$ over multiple periods $t=1,...,T$:
\begin{align*}
	Y_{i,t} = g_k\left(\mu_{k}(X_{i,t},v_{k,t}),X_{i,t},U_{i,t}\right), \quad i \in N_{k,t},
\end{align*}
where $\mu_k(X_{i,t},v_{k,t}) = \mu(X_{i,t},v_{k,t};\beta_k)$ for a parametric function $\mu(\cdot,\cdot;\beta_k)$. Here $X_{i,t}$ denotes a vector of individual covariates and $v_{k,t}$ the vector of time-varying observed aggregate variable for population $k$. We assume that there are not many time periods in the sample, and hence, any aggregate observed variables are regarded as ``observed constants.'' 

We are interested in the average outcome of $Y_{i,T}$ at time $T$ in the target population $0$ after a new policy that sets the vector $v_{0,T}$ to $v_{0,T}^*$. Thus our target parameter is
\begin{align*}
	\theta_0 = \mathbf{E}_0\left[g_0\left(\mu_{0}\left(X_{i,T},v_{0,T}^*\right),X_{i,T},U_{i,T}\right)\right].
\end{align*}
The quantity $\theta_{0}$ represents the average outcome when the policy changes the variable $v_{0,T}$ into $v_{0,T}^*$. While we observe $X_{i,T}$ for the target region, we do not observe the outcome $Y_{i,T}$ after the policy. This lack of observation presents the primary challenge in identifying $\theta_0$.\footnote{When we do not observe $X_{i,T}$, the researcher can choose a reference distribution of covariates to define the target parameter $\theta_0$. For example, one can use the distribution of the covariates for the entire years or just the most recent year in the sample. Alternatively, one can choose some predicted distribution of the covariates in the target year. Extending the method to such settings is straightforward, and we do not discuss it further here.}

We introduce the following assumption.

\begin{assumption}
	For each $k=0,1,...,K$, the conditional distribution of $U_{i,t}$ given $X_{i,t} = x$ is identical across $i \in N_{k,t}$ and time-invariant for all $x$ in the support of $X_{i,t}$.
\end{assumption}

While this assumption is not innocuous, it does not seem overly restrictive in practice because we already accommodate the observed time-varying aggregate variables $v_{k,t}$. Note that the time-invariance of the conditional distribution of $U_{i,t}$ given $X_{i,t} = x$ within-group is also used in \cite{Athey/Imbens:06:ECMA} (see Assumption 3.3 there).

Let us introduce conditions corresponding to Assumptions \ref{assump: support target}, \ref{assump: support source}, and \ref{assump: synthetic transferability}. We define the matched group in the target period $T$ as 
\begin{align*}
	\mathcal{X}_{0,T}^\mathsf{M} = \left\{ x \in \mathcal{X}_{0,T}: \mu_{0}(x,v_{0,T}^*) = \mu_{0}(\tilde x,v_{0,T}) \text{ for some } \tilde x \in \mathcal{X}_{0,T} \right\},
\end{align*}
where $\mathcal{X}_{0,T}$ denotes the support of $X_{i,T}$ in the target population. Similarly as before, we introduce the average response function for region $k=0,1,...,K$ as follows: 
\begin{align*}
	m_{k}(\overline \mu,x) =  \int g_k\left(\overline \mu,x,u\right)dP_{k}(u \mid x),
\end{align*}
where $P_{k}(\cdot \mid x)$ denotes the conditional distribution of $U_{i,t}$ given $X_{i,t} = x$ for $i \in N_{k,t}$. Thus we can write the target parameter as
\begin{align*}
	\theta_{0} = \int m_{0}\left(\mu_{0}\left(x,v_{0,T}^*\right),x\right) d P_{0,T}(x),
\end{align*}
where $P_{0,T}$ denotes the distribution of $X_{i,T}$ in the target population.

The following assumption is an analogue of Assumption \ref{assump: support target}.

\begin{assumption}
	\label{assump: support target2}
	(i) The set $\mathcal{X}_{0,T}^\mathsf{M}$ is nonempty and identified.

	(ii) $m_0(\mu_0(\overline \mu,v_{0,T}^*),x)$ is identified for all $x \in \mathcal{X}_{0,T}^\mathsf{M}$.
\end{assumption}

The matched group $\mathcal{X}_{0,T}^\mathsf{M}$ is identified, if the union of the supports of $X_{i,t}$, $i \in N_{0,t}$, $t = 1,...,T$, contains the support $\mathcal{X}_{0,T}$, and $\mu_0(x,v_{0,T}^*)$ and $\mu_0(x,v_{0,T})$ are identified for each $x$ on the matched group $\mathcal{X}_{0,T}^\mathsf{M}$. Let us turn to an analogue of Assumption \ref{assump: support source}. The analogue requires that $m_k(\mu_k(x,v_{0,T}^*),x)$ be identified for all $x$ in the support $\mathcal{X}_{0,T}$ and all $k=1,...,K$. Below we provide low level conditions that ensure this.

\begin{assumption}
	\label{assump: support source2}
	(i) For $k=1,2,...,K$, $m_k(\overline \mu,x)$ is identified at each point $(\overline \mu,x)$ in the union of the supports of $(\mu_k(X_{i,t},v_{0,T}^*),X_{i,t})$, $t=1,...,T$, in population $k$.
	
	(ii) For each $k=1,2,...,K$,
	\begin{align*}
		\left\{ \mu_k(x,v_{0,T}^*): x \in \mathcal{X}_{0,T} \right\} \subset \bigcup_{t=1}^{T} \left\{ \mu_k(x,v_{0,T}^*): x \in \mathcal{X}_{k,t} \right\},
	\end{align*}
	where $\mathcal{X}_{k,t}$ denotes the support of $X_{i,t}$ in population $k$ for $i \in N_{k,t}$ in period $t$.
\end{assumption}

To understand Assumption \ref{assump: support source2}(i), note that due to the time invariance of the map $\mu_k$, we can identify this map using the combined distribution of $(Y_{i,t},X_{i,t})$ across time $t=1,...,T$. Regarding Assumption \ref{assump: support source2}(ii), as $T$ increases, the union of the supports of $\mu_k(x,v_{0,T}^*)$, $k=1,...,K$, also expands, making it more feasible to find source populations that satisfy this assumption in practice.

Let us see how our method applies in this setting with multi-period observations. We define the measure of goodness-of-fit as follows:
\begin{align}
	\label{obj22}
	 \rho^2(\boldsymbol{w}) = \int_{\mathcal{X}_{0,T}^\mathsf{M}} \left( m_0\left(\mu_0(x,v_{0,T}^*),x\right) - \sum_{k=1}^K m_k \left(\mu_k(x,v_{0,T}^*),x \right)w_k \right)^2 dP_{0,T}(x),
\end{align}
and define $\mathbb{W}_0 = \argmin_{\boldsymbol{w} \in \Delta_{K-1}} \rho^2(\boldsymbol{w})$, as before. The synthetic transferability condition is given as follows.

\begin{assumption}[Synthetic Transferability]\label{assump: synthetic transferability2} There exists $\boldsymbol{w} \in \mathbb{W}_0$ such that for each $x \in \mathcal{X}_{0,T}$, we have 
\begin{align}
	\label{eq}
	m_{0}\left(\mu_{0}(x,v_{0,T}^*),x\right) = \sum_{k=1}^K m_k\left(\mu_k(x,v_{0,T}^*),x\right) w_{k}.
\end{align}
\end{assumption}
Note that we require the condition (\ref{eq}) to hold only for $x \in \mathcal{X}_{0,T}$, not for all $x$ in the union of $\mathcal{X}_{k,t}$ $k=1,...,K$, $t=1,...,T$. As before, the optimization problem can be written as a simple quadratic programming problem. Under Assumptions \ref{assump: support source2}-\ref{assump: synthetic transferability2}, the identified set for $\theta_0$ is given by $\left\{\theta(\boldsymbol{w}): \boldsymbol{w} \in \mathbb{W}_0 \right\}$. In particular, when $H$ is invertible, $\mathbb{W}_0 = \{\boldsymbol{w}_0\}$ for some $\boldsymbol{w}_0 \in \Delta_{K-1}$, and $\theta_0 = \theta(\boldsymbol{w}_0)$.

\section{Estimation and Confidence Intervals}
\label{sec:estimation}
\subsection{Estimation}
We focus on the multi-period setting in Section \ref{subsec: aggregate shocks} and assume that the observations are made over multiple periods $t=1,...,T$, with repeated cross-sections. (Needless to say, it is straightforward to adapt the method for a setting with single-period observations or panel observations.) We focus on the case where $\mathbb{W}_0 = \{\boldsymbol{w}_0\}$ for some $\boldsymbol{w}_0 \in \Delta_{K-1}$. The case with $\mathbb{W}_0$ being a non-singleton set is discussed in the Supplemental Note. Let us first consider the estimation of $\boldsymbol{w}_0$ and $\theta(\boldsymbol{w}_0)$. As for the estimation of $\boldsymbol{w}_0$, we first make use of the characterization (\ref{characterization}) and consider its sample counterpart. 

For each region $k=1,...,K$, we define 
\begin{align}
    \label{N_k}
	N_k = \bigcup_{t=1}^{T} N_{k,t} \text{ and } N_0 = \bigcup_{t=1}^{T-1} N_{0,t},
\end{align}
and let $n_k = |N_k|$ and $n_0 = |N_0|$. We let $\{(\tilde Y_i,\tilde X_i): i \in N_{0} \}$ be the set of $(Y_{i,t},X_{i,t})$ for $i \in N_{0,t}$, $t=1,...,T-1$. Similarly, for each $k=1,...,K$, we let $\{(\tilde Y_i,\tilde X_i): i \in N_{k} \}$ be the set of $(Y_{i,t},X_{i,t})$ for $i \in N_{k,t}$ and $t=1,...,T-1$. Recall that we do not observe $Y_{i,T}$ because the policy $v_{0,T}^*$ is not implemented in period $T$ yet, but we still observe $X_{i,T}$. The estimation of $\theta_0(\boldsymbol{w})$ can be proceeded as follows.\medskip

(\textbf{Step 1}) Obtain the estimators $\hat \mu_0(x,v_{0,T}^*)$, $\hat \mu_0(x,v_{0,T})$, and $\hat m_0(\hat \mu_0(x,v_{0,T}^*),x)$, for each $x$ in the support of $\tilde X_i$ using the sample $\{(\tilde Y_i,\tilde X_i): i \in N_{0} \}$.\medskip

(\textbf{Step 2}) Construct $\mathcal{\hat X}_{0,T}^\mathsf{M}$ and $\mathcal{\hat X}_{0,T}^\mathsf{U}$ as the estimated sets of $\mathcal{X}_{0,T}^\mathsf{M}$ and $\mathcal{X}_{0,T}^\mathsf{U}$ using $\hat \mu_0(\cdot,v_{0,T}^*)$, $\hat \mu_0(\cdot,v_{0,T})$, and the sample $\{X_{i,T}: i \in N_{0,T}\}$.\medskip

(\textbf{Step 3}) For each $k=1,...,K$, obtain the estimators $\hat m_k(\hat \mu_k(x,v_{0,T}^*),x)$ for each $x \in \mathcal{\hat X}_{0,T}$, using the sample $\{(\tilde Y_i,\tilde X_i): i \in N_{k} \}$.\medskip

(\textbf{Step 4}) Using the estimates, we construct the sample version of $H$ and $\boldsymbol{h}$ as follows:
\begin{align}
	\label{H and h}
	\hat H &= \frac{1}{n_{0,T}}\sum_{i \in N_{0,T}} \boldsymbol{\hat m}(X_{i,T})\boldsymbol{\hat m}(X_{i,T})^{\prime} 1\left\{X_{i,T} \in \mathcal{\hat X}_{0,T}^\mathsf{M}\right\}, \text{ and }\\ \notag
	\boldsymbol{\hat h} &=  \frac{1}{n_{0,T}}\sum_{i \in N_{0,T}} \boldsymbol{\hat m}(X_{i,T}) \hat m_0\left(\hat \mu_0(X_{i,T},v_{0,T}^*),X_{i,T} \right) 1\left\{X_{i,T} \in \mathcal{\hat X}_{0,T}^\mathsf{M}\right\},
\end{align}
where $\boldsymbol{\hat m}(x) = [\hat m_1(\hat \mu_1(x,v_{0,T}^*),x),...,\hat m_K(\hat \mu_K(x,v_{0,T}^*),x)]^{\prime}$.\medskip

(\textbf{Step 5}) Using $\hat H$ and $\boldsymbol{\hat h}$, we obtain
\begin{align*}
	\boldsymbol{\hat w} = \argmin_{\boldsymbol{w} \in \Delta_{K-1}} \enspace \boldsymbol{w}' \hat H  \boldsymbol{w} - 2 \boldsymbol{w}' \boldsymbol{\hat h}.
\end{align*}

(\textbf{Step 6}) Finally, we obtain the estimator of $\theta(\boldsymbol{w})$ as follows:
\begin{align}
	\label{hat theta}
	\hat \theta(\boldsymbol{\hat w}) &=  \frac{1}{n_{0,T}}\sum_{i \in N_{0,T}} \hat m_0\left(\hat \mu_0(X_{i,T},v_{0,T}^*),X_{i,T} \right) 1\left\{X_{i,T} \in \mathcal{\hat X}_{0,T}^\mathsf{M} \right\} \\ \notag
	&\quad + \sum_{k=1}^K \frac{1}{n_{0,T}}\sum_{i \in N_{0,T}} \hat m_k\left(\hat \mu_k(X_{i,T},v_{0,T}^*),X_{i,T}\right) 1\left\{X_{i,T} \in \mathcal{\hat X}_{0,T}^\mathsf{U} \right\}\hat w_{k}.
\end{align}

In the Supplemental Note, we show that $\boldsymbol{\hat w}$ is $\sqrt{n_{0,T}}$-consistent for $\boldsymbol{w}_0$. (As we formally state later, the sample size of each population is assumed to be asymptotically comparable, i.e., there exists $r_k > 0$ such that $n_k/n_{0,T} \rightarrow r_k$ as $n_{0,T},n_k \rightarrow \infty$ for each $k=1,...,K$.)

\subsection{Confidence Intervals for $\theta(\boldsymbol{w}_0)$}

Let us construct confidence intervals for $\theta(\boldsymbol{w}_0)$. First, we construct a confidence set of $\boldsymbol{w}_0$ following the proposal by \cite{Canen/Song:25:arXiv}. More specifically, we first take 
\begin{align*}
	\hat \varphi(\boldsymbol{w}) = \hat H \boldsymbol{w} - \boldsymbol{\hat h}, \enspace \quad \boldsymbol{w} \in \Delta_{K-1}.
\end{align*}
Let $I_K - \mathbf{1}\mathbf{1}'/K = U D U'$ be the spectral decomposition and let $B_2$ be the $K \times (K-1)$ matrix after removing the eigenvector from $U$ that corresponds to the zero diagonal element of $D$. First, we define 
\begin{align*}
	\hat \lambda(\boldsymbol{w}) = \argmin_{\lambda} \left( \hat \varphi(\boldsymbol{w}) - \lambda \right)'B_2 \hat \Omega^{-1} B_2' ( \hat \varphi(\boldsymbol{w}) - \lambda ),
\end{align*}
where the minimization over $\lambda$ is done under the constraints: $\boldsymbol{w}'\lambda = 0$ and $\lambda \ge 0$ and $\hat \Omega$ is a matrix for normalization that we explain below. 

Let $\hat d(\boldsymbol{w})$ be the number of zeros in the vector $B_2 \hat \Omega^{-1} B_2'( \hat \varphi(\boldsymbol{w}) - \hat \lambda(\boldsymbol{w}))$, and let $\hat c_{1-\alpha}(\boldsymbol{w})$ be the $1-\alpha$ percentile of the $\chi_{\hat k(\boldsymbol{w})}^2$ distribution, where $\chi_k^2$ denotes the chi-squared distribution with $k$ degrees of freedom, and
\begin{align*}
	\hat k(\boldsymbol{w}) = \max \left\{K-1 - \hat d(\boldsymbol{w}),1 \right\}.
\end{align*} 
Now, we construct the $(1-\alpha)$-level confidence set for $\boldsymbol{w}_0$ as follows: 
\begin{align}
	\label{tilde C}
	\tilde C_{1-\alpha} = \left\{ \boldsymbol{w} \in \Delta_{K-1}: T(\boldsymbol{w}) \le \hat c_{1-\alpha}(\boldsymbol{w}) \right\},
\end{align}
where 
\begin{align}
	\label{T(u)}
	T(\boldsymbol{w}) = n_{0,T}\left( \hat \varphi(\boldsymbol{w}) - \hat\lambda(\boldsymbol{w}) \right)'B_2 \hat \Omega^{-1}B_2' \left( \hat \varphi(\boldsymbol{w}) - \hat\lambda(\boldsymbol{w}) \right).
\end{align}

We could take the normalization matrix $\hat \Omega = B_2' \hat V(\boldsymbol{\hat w}) B_2$, where $V(\boldsymbol{w}_0)$ is the asymptotic variance as below:
\begin{align*}
	\sqrt{n_{0,T}}\left(\hat \varphi(\boldsymbol{w}) - \varphi(\boldsymbol{w}) \right) \rightarrow_d N(0, V(\boldsymbol{w})),
\end{align*}
and $\hat V(\boldsymbol{w})$ is the consistent estimator of $V(\boldsymbol{w})$. In this paper, we pursue a bootstrap approach that does not require the researcher to find the asymptotic variance $V(\boldsymbol{w})$ analytically. Since each population has a different distribution, we need to resample (with replacement) from each region. For each region $k=0,1,...,K$, let $\{W_{i}^*: i \in N_k\}$, $W_i^* = (\tilde Y_{i}^*,\tilde X_{i}^{*})$, be the bootstrap sample from the sample $\{W_i: i \in N_k\}$, where $W_i = (\tilde Y_i, \tilde X_i)$, and $\{(\tilde Y_i, \tilde X_i): i \in N_{k}\}$ is the set of $(Y_{i,t},X_{i,t})$, with $i \in N_{k,t}$ and $t=1,...,T$. We also let $\{W_{i}^*: i \in N_0\}$, $W_i^* = (\tilde Y_{i}^*,\tilde X_{i}^{*})$, be the bootstrap sample from the sample $\{W_i: i \in N_{0}\}$, where $W_i = (\tilde Y_i, \tilde X_i)$ and $\{(\tilde Y_i,\tilde X_i): i \in N_0\}$ is the set of $Y_{i,t}$, with $i \in N_{0,t}$ and $t=1,...,T$. Then, for each $k=0,1,...,K$, we construct the bootstrap version of the conditional average outcome, $\hat m_{k}^*(\hat \mu_k^*(\cdot,v_{0,T}^*),\cdot)$, using the bootstrap sample from the region $k$, and define
\begin{align*}
	\boldsymbol{\hat m}^*(\cdot) = \left[\hat m_{1}^*\left(\hat \mu_1^*(\cdot,v_{0,T}^*),\cdot\right),...,\hat m_{K}^*\left(\hat \mu_K^*(\cdot,v_{0,T}^*),\cdot \right) \right]^{\prime},
\end{align*}
and let
\begin{align*}
	\hat H^* &= \frac{1}{n_{0,T}}\sum_{i \in N_{0,T}} \boldsymbol{\hat m}^*(X_{i,T}^*)\boldsymbol{\hat m}^*(X_{i,T}^*)^{\prime}1\left\{X_{i,T}^* \in \mathcal{\hat X}_{0,T}^{\mathsf{M} *} \right\}, \text{ and } \\
	\boldsymbol{\hat h}^* &= \frac{1}{n_{0,T}}\sum_{i \in N_{0,T}} \boldsymbol{\hat m}^*(X_{i,T}^*) \hat m_{0}^*\left(\hat \mu_0^{*}(X_{i,T}^*,v_{0,T}^*),X_{i,T}^* \right) 1\left\{X_{i,T}^* \in \mathcal{\hat X}_{0,T}^{\mathsf{M} *}\right\},
\end{align*}
where $\mathcal{\hat X}_{0,T}^{\mathsf{M} *}$ denotes the estimated set of $\mathcal{X}_{0,T}^{\mathsf{M}}$ using the bootstrap sample. Then, we define
\begin{align}
	\label{hat gamma star}
	\boldsymbol{\hat \gamma}^* = \sqrt{n_{0,T}} \left(\hat H^* - \hat H \right) \boldsymbol{\hat w} - \sqrt{n_{0,T}} \left( \boldsymbol{\hat h}^* - \boldsymbol{\hat h}\right),
\end{align}
where $\{X_{i,T}^*\}_{i \in N_{0,T}}$ denotes the bootstrap sample from $\{X_{i,T}\}_{i \in N_{0,T}}$, and construct\footnote{The use of the bootstrap variance estimator does not affect the asymptotic validity of the inference based on it, but it can make the inference conservative. See \cite{Hahn/Liao:21:Ecma} for this point.}
\begin{align*}
	\hat \Omega = B_2' \hat V B_2,
\end{align*}
with
\begin{align}
	\label{hat Omega}
	\hat V = \mathbf{E}^*\left[ \boldsymbol{\hat \gamma}^* \boldsymbol{\hat \gamma}^{* \prime} \right] - \mathbf{E}^*\left[ \boldsymbol{\hat \gamma}^* \right] \mathbf{E}^*\left[ \boldsymbol{\hat \gamma}^{* \prime} \right],
\end{align}
where $\mathbf{E}^*$ denotes the expectation with respect to the bootstrap distribution. We may be interested in checking whether data support the synthetic transferability condition in (\ref{tilde C}). Consider testing the following implication from the synthetic transferability condition:\medskip

$H_0$: There exists $\boldsymbol{w} = (w_1,...,w_K) \in \Delta_{K-1}$ such that for all $x \in \mathcal{X}_{0,T}^\mathsf{M}$,
	\begin{align*}
		m_0\left( \mu_0(x,v_{0,T}^*),x \right) = \sum_{k=1}^K m_k\left( \mu_k(x,v_{0,T}^*),x \right)w_k.
	\end{align*}

$H_1$: $H_0$ is false.\medskip

We construct $\tilde C_{1-\kappa}$ as in (\ref{tilde C}), with $\kappa$ set to be the level $\alpha$ of the test and perform the following procedure: if $\tilde C_{1-\alpha} = \varnothing$, we reject $H_0$ at level $\alpha$. Otherwise, we do not reject $H_0$ at level $\alpha$. It is not hard to see that this test is asymptotically valid as long as the confidence set $\tilde C_{1-\alpha}$ is asymptotically valid at level $\alpha$. The asymptotic validity of the confidence set $\tilde C_{1-\alpha}$ is established using a general result in \cite{Canen/Song:25:arXiv}.

Now, let us construct a confidence interval for $\theta(\boldsymbol{w}_0)$. First, we can show that
\begin{align}
	\label{conv}
	\frac{n_{0,T} (\hat \theta(\boldsymbol{w}_0) - \theta(\boldsymbol{w}_0))^2}{\hat \sigma^2} \rightarrow_d \chi_1^2,
\end{align}
for an appropriate scale normalizer $\hat \sigma$. To construct $\hat \sigma$, we use a bootstrap interquartile range as proposed by \cite{Machado/Parente:05:EJ}. More specifically, we define
\begin{align}
	\label{hat theta star}
	\hat \theta^*(\boldsymbol{w}) &= \frac{1}{n_{0,T}}\sum_{i \in N_{0,T}} \hat m_{0}^*\left(\hat \mu_0^{*}(X_{i,T}^*,v_{0,T}^*),X_{i,T}^*\right) 1\left\{X_{i,T}^* \in \mathcal{\hat X}_{0,T}^{\mathsf{M} *}\right\} \\ \notag
	& \quad \quad + \sum_{k=1}^K \frac{1}{n_{0,T}}\sum_{i \in N_{0,T}} \hat m_k^{*}\left(\hat \mu_k^{*}(X_{i,T}^*,v_{0,T}^*),X_{i,T}^*\right) 1\left\{X_{i,T}^* \in \mathcal{\hat X}_{0,T}^{\mathsf{U} *}\right\} w_k.
\end{align}
We take the bootstrap statistic: 
\begin{align*}
	\hat \tau^* = \sqrt{n_{0,T}}\left(\hat \theta^*(\boldsymbol{\hat w}) - \hat \theta(\boldsymbol{\hat w})\right),
\end{align*}
and read the $0.75$ quantile and $0.25$ quantile of the bootstrap distribution of $\{\hat \tau^* : b = 1,...,B\}$, denoting them to be $\hat q_{0.75}$ and $\hat q_{0.25}$, respectively. Define
\begin{align}
	\label{hat sigma}
	\hat \sigma = \frac{\hat q_{0.75} - \hat q_{0.25}}{z_{0.75} - z_{0.25}},
\end{align}
where $z_{0.75}$ and $z_{0.25}$ are the $0.75$- and $0.25$-quantiles of $N(0,1)$. Define
\begin{align*}
	\hat \tau(\boldsymbol{w}, \theta) = \frac{\sqrt{n_{0,T}} (\hat \theta(\boldsymbol{w}) - \theta)}{\hat \sigma}.
\end{align*}
We construct the $(1-\alpha)$-level confidence interval using the Bonferroni approach as follows:
\begin{align}
	\label{conf interval}
	C_{1- \alpha} = \left\{ \theta \in \Theta: \inf_{\boldsymbol{w} \in \tilde C_{1- \kappa}} \hat \tau^2(\boldsymbol{w}, \theta) \le c_{1 - \alpha + \kappa}(1) \right\},
\end{align}
where $\kappa>0$ is a small constant, such as $\kappa = 0.005$, and $c_{1 - \alpha + \kappa}(1)$ denotes the $(1 - \alpha + \kappa)$-quantile of the $\chi_1^2$ distribution. 

\subsection{Uniform Asymptotic Validity}

We summarize the conditions that we use to establish the uniform asymptotic validity of the confidence set $C_{1- \alpha}$. Here, we state the conditions verbally. The formal statements are found in the Supplemental Note.

\begin{assumption}
    \label{assump: main, iid}
    (i) The random vectors $(X_{i,t},U_{i,t})$ are independent across all sample units $i$ and time $t$, and identically distributed within each population.
	
	(ii) For $k=0,1,...,K$, there exists a constant $r_k >0$ such that $n_k/n_{0,T} \rightarrow r_k$ as $n_{0,T}, n_k \rightarrow \infty$.
\end{assumption}

Assumption \ref{assump: main, iid}(i) says that the samples are independent across the sample units $i \in N$ and time periods and are identically distributed across individuals $i \in N_{k,t}$ and time periods $t=1,...,T$, within each population. This assumption is suitable as we assume repeated cross-sections over time. Assumption \ref{assump: main, iid}(ii) excludes a setting where the sample size of some source population is asymptotically negligible compared to the target population. We can relax this assumption as long as the sample sizes of all the populations are large enough. This assumption is made for convenience, as the effective sample size can now be written in terms of $n_{0,T}$.

\begin{assumption}
    \label{assump 2}
    (i) The conditional average outcomes in the target and source populations have the $(4+\delta)$-th moment bounded uniformly over $P$.
	
	(ii) The estimated conditional average outcomes and their bootstrap versions have an asymptotic linear representations uniform over $P$, with the influence function having the $(4+\delta)$-th moment bounded uniformly over $P$.
\end{assumption}

The moment condition is a technical condition that is often used in asymptotic inference. The asymptotic linear representation is often part of the proofs that show asymptotic normality of an estimator. Its derivation is standard in many examples.

\begin{assumption}
    \label{assump 3}
    The matrix $H$ and the population version of $\hat V$ have minimum eigenvalues bounded from below uniformly over $P$ and over the sample sizes.
\end{assumption}

This assumption requires that the conditional average outcomes are not redundant, and ensures that $\mathbb{W}_0 = \{\boldsymbol{w}_0\}$ for some $\boldsymbol{w}_0 \in \Delta_{K-1}$. As mentioned before, we can relax this assumption once we modify the procedure. Details are found in the Supplemental Note.

Under these conditions, the confidence interval $C_{1- \alpha}$ is asymptotically valid uniformly over $P$ as shown below.

\begin{theorem}
	\label{thm: validity0}
	Suppose that Assumptions \ref{assump: main, iid}-\ref{assump 3} hold. Then, for each $\alpha \in (0,1)$, the confidence interval $C_{1- \alpha}$ is asymptotically valid uniformly over $P$.
\end{theorem}

The proof of the theorem is found in the Supplemental Note.

\section{Empirical Application: Minimum Wages and Labor Supply}
\label{sec: empirical app}

\subsection{Background}

Minimum wages have been among the most studied and debated policies for the labor market, spurring an immense literature in economics. The predominant paradigm in empirical work is to study their effects on employment or other outcomes by leveraging their state-level variation. This includes difference-in-difference designs with Two-Way Fixed Effects models (which \cite{Neumark:19:GER} summarizes as the workhorse approach), synthetic control (see \cite{Allegretto/Dube/Reich/Zipperer:17:ILR,Neumark/Wascher:17:ILR} for extensive discussions), decomposition methods (\cite{DiNardo/Fortin/Lemieux:96:Eca}), cross-border comparisons (\cite{Dube/Lester/Reich:10:ReStat}), among others.

While this literature can evaluate minimum wage increases that have already been implemented, they are by-and-large inappropriate to predict the effects of policies yet to occur, including increases in minimum wages beyond the support of historical variations. Indeed, even simple theoretical models predict highly nonlinear effects of minimum wages (e.g., \cite{Flinn:06:Ecma, Gorry/Jackson:17:CEP}).\footnote{This is best summarized by \cite{Neumark:19:GER} who writes in a recent review that, ``even if one has a strong view of what the U.S. literature says about the employment effects of past minimum wage increases, this may provide much less guidance in projecting the consequences of much larger minimum wage increases than those studied in the prior literature. Predicting the effects of minimum wage increases of many dollars, based on research studying much smaller increases, is inherently risky for the usual statistical reasons. But the problem is potentially exacerbated because the reduced form estimates on which the prior literature is based may fail to capture changes in underlying behavior as high minimum wages affect a far greater share of workers.'' (p.294)} The synthetic decomposition method presented in this paper is able to address such policy questions.

As foreshadowed in Section \ref{sec: example_minwage}, our empirical illustration studies a (counterfactual) increase in minimum wage in Texas beyond federally mandated levels and how it affects teenage employment. The focus on Texas, while used as an illustration, is of both academic and policy interest. Texas is the largest state in the U.S. with minimum wages set at the federal level (constant since 2009). Raising the minimum wages has also been a policy of the 2022 Democratic gubernatorial candidate. We illustrate our method by investigating the effects of an increase in minimum wages in Texas from US\$7.25 to US\$9.00, on teenage employment. We follow the structural labor economics literature in basing such predictions on an equilibrium search and matching model of labor markets (e.g., \cite{Flinn:06:Ecma, Flinn/Mullins:15:IER} and \cite{Ahn/Arcidiacono/Wessels:11:JBES}, in particular). However, in contrast to such papers, we construct a synthetic comparison using other states beyond Texas where the policy has been observed (e.g., California, Washington, etc.).

This setting suits the synthetic decomposition method very well. There are two main sources of heterogeneity across regions. First, the population characteristics differ. For example, states are heterogeneous in workers' education, age and skill, among others, all of which may matter for the effects of minimum wages (\cite{Neumark:19:GER}, and seen in the data below). More importantly, the causal structure $g_0$ for the source region could be different than those for other states, $g_k$, even those from neighboring states. Intuitively, even if California and other states had similar characteristics to Texas, they may have very different labor market environments (e.g., state income taxation, different labor laws, etc.). In fact, \cite{Flinn:02:ReStud} argues that estimated structural parameters are very different across submarkets.  The synthetic decomposition method respects such heterogeneity across regions. It assigns weights to those source states to form the best comparison units in terms of their pre-policy predictions.

\subsection{An Empirical Model of Labor Markets with Minimum Wages}

\subsubsection{A Two-Sided Search Model of Labor Markets with Minimum Wages}

We follow \cite{Ahn/Arcidiacono/Wessels:11:JBES} and consider the following static model of two-sided matching between firms and workers. For each population $k=0,1,...,K$, we let $\overline N_k$ be the total measure of the workers and $\overline J_k$ the total measure of the firms in the population $k$. Each worker-firm pair $(i,j)$ is drawn, and then for each worker $i$, $(R_i,K_i)$ is drawn, where $R_i$ is the reservation wage of worker $i$ and $K_i$ the cost of searching for the worker $i$. The worker-firm pair is given the offer of matching with a contact rate $\lambda_k >0$. The timing of the events for the worker-firm pair given the offer of the match proceeds as follows.
\begin{enumerate}
\item The worker decides to search for a match with a firm. Once the worker decides to search, the worker pays the search cost $K_i$ and receives an offer of match with a firm $j$ with probability $\lambda_k>0$. If the worker decides not to search for a firm, the worker receives zero payoff.
\item
The worker decides whether to accept the match offer or not. If the worker rejects the offer of the match, the worker receives a reservation wage $R_i$. If the worker accepts the offer, the worker-firm pair $(i,j)$ jointly produces output $M_{i,j}$.
\item 
Once the output $M_{i,j}$ is realized, the firm and the worker enter a Nash bargaining to determine the wage, $W_{i,j}$ under the minimum wage constraint.
\item
After the wage $W_{i,j}$ is determined, the firm decides whether to retain the worker or not. If the firm retains the worker, the firm obtains the profit $M_{i,j} - W_{i,j}$ and the worker receives the wage $W_{i,j}$. If the firm does not retain the worker, the firm and the worker receive payoff equal to zero.
\item
After these events are completed, the econometrician observes a random sample of the workers, their employment status and wages, and their observed characteristics.
\end{enumerate}

To close the model, we need to state the equilibrium constraints. First, it is profitable for worker $i$ to accept the offer from the match with firm $j$ if and only if
\begin{align}
	\label{ineq34}
	\mathbf{E}_k[1\{M_{i,j} \ge W_{i,j}\} W_{i,j} \mid R_i, K_i] \ge R_i,
\end{align}
where the conditional expectation $\mathbf{E}_k$ is with respect to the distribution in population $k$. Then, it is profitable for the worker to search for a job if and only if
\begin{align*}
	\lambda_k \mathbf{E}_k[\max\{1\{M_{i,j} \ge W_{i,j}\} W_{i,j},R_i\} \mid R_i,K_i] \ge K_i.
\end{align*}
For the firm, it is profitable for it to retain the worker if and only if $M_{i,j} \ge W_{i,j}$. Finally, we assume that the contact rate $\lambda_k$ is endogenously determined as a fixed point as follows:
\begin{align*}
	\lambda_k = \frac{\mathcal{M}_k(\lambda_k, \overline J_k, \overline N_k)}{\zeta_k(\lambda_k) \overline N_k},
\end{align*}
where $\mathcal{M}_k(\lambda_k, \overline J_k, \overline N_k)$ denotes the matching technology, representing the total measure of matched workers, and 
\begin{align*}
	\zeta_k(\lambda_k) = P\{ \lambda_k \mathbf{E}_k[\max\{1\{M_{i,j} \ge W_{i,j}\} W_{i,j},R_i\} \mid R_i,K_i] \ge K_i \},
\end{align*}
i.e., the probability of the worker deciding to search for a firm. Hence, $\zeta_k(\lambda_k) \overline N_k$ represents the total measure of workers searching for a match with a firm. 

As for the wage determination through Nash bargaining, we follow \cite{Ahn/Arcidiacono/Wessels:11:JBES} and obtain the following wage generation: for $M_{i,j} \ge \underline W_k$,
\begin{align*}
	W_{i,j} = \max\{\delta_k M_{i,j},R_i,\underline W_k\},
\end{align*}
where $\delta_k \in (0,1)$ is a parameter that represents worker $i$'s bargaining strength. We also follow \cite{Ahn/Arcidiacono/Wessels:11:JBES} in simplifying the procedure by assuming that (\ref{ineq34}) is satisfied for all the workers such that $R_i \le \underline W_k$. Then the wage is generated only for those workers with $R_i \le \underline W_k$, and hence, the wage generation is simplified as follows: for $M_{i,j} \ge \underline W_k$,
\begin{align}
	\label{wage}
	W_{i,j} = \max\{\delta_k M_{i,j},\underline W_k\}.
\end{align}
The employment indicator $Y_{i,j} \in \{0,1\}$ is also given as follows:
\begin{align}
	\label{employment}   
	Y_{i,j} = 1\{M_{i,j} \ge W_{i,j}\} = 1\{M_{i,j} \ge \underline W_k\},
\end{align}
where the last equality follows from (\ref{wage}) and $\delta_k \in (0,1)$. 

\subsubsection{Building an Empirical Model}

We now build up an empirical model. To do so, we first explain the data structure in this setting. As for each source population $k=1,...,K$, we observe $(Y_{i,t},X_{i,t},W_{i,t})$, the repeated cross-sections of individuals and the minimum wages $\underline W_{k,t}$, over $t=1,...,T$, where $Y_{i,t}$ represents the employment status of an individual $i$ in period $t=1$. For the target population, we observe similarly $(Y_{i,t},X_{i,t},W_{i,t})$, the repeated cross-sections of individuals and the minimum wages $\underline W_{0,t}$, over $t=1,...,T-1$, and also observe $\{X_{i,T}\}$. 

The counterfactual policy of interest is to set the minimum wage for population 0 (Texas) from $\underline W_{0,T}$ to $\underline W_0^* = US\$9$. We aim to predict the employment rate after the minimum wage changes.

We specify the match output $M_{i,j,t}$ in time $t$ as follows:
\begin{align*}
	\log M_{i,j,t} &= X_{i,t}^{\prime} \gamma_k + U_{i,j,t},
\end{align*}
where $X_{i,t}$ denotes the observed characteristics of worker $i$, $U_{i,j,t}$ represents a match component that is unobserved by the econometrician, and $\gamma_k$ is a parameter vector. We assume that $U_{i,j,t}$'s are i.i.d., independent of $(X_{i,t}, \underline W_{k,t})$, $i \in N_{k,t}$, and all firms $j$, and follow the distribution with the CDF, $F_k$. Unlike \cite{Ahn/Arcidiacono/Wessels:11:JBES}, we leave $F_k$ as nonparametrically specified. Since we do not restrict $U_{i,j,t}$ to have mean zero, we lose no generality by assuming that the vector $X_{i,t}$ does not include an intercept term. It follows from this parametrization and (\ref{employment}) that: for each $i \in N_{k,t}$,
\begin{align}
	\label{employment2}
	Y_{i,t} &= 1\{M_{i,j(i),t} \ge \underline W_{k,t}\} = 1\{X_{i,t}^{\prime} \gamma_k + U_{i,j(i),t} \ge \log \underline W_{k,t}\},
\end{align}
where $j(i)$ denotes the firm matched with worker $i$ in period $t$. 

In order to check the applicability of the synthetic decomposition method, we consider the support conditions required in this setting. First, we define our policy components
\begin{align*}
	\mu_k(X_{i,t}, \underline W_{k,t}) = X_{i,t}' \gamma_k - \log \underline W_{k,t}.
\end{align*}
We take
\begin{align}
	\mathcal{X}_{0,T}^\mathsf{M} = \{x \in \mathcal{X}_{0,T}: x^{\prime} \gamma_0 - \log \underline W_{0,T} = \tilde x^{\prime} \gamma_0 -\log \underline W_{0}^*, \text{ for some } \tilde x \in \mathcal{X}_{0,T}\},\label{support_emp2}
\end{align}
where we denote the support of $X_{i,T}$ in the target population by $\mathcal{X}_{0,T}$. Due to the independence between $U_{i,j,t}$'s and $(X_{i,t}, \underline W_{k,t})$'s, the average response function $m_k(\overline \mu,x)$ does not depend on the second argument, and we simply write $m_k(\overline \mu)$. In this empirical model, the conditional average outcome takes the following form:
\begin{align}
	\label{conditional average outcome_emp}
	m_k(\mu_k(X_{i,t},\underline W_{k,t})) &= \int g_k(\mu_k(X_{i,t},\underline W_{k,t}),u) dF_k(u) \\ \notag
	&= \mathbf{E}_k\left[ Y_{i,t} \mid \mu_k(X_{i,t},\underline W_{k,t}) = \overline \mu \right],
\end{align}
where $g_k(\overline \mu,u) = 1\{\overline \mu +u \ge 0\}.$ The synthetic prediction is obtained by using the weights $w_k$'s which minimize the $L^2$-distance between 
\begin{align*}
	m_0(\mu_0(x,\underline W_0^*)) \text{ and } \sum_{k=1}^K m_k(\mu_k(x,\underline W_0^*))w_k,
\end{align*}
over $x \in \mathcal{X}^\mathsf{M}_{0,T}$.

As for estimation, we rewrite the equilibrium wage generation in (\ref{wage}) as follows: for each $i \in N_{k,t}$, we have
\begin{align}
	\label{censored regression}
	\log W_{i,t} = \max\{ \log \delta_k + X_{i,t}^{\prime} \gamma_k + U_{i,j(i),t}, \log \underline W_{k,t} \}.
\end{align}
Thus the log wage follows a semiparametric censored regression model. We estimate $\gamma_k$ using the pairwise differencing method of \cite{Honore/Powell:94:JOE}. We plug them into $\mu_k(X_{i,t},\underline W_{k,t})$ and estimate $m_k$ nonparametrically using a kernel regression estimation method and a cross-validated bandwidth. Details are provided in the Supplemental Note. We use $B=200$ bootstrap draws, set $\kappa = 0.005$ and $\alpha = 0.05$. We draw a fine grid of $\boldsymbol{w}$ uniformly over its simplex, using a procedure based on \cite{Rubin:81:AoS}.\footnote{To construct each gridpoint, we first draw a vector of dimension $K-1$, where each element is drawn i.i.d. from the uniform distribution with support $[0,1]$. Then, we include 0 and 1 into that drawn vector, which is then sorted. The grid point is the vector of differences across adjacent elements of $\boldsymbol{w}$ (which are all nonnegative and must sum up to 1 by construction).}

\subsection{Data}
	
We use the dataset from \cite{Allegretto/Dube/Reich/Zipperer:17:ILR} for our exercises, which is drawn from the Current Population Survey (CPS), a repeated cross-section. Following the authors, among many others, we focus on teenagers and use their individual-level employment status as the outcome, $Y_{i,j} \in \{0,1\}$, individual-level characteristics as $X_{i}$ (age, sex, marriage status, whether they are Hispanic, whether they are African-American or another non-white race). We further observe wages for an employed sample, $W_{i,j}$, and each state's minimum wages. Our sample is restricted from 2002 to 2014, so that it does not start during the 2001 recession (see \cite{Neumark/Wascher:17:ILR} for a discussion). We then merge this dataset with quarterly state inflation data from \cite{Hazell_et_al:22:QJE} which is used as an additional aggregate measure of economic activity. 
	
	The counterfactual policy of interest sets the minimum wage in Texas (US\$7.25 in 2014) to US\$9 in 2014 (i.e., to US\$11.87 in 2024 dollars).\footnote{Thus, our empirical setting is that with multi-period observations, as discussed in Section \ref{subsec: aggregate shocks}. The target year here is set to be $T=2014$.} Our parameter of interest, $\theta_0$ is the average teenage employment in Texas in 2014 (for Texas' 2014 population) had the minimum wage been US\$9. We compare this to teenage employment for those in Texas in 2014 with the prevailing minimum wage.
	
To make this comparison, we consider two sets of source regions. First, we use the states with the highest prevailing minimum wages within our sample, which are California, Connecticut and Washington.\footnote{D.C. and Vermont also satisfy this restriction, but we drop them as their sample is too small to provide meaningful variation for estimation of D.C./Vermont-specific parameters. We also tested specifications with Oregon, another Pacific Northwest state satisfying the restrictions on minimum wages. However, it suffered from multicollinearity in the wage equation when aggregate variables were included. Furthermore, its estimated weight was 0 for the other specifications, so we opt not to report it.} We note that the support conditions (\ref{support_emp}) can include more states because it is a condition on the support of the conditional average outcome and not on the policy itself. Hence, in a second exercise, we further include Florida (a large state close to Texas) and Louisiana (a neighboring state). For illustration purposes, we use a 10\% random sample of the data for each region. This shows the performance of our estimator with reasonably standard sample sizes. 

\begin{figure}[htbp!]
		\begin{center}
					\caption{The Variation in Minimum Wages Over Time Across Regions in Our Sample}
			\includegraphics[scale=1]{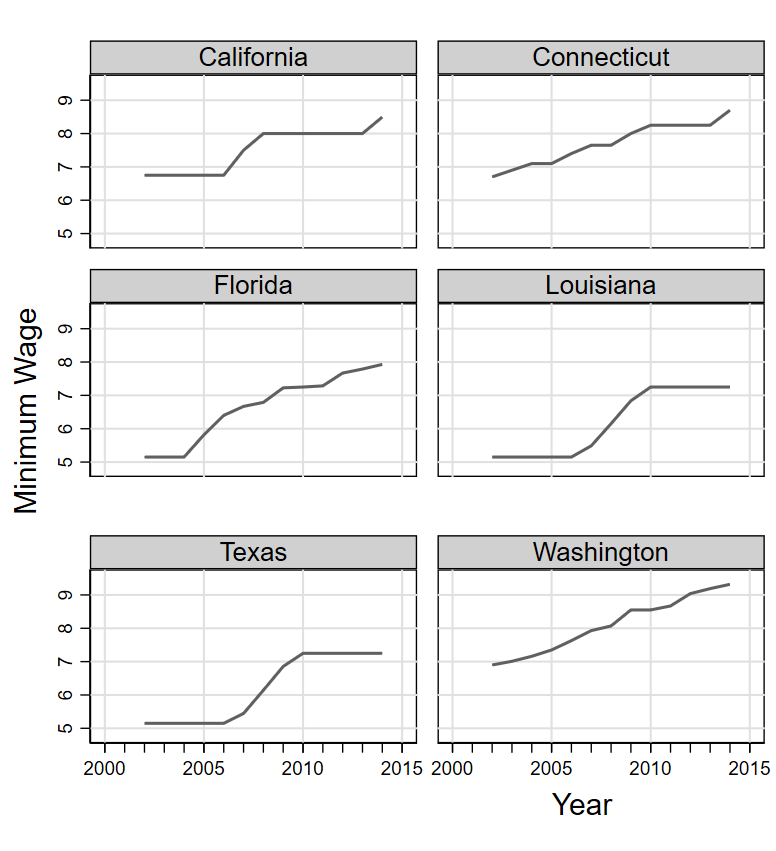}
			\label{mw_variation}
		\end{center}
		\parbox{6.4in}{\footnotesize
			Notes: The panels depict the minimum wages during the sample periods (2002-2014) in some selected states in the U.S.}
	\end{figure}

Summary statistics are provided in Table \ref{sum_stats}, while the variation in minimum wages across all source and target regions is shown in Figure \ref{mw_variation}. In terms of demographics (e.g., the share of teenage Hispanics and African-Americans), Texas most resembles California. However, it is more similar to Florida and Louisiana in terms of average teenage employment and in wages. On the other hand, Louisiana's minimum wage policies are very similar to Texas'.

	\begin{table}[t]
		\caption{\small Summary Statistics for the Whole Sample (2002-2014)}
				\label{sum_stats}
			\begin{centering}
			\small
			\begin{tabular}{c c c c c c c c}
				\hline 
				\hline
				\\ 
				& CA & CT  & FL & LA & TX & WA \\
				\\
				\hline 
    				\\
				(Teenage) Employment & 0.248 & 0.359 &	0.311 &0.275 &	0.295 &	0.326
				\\
				Wages (US\$) & 8.74 &	8.76 &	7.67 &	7.40 &		7.51 &	8.81\\
				\\
				Age & 17.45 &	17.44 &	17.45 &	17.41 &	 17.42 &	17.36
				\\
				Male &  0.510&	0.513&	0.507&0.489&		0.516&	0.512
				\\
				Married & 0.012 &	0.004 &	0.012 &	0.009 &		0.020 &	0.010
				\\
				Hispanic & 0.468&	0.145&	0.244&	0.038&		0.459&	0.122
				\\
				Black & 0.066 &	0.107 &	0.198 &	0.384 &		0.123 &	0.036\\
				\\
				Share of Teenagers&  0.075 & 0.073  & 0.062 & 0.077 & 0.077&0.069
				\\
				Average State & 2.39  & 2.05	 &	2.43&	2.13  & 2.35	& 2.58	\\
				Inflation ($\%$) & & & & & & & \\
					\hline 
				\multicolumn{1}{c}{} &      \tabularnewline
			\end{tabular}
			\par\end{centering}
		\parbox{6.4in}{\footnotesize
			Notes: The table presents summary statistics for the variables used in the main specification. This includes the labor market outcomes (employment and wages for the employed) and observable characteristics. Note that these statistics are averaged over all years in the sample period (2002-2014), while our main comparison is to Texas in 2014.}
	\end{table}
	
	\subsection{Main Results}
		\begin{table}[t]
		\caption{\small Confidence Intervals for $\theta_0$: Predicted Average (Teenage) Employment in Texas After a Counterfactual Minimum Wage Increase}
		
		\label{emp}
		
		\small 
		
		\begin{centering}
			\small
			\begin{tabular}{c|cccc}
				\hline 
				\hline
				\\ 
				& \multicolumn{4}{c}{Increase to US\$9}\\
				\\
				\hline 
				\\
				$\theta_0$ &     0.195  &     0.215  & 0.196 & 0.215\tabularnewline 
				&      [0.140, 0.250]&     [0.168, 0.263]  & [0.145,  0.247]&      [0.169, 0.261]\tabularnewline 
				\\
				$\boldsymbol{w}_0 = \begin{pmatrix} CA \\ CT  \\ FL \\ LA  \\ WA\end{pmatrix}$  & $\begin{pmatrix} 0.753 \\ 0  \\ - \\ - \\ 0.247\end{pmatrix}$ & $\begin{pmatrix} 0.496 \\ 0  \\ - \\ - \\ 0.504\end{pmatrix}$ & $\begin{pmatrix} 0.332 \\ 0 \\  0.389 \\ 0  \\ 0.279\end{pmatrix}$ & $\begin{pmatrix} 0.496 \\ 0 \\ 0 \\ 0 \\ 0.504\end{pmatrix}$\\
				\\
				\hline
				\\
				Teenage Employment in& 0.283 & 0.283 & 0.283 & 0.283\\
				Texas in 2014\\
				\\
				Effect of Minimum Wage   &  -8.85 p.p. &    -6.80 p.p. & -8.70 p.p. & -6.80 p.p.  \\
				Increase on Employment &  (or -31.3\%) & (or -24.0\%) & (or -30.7\%) & (or -24.0\%)\\
				
				\\
				Aggregate Variables &  & \checkmark & & \checkmark\\
				\\
			  More Source Regions &  & &  \checkmark & \checkmark\\

				\\
				\hline 
				\multicolumn{1}{c}{} &  &    \tabularnewline
			\end{tabular}
			\par\end{centering}
		\parbox{6.4in}{\footnotesize
			Notes: The table presents the results from synthetic decomposition for increasing minimum wages in Texas to US\$9 on (teenage) employment in 2014. $\theta_0$ represents our parameter of interest, which is the predicted average (teenage) employment after the policy, keeping the population in Texas in 2014 the same. This prediction uses information from the target region (Texas) and source regions. We present its estimates across two specifications: one using only individual-level covariates (age, sex, married, hispanic, black, other race), and another which further includes aggregate variables (teen share in the state, average inflation rate in the state). We then use two different sets of source regions. The confidence interval for $\theta_0$ is presented in brackets. For comparability, we also present the empirical average (pre-policy), and how the estimated $\theta_0$ translates to changes in employment relative to the data (the baseline employment in Texas is 0.292). We also present estimates for the weights, $\boldsymbol{w}_0$.}
	\end{table}
	
	Table \ref{emp} presents the results of the estimation. We present two specifications per exercise, which only differ in whether they accommodate aggregate variables: the share of teenagers in the state population and the average inflation in the state. 
	
	Our estimates suggest that an increase in the minimum wage decreases predicted average (teenage) employment: our estimates of $\theta_0$ and all upper bounds of their associated confidence intervals are all below the observed employment rate of 0.283. In particular, the counterfactual employment is estimated between 0.195-0.215, implying a decrease in average (teenage) employment between 6.8-8.9 percentage points. This is robust across specifications and consistent with the labor economics literature finding such negative effects (see \cite{Neumark:19:GER} for a review). In terms of magnitudes, it is also very similar to those found in \cite{Flinn:06:Ecma} with a similar proportional increase in minimum wages from US\$5 to US\$6 -- see his Figure 4. 

Our synthetic comparison is predominantly based on California and Washington. This seems intuitive, as California best approximates the demographics of Texas. However, our estimates also suggest that accounting for common shocks/aggregate variables is important. Absent state-specific economic trends, we would have estimated the effects of minimum wages on employment to be about 2 percentage points higher, thereby overestimating its negative effects. The aggregate variables also matter for the weights given to source regions: because state-level variables change the model's causal structure, as well as the characteristics of those states, there is no reason why each region would remain equally comparable to Texas with/without them. In fact, we find that California receives lower weights when including such variables. This is because its state unemployment levels are much larger than Texas's which, in turn, is more similar to Washington's.

To further validate our exercise, we implement the test implied by the synthetic transferability condition, based on equation (\ref{tilde C}). We easily find that $\tilde{C}_{1-\alpha} \neq \varnothing$ for $\alpha = 0.05$ : for example, values of $w$ close to the estimates in Table \ref{emp} are not rejected. Thus, we do not find evidence against the synthetic transferability condition.

\begin{figure}[t]
	\begin{center}
				\caption{Overlap in the Estimated Supports of $\mu_0(X_{i,T}, \underline W_{0,T})$ and $\mu_0(X_{i,T}, \underline W_0^*)$}
		\includegraphics[scale=0.27]{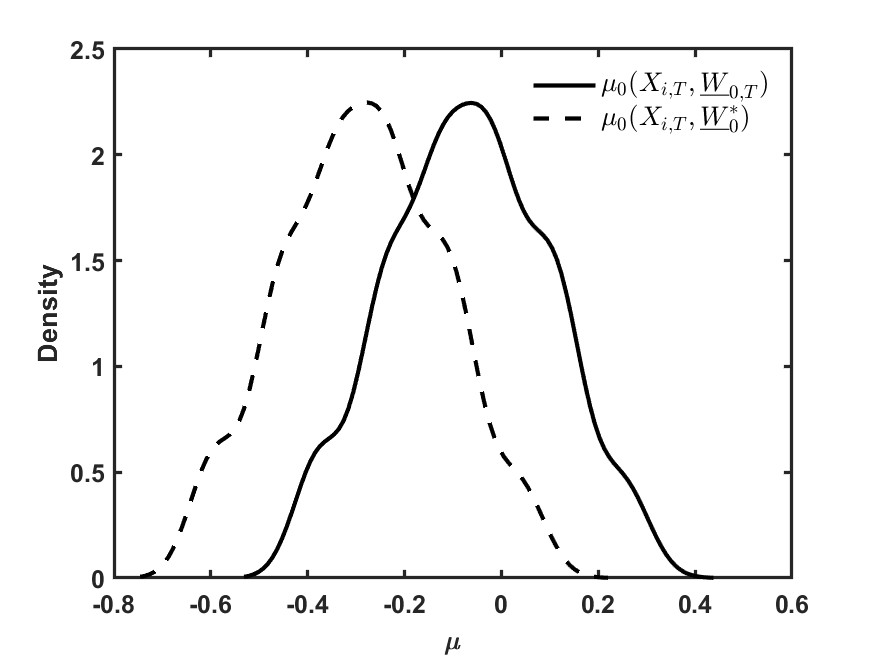}
		~
		\includegraphics[scale=0.27]{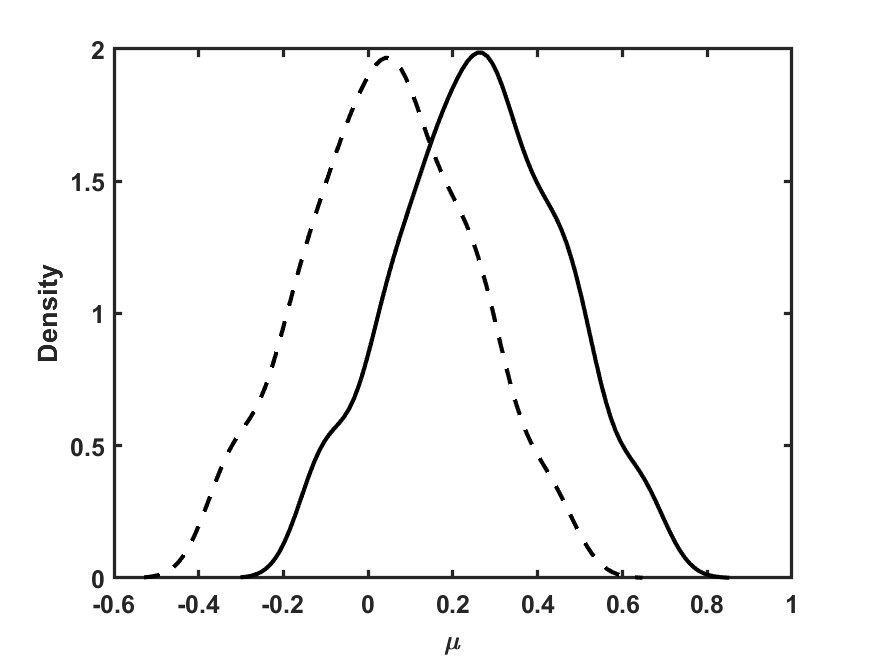}
		\label{conditiona_support}
	\end{center}
	\parbox{6.4in}{\footnotesize
		Notes: The figures present the estimated distribution of $\mu_0(X_{i,T}, \underline W_{0,T})$ and $\mu_0(X_{i,T}, \underline W_{0}^*)$ for Columns 1 and 3 of Table \ref{emp} (left) and Columns 2 and 4 of Table \ref{emp} (right). The policy location shifts the density of $\mu_0(X_{i,T}, \underline W_{0,T})$ by the change in minimum wages, $\log(7.25)-\log(9)$. We can see that the matched group has a substantial overlap.}
\end{figure}

Finally, we remark that we find extensive evidence upholding the support conditions in Assumptions \ref{assump: support target} and \ref{assump: support source} in this setting. First, for the support condition in Assumption \ref{assump: support target}, we estimate that the overlap of $\mu_0(X_{i,t}, \underline W_{0,T})$ and that of $\mu_0(X_{i,t}, \underline W_0^*)$ in our main specifications is between 67.6\% (for Columns 1 and 3) and 72.9\% (Columns 2 and 4 of Table \ref{emp}). This is represented graphically in Figure \ref{conditiona_support}, where we plot the densities of both distributions across $X_{i,t}$. Our policy is represented by a shift in the distribution of $\mu_0$ by $\log(7.25)-\log(9)$ (the change in the minimum wages).

As for the support condition for the source populations in Assumption \ref{assump: support source}, Table \ref{support_b} in the Supplemental Note provides evidence confirming the plausibility of the support condition in our context. As we can see, for most source regions and most specifications, the support condition is satisfied empirically. Even when the estimated support $\left\{\mu_k(x, \underline W_0^*): x \in \mathcal{X}_{0,T}\right\}$ is not fully contained in the estimated $\left\{\mu_k(x, \underline W_0^*): x \in \mathcal{X}_{k,t}, t=1,...,T\right\}$, the share of observations that are outside of the support is very small. Of course, this is not immediately evidence of violation of the support condition, because the reported sets are subject to estimation errors.

\section{Conclusion}

In this paper, we propose a novel way to utilize data from other populations to generate counterfactual predictions for a target population, when we do not have enough data for the latter. We explore ways to utilize data from other populations (``source populations''), motivated by a synthetic transferability condition. This hypothesis generalizes existing invariance conditions for extrapolation of causal effects and allows us to build predictions based on a synthetic causal structure, chosen to be as close as possible to the target conditional average outcome under a certain metric. Our approach is quite general and applies to various policy settings where the researcher may have multiple source populations, regardless of how the reduced forms are originated structurally.

There are further extensions that one can explore from this research. First, it is possible that, just like in synthetic control methods, using many source populations may cause overfitting. As in synthetic control, a judicious selection of source populations based on the domain knowledge of the context of application is important in practice. We believe that a decision-theoretic guidance in this selection would be helpful, although to the best of our knowledge, the predominant portion of the literature focuses on a decision setting under a single population. Second, it would be useful to statistically gauge the plausibility of the synthetic transferability condition. For this, we may need to sacrifice the generality of this paper's setting and make use of further restrictions on the conditional average outcomes, such as continuity or shape constraints of the conditional average outcomes, depending on the application of focus. Finally, the current paper has assumed that the policy is known to the researcher. However, in practice, the precise form of the policy may be unknown. The researcher may face a range of policies under consideration, or may not have precise knowledge of how the policy alters the reduced form, and may need to estimate it using additional data. This question seems relevant in practice.

\putbib[SyntheticDecomp]
\end{bibunit}

\begin{bibunit}[econometrica] 
	\pagebreak
	\renewcommand\thesection{\Alph{section}}

\begin{center}
	\Large \textsc{Supplemental Note to ``Synthetic Decomposition for Counterfactual Predictions''}
\end{center}

\date{%
	\today%
}

\setcounter{section}{0}
\setcounter{subsection}{0}
\setcounter{equation}{0}
\setcounter{lemma}{0}

\vspace*{3ex minus 1ex}
\begin{center}
	Nathan Canen and Kyungchul Song\\
	\textit{University of Warwick \& CEPR and University of British Columbia}
	\bigskip
\end{center}

The supplemental note provides the proofs of the asymptotic validity of inference proposed in the main text, and some details on the empirical application.

\section{Uniform Asymptotic Validity}
\setcounter{equation}{0}
\renewcommand\theequation{A.\arabic{equation}}

\subsection{Notation}

Let us introduce notation that is used throughout the proofs. Recall that we assume the multi-period setting in Sections \ref{subsec: aggregate shocks} and \ref{sec:estimation}, where the observations are made over multiple periods $t=1,...,T$, with repeated cross-sections. Also, recall the notation, $N_k$, which denotes the set of the sample units for each region $k=0,1,...,K$, in (\ref{N_k}) in Section \ref{sec:estimation}. We define 
\begin{align}
    \label{N_k app}
	N = N_{0,T} \cup \bigcup_{k=0}^K N_k.
\end{align}
We let $n_{k,t} = |N_{k,t}|$ be the size of the sample from region $k$ in period $t$ and let $n_k = |N_k|$ and $n = |N|$.

The random quantities constructed from data involve a multi-dimensional sample size: 
\begin{align*}
	\boldsymbol{n} = (n_0,n_1,...,n_K) \in \mathbb{N}^{K+1},
\end{align*}
where $\mathbb{N} = \{1,2,...\}$, the set of natural numbers. Our asymptotic theory is based on the joint asymptotic behavior of the sample sizes, $\boldsymbol{n}$, as $n_0,n_1,...,n_K \rightarrow \infty$. (Recall that we assume that $T$ is fixed.) Let $\mathbf{R}_{++} = \{x \in \mathbf{R}: x >0\}$. For each $\boldsymbol{r} = (r_1,...,r_K)\in \mathbf{R}_{++}^K$ and $\epsilon>0$, we also define 
\begin{align*}
	\mathbb{N}_{K+1}(\boldsymbol{r},\epsilon) = \left\{ \boldsymbol{n} \in \mathbb{N}^{K+1}: \max_{0 \le k \le K} \left|n_k/n_{0,T} - r_k\right| \le \epsilon\right\}.
\end{align*}
Now, we define a collection of sets of multi-sample sizes that proceed to infinity: 
\begin{align*}
	\mathcal{N}_{K+1} = \left\{A \subset \mathbb{N}^{K+1}: |A \cap \mathbb{N}_{K+1}(\boldsymbol{v})| = \infty, \text{ for each } \boldsymbol{v} \in \mathbb{N}^{K+1}\right\},
\end{align*}
where 
\begin{align*}
	\mathbb{N}_{K+1}(\boldsymbol{v}) = \left\{ \boldsymbol{n} \in \mathbb{N}^{K+1}: n_k > v_k, \forall k=0,1,...,K\right\},
\end{align*}
and $\boldsymbol{v} = (v_0,...,v_K)$. For each $\boldsymbol{r}\in \mathbf{R}_{++}^K$, let
\begin{align*}
	\mathcal{N}_{K+1}(\boldsymbol{r}) = \left\{A \in \mathcal{N}_{K+1}: A \cap \mathbb{N}_{K+1}(\boldsymbol{r},\epsilon) \ne \varnothing \text{ for all } \epsilon > 0 \right\}.
\end{align*}
The set $\mathcal{N}_{K+1}(\boldsymbol{r})$ is the collection of infinite sets of multi-sample sizes $\{\boldsymbol{n}\}$ such that $n_k/n_{0,T} \rightarrow r_k$ in the set, as $n_0,n_1,...,n_K \rightarrow \infty$. From now, we denote a generic element of $\mathcal{N}_{K+1}(\boldsymbol{r})$ by $\{\boldsymbol{n}\}$. 

For each $\boldsymbol{r} = (r_1,...,r_K)\in \mathbf{R}_{++}^K$, we introduce a joint asymptotic scheme.\footnote{Note that \cite{Phillips/Moon:99:Eca} considered unrestricted joint asymptotics for the cross-section dimension and the time dimension in the context of nonstationary panel data setting.} In this scheme, for any set $\{a_{\boldsymbol{n}} \in \mathbf{R}^d: \boldsymbol{n} \in \mathbb{N}^{K+1}\}$, if we write 
\begin{align*}
	a_{\boldsymbol{n}} \rightarrow a, \enspace (\boldsymbol{n} \rightarrow \infty)_{\boldsymbol{r}}, 
\end{align*}
we mean that along any increasing sequence of multi-sample sizes $\{\boldsymbol{n}_\ell: \ell = 1,2,...\} \in \mathcal{N}_{K+1}(\boldsymbol{r})$, we have $a_{\boldsymbol{n}_\ell} \rightarrow a, \text{ as } \ell \rightarrow \infty.$ Similarly, if we write  
\begin{align*}
	\enspace \liminf_{(\boldsymbol{n} \rightarrow \infty)_{\boldsymbol{r}}} a_{\boldsymbol{n}} = a, \text{ or } \limsup_{(\boldsymbol{n} \rightarrow \infty)_{\boldsymbol{r}}} a_{\boldsymbol{n}} = a,
\end{align*}
we mean that $\liminf_{\ell \rightarrow \infty} a_{\boldsymbol{n}_\ell} = a$, or $\limsup_{\ell \rightarrow \infty} a_{\boldsymbol{n}_\ell} = a$, respectively. 

Since we consider short time periods for all populations, we regard the aggregate variables $v_{k,t}$ and the counterfactual quantity $v_{0,T}^*$ as constants. For each $k=0,1,...,K$, we let $\mathcal{P}_k$ be the space of probability distributions of $(X_{i,t},U_{i,t})$, $i \in N_k$, and $t=1,...,T$, which satisfy Assumptions \ref{assump: nontrivial smple size}-\ref{assump: lamda min} below. Then the set of joint distributions of $(X_{i,t},U_{i,t})$, $i \in N$, and $t=1,...,T$, is denoted by $\mathcal{P}$. We assume that these distributions are induced from the probability measures on the sample space that is common for all $\boldsymbol{n} \in \mathbb{N}^{K+1}$. Note that due to the assumption that the random variables are i.i.d.\ across $i$ and $t$ within each region, the collection $\mathcal{P}$ depends on $K$ but does not depend on $\boldsymbol{n}$ or $T$.

For any $\boldsymbol{n}$-indexed random vectors $Z_{\boldsymbol{n}}$ and $W_{\boldsymbol{n}}$ in $\mathbf{R}^d$ as functions of $(X_{i,t},U_{i,t})$, $i \in N$, and $t=1,...,T$, we denote 
\begin{align*}
	Z_{\boldsymbol{n}} = W_{\boldsymbol{n}} + o_{\mathcal{P}}(1), \enspace (\boldsymbol{n} \rightarrow \infty)_{\boldsymbol{r}},
\end{align*}
if, for each $\epsilon>0$,
\begin{align*}
	\sup_{P \in \mathcal{P}} P\left\{ \| Z_{\boldsymbol{n}} - W_{\boldsymbol{n}} \| > \epsilon \right\} \rightarrow 0, \enspace (\boldsymbol{n} \rightarrow \infty)_{\boldsymbol{r}}.
\end{align*}
Similarly, we write  
\begin{align*}
	Z_{\boldsymbol{n}} = W_{\boldsymbol{n}} + O_{\mathcal{P}}(1), \enspace (\boldsymbol{n} \rightarrow \infty)_{\boldsymbol{r}},
\end{align*}
if, for each $\epsilon>0$, there exists $M_\epsilon>0$ such that
\begin{align*}
	\limsup_{(\boldsymbol{n} \rightarrow \infty)_{\boldsymbol{r}}}\sup_{P \in \mathcal{P}} P\left\{ \| Z_{\boldsymbol{n}} - W_{\boldsymbol{n}} \| > M_\epsilon \right\} < \epsilon.
\end{align*}

For the proofs below, recall that we focus on the case where $\mathbb{W}_P = \{\boldsymbol{w}_0\}$ for some $\boldsymbol{w}_0 \in \Delta_{K-1}$. To facilitate the presentation of the uniform asymptotic validity, we make explicit the dependence of $\boldsymbol{w}_0$, $\theta(\boldsymbol{w}_0)$, and $\Omega$ on $P \in \mathcal{P}$ by rewriting them as $\boldsymbol{w}_P$, $\theta_P(\boldsymbol{w}_P)$ and $\Omega_P$. Similarly we write $H_P$ and $\boldsymbol{h}_P$ instead of $H$ and $\boldsymbol{h}$, and write $\mu_{k,P}$, $\mu_{k,P}(\cdot,v_{0,T}^*)$ and $m_{k,P}$ instead of $\mu_{k}(\cdot)$, $\mu_{k}(\cdot,v_{0,T}^*) $, and $m_k$. We also write $\mathbf{E}_P$ to make explicit that the expectation is taken with respect to the distribution $P \in \mathcal{P}$.

\subsection{Assumptions and the Results}

As for the random vectors $X_{i.t}$ and $U_{i,t}$, we make the following assumptions.

\begin{assumption}
	\label{assump: nontrivial smple size}
	(i) The random vectors, $(X_{i,t},U_{i,t})$, are independent across $i \in N$ and $t=1,2,...,T$.
	
	(ii) For each $k=0,1,...,K$, the random vectors, $(X_{i,t},U_{i,t})$, are identically distributed across $i \in N_{k,t}$ and $t=1,...,{T}$.
\end{assumption}

The nonstandard aspect of uniform asymptotic validity in our setting comes from the fact that $\sqrt{n_{0,T}}(\boldsymbol{\hat w} - \boldsymbol{w}_P)$ exhibits discontinuity in its pointwise asymptotic distribution. Hence, our proof focuses on dealing with this aspect, invoking high level conditions for other aspects that can be handled using standard arguments.

\begin{assumption}
	\label{assump: conditional average outcome bound}
	For each $k=0,1,...,K$, there exists $\delta>0$ such that
	\begin{align*}
		&\sup_{P \in \mathcal{P}} \mathbf{E}_P\left[ \left|m_{k,P}\left(\mu_{k,P}(X_{i,T},v_{0,T}), X_{i,T}\right)\right|^{4+ \delta}\right] < \infty \text{ and } \\
		&\sup_{P \in \mathcal{P}} \mathbf{E}_P\left[ \left|m_{k,P}(\mu_{k,P}(X_{i,T},v_{0,T}^*), X_{i,T}) \right|^{4+ \delta}\right] < \infty.
	\end{align*}
\end{assumption}

Assumption \ref{assump: conditional average outcome bound} requires that the conditional average outcomes have some moment bounded uniformly over $P \in \mathcal{P}$.

\begin{assumption}
	\label{assump: lamda min}
	There exists $\eta>0$ such that
	\begin{align*}
		\inf_{P \in \mathcal{P}} \lambda_{\min}(H_P) > \eta,
	\end{align*}
where $\lambda_{\min}(H_P)$ denotes the smallest eigenvalue of $H_P$.
\end{assumption}

Assumption \ref{assump: lamda min} requires that the matrix $H_P$ has eigenvalues bounded away from zero uniformly over $P \in \mathcal{P}$. The assumption excludes a setting where $\boldsymbol{w}_P$ is weakly identified. Later in Section \ref{subsec: without identifiability}, we will discuss how this assumption can be relaxed.

Recall the definition $W_i = (\tilde Y_i, \tilde X_i)$, where $(\tilde Y_i, \tilde X_i)$ is defined in Section \ref{sec:estimation} in the main text. For each $k = 0,1,...,K$, let us define
\begin{align*}
	q_{k,0,P}(X_{i,T}) &= m_{k,P}\left(\mu_{k,P}(X_{i,T},v_{0,T}^*),X_{i,T} \right) 1\left\{X_{i,T} \in \mathcal{X}_{0,T}^\mathsf{M} \right\} \text{ and }\\
	\hat q_{k,0}(X_{i,T}) &= \hat m_{k}\left(\hat \mu_k(X_{i,T},v_{0,T}^*),X_{i,T} \right) 1\left\{X_{i,T} \in \mathcal{\hat X}_{0,T}^\mathsf{M} \right\}.
\end{align*} 
Similarly, we define $q_{k,1,P}(X_{i,T})$ and $\hat q_{k,1}(X_{i,T})$ to be the same as $q_{k,0,P}(X_{i,T})$ and $\hat q_{k,0}(X_{i,T})$ except that $1\{X_{i,T} \in \mathcal{X}_{0,T}^\mathsf{M}\}$ and $1\{X_{i,T} \in \mathcal{\hat X}_{0,T}^\mathsf{M}\}$ are replaced by $1\{X_{i,T} \in \mathcal{X}_{0,T}^\mathsf{U}\}$ and $1\{X_{i,T} \in \mathcal{\hat X}_{0,T}^\mathsf{U}\}$ respectively.

The following assumption requires the asymptotic linear representation of the estimated conditional average outcomes.

\begin{assumption}
	\label{assump: asym lin}
	Suppose that for each $k=0,1,...,K$, $\ell=0,1$, $\varphi_{k,\ell,P}(\cdot)$ is equal to $q_{k,\ell,P}(\cdot)$ or a constant function taking number one. Then, for each $j,k=0,1,...,K$, $\ell=0,1$, and $\boldsymbol{r} \in \mathbf{R}_{++}^{K}$,
	\begin{align}
		\label{asym lin}
		&\frac{1}{\sqrt{n_{0,T}}} \sum_{i \in N_{0,T}} \left(\hat q_{j,\ell}(X_{i,T}) -  q_{j,\ell,P}(X_{i,T})\right)\varphi_{k,\ell,P}(X_{i,T}) \\ \notag
		&\quad = \frac{1}{\sqrt{n_j}} \sum_{i \in N_j} \psi_{j,\ell,P}(W_i;\varphi_{k,\ell,P})+ o_{\mathcal{P}}(1), \enspace (\boldsymbol{n} \rightarrow \infty)_{\boldsymbol{r}}, \text{ and }\\ \notag
		&\frac{1}{\sqrt{n_{0,T}}} \sum_{i \in N_{0,T}} \left(\hat q_{j,\ell}(X_{i,T}) -  q_{j,\ell,P}(X_{i,T})\right)\left(\hat q_{k,\ell}(X_{i,T}) - q_{k,\ell,P}(X_{i,T})\right) = o_{\mathcal{P}}(1), \enspace (\boldsymbol{n} \rightarrow \infty)_{\boldsymbol{r}},
	\end{align}
	where $\psi_{j,\ell,P}(W_i;\varphi_{k,\ell,P})$ is a mean zero random variable such that for some $\delta>0$,
	\begin{align}
		\label{psi bound}
		\sup_{P \in \mathcal{P}} \mathbf{E}_P\left[\psi_{j,\ell,P}^{4+\delta}(W_i;\varphi_{k,\ell,P})\right] < \infty,
	\end{align}
	for all $j,k=0,1,...,K$ and $\ell = 0,1$.
\end{assumption}

To understand the plausibility of the assumption, consider the sum: 
\begin{align*}
    \frac{1}{\sqrt{n_{0,T}}} \sum_{i \in N_{0,T}} \left(\hat q_{j,\ell}(X_{i,T}) -  q_{j,\ell,P}(X_{i,T})\right)\varphi_{k,\ell,P}(X_{i,T}).
\end{align*} 
Note that the estimation error in $\hat q_{j,\ell}(\cdot)$ comes from the sample in region $j$, whereas the summation is over the sample in region 0. The influence function is driven by the randomness in the estimation error $\hat q_{j,\ell}(\cdot) - q_{j,\ell,P}(\cdot)$, which comes from the sampling error in region $j$. This is why the asymptotic linear representation on the right hand side of the first equation in (\ref{asym lin}) involves sample units in $N_j$. 

We define the bootstrap analogues: for $k=0,1,...,K$,
\begin{align*}
	\hat q_{k,0}^*(X_{i,T}^*) &= \hat m_{k}^*\left(\hat \mu_k^*(X_{i,T}^*,v_{0,T}^*),X_{i,T}^* \right) 1\left\{X_{i,T}^* \in \mathcal{\hat X}_{0,T}^{\mathsf{M} *} \right\},
\end{align*} 
and similarly define $\hat q_{k,1}^*(X_{i,T}^*)$ to be the same as $\hat q_{k,0}^*(X_{i,T}^*)$ except that $1\{X_{i,T}^* \in \mathcal{\hat X}_{0,T}^{\mathsf{M} *}\}$ are replaced by $1\{X_{i,T}^* \in \mathcal{\hat X}_{0,T}^{\mathsf{U}*}\}$. We make the following assumption for the bootstrap version of the estimators.

\begin{assumption}
	\label{assump: asym lin boot}
	Suppose that for each $k=0,1,...,K$, $\ell=0,1$, $(\hat \varphi_{k,\ell}(\cdot),\varphi_{k,\ell,P}(\cdot))$ is equal to $(\hat q_{k,\ell}(\cdot),q_{k,\ell,P}(\cdot))$ or a pair of constant functions taking number one. Then, for each $j,k=0,1,...,K$, $\ell=0,1$, and $\boldsymbol{r} \in \mathbf{R}_{++}^{K}$, the following statements hold.

	(i)
	\begin{align*}
		&\frac{1}{\sqrt{n_{0,T}}} \sum_{i \in N_{0,T}} \left(\hat q_{j,\ell}^*(X_{i,T}^*) -  \hat q_{j,\ell}(X_{i,T}^*)\right)\hat \varphi_{k,\ell}(X_{i,T}^*) \\
		&\quad = \frac{1}{\sqrt{n}_j} \sum_{i \in N_j} \hat \psi_{j,\ell,P}(W_{i}^*;\varphi_{k,\ell,P}) + o_{\mathcal{P}}(1), \enspace (\boldsymbol{n} \rightarrow \infty)_{\boldsymbol{r}}, \\
		&\frac{1}{\sqrt{n_{0,T}}} \sum_{i \in N_{0,T}} \left(\hat q_{j,\ell}^*(X_{i,T}^*) -  \hat q_{j,\ell}(X_{i,T}^*)\right)\left(\hat q_{k,\ell}^*(X_{i,T}^*) -  \hat q_{k,\ell}(X_{i,T}^*)\right) = o_{\mathcal{P}}(1), \enspace (\boldsymbol{n} \rightarrow \infty)_{\boldsymbol{r}}, \text{ and }\\
		&\frac{1}{\sqrt{n_{0,T}}} \sum_{i \in N_{0,T}} \left(\hat q_{j,\ell}(X_{i,T}^*) -  q_{j,\ell}(X_{i,T}^*)\right)\left(\hat q_{k,\ell}(X_{i,T}^*) -  q_{k,\ell}(X_{i,T}^*)\right) = o_{\mathcal{P}}(1), \enspace (\boldsymbol{n} \rightarrow \infty)_{\boldsymbol{r}}.
	\end{align*}
	where 
    \begin{align*}
        \hat \psi_{j,\ell,P}(W_{i}^*;\varphi_{k,\ell,P}) = \psi_{j,\ell,P}(W_i^*;\varphi_{k,\ell,P}) - \frac{1}{n_j} \sum_{i \in N_j} \psi_{j,\ell,P}(W_i;\varphi_{k,\ell,P}),
    \end{align*}
    and $\psi_{j,\ell,P}(\cdot;\varphi_{k,\ell,P})$ is the influence function in Assumption \ref{assump: asym lin}.

	(ii)
	\begin{align*}
		&\frac{1}{\sqrt{n_{0,T}}} \sum_{i \in N_{0,T}} \left(\hat q_{j,\ell}(X_{i,T}^*) \hat q_{k,\ell}(X_{i,T}^*) - \frac{1}{n_{0,T}}\sum_{i \in N_{0,T}} \hat q_{j,\ell}(X_{i,T}) \hat q_{k,\ell}(X_{i,T})\right) \\
		&\quad \quad = \frac{1}{\sqrt{n_{0,T}}} \sum_{i \in N_{0,T}} \left(q_{j,\ell}(X_{i,T}^*) q_{k,\ell}(X_{i,T}^*) - \frac{1}{n_{0,T}}\sum_{i \in N_{0,T}} q_{j,\ell,P}(X_{i,T}) q_{k,\ell,P}(X_{i,T})\right) + o_{\mathcal{P}}(1), \enspace (\boldsymbol{n} \rightarrow \infty)_{\boldsymbol{r}}.
	\end{align*}
\end{assumption}

Condition (i) in Assumption \ref{assump: asym lin boot} is a bootstrap analogue of Assumption \ref{assump: asym lin}. Condition (ii) follows from the stochastic equicontinuity of a bootstrap empirical process.

Define
\begin{align}
	\label{Omega_{n,P}}
	 V_{\boldsymbol{n},P} = \sum_{i \in N} \mathbf{E}_P\left[ \boldsymbol{\tilde \psi}_{i,P} \boldsymbol{\tilde \psi}_{i,P}^{\prime}\right],
\end{align}
where $\boldsymbol{\tilde \psi}_{i,P} = \Psi_{i,P}\boldsymbol{w}_P - \boldsymbol{\psi}_{i,P}$ and $\Psi_{i,P}$ and $\boldsymbol{\psi}_{i,P}$ are defined in Lemma \ref{lemm: asym lin} below. Inspection of $V_{\boldsymbol{n},P}$ shows that it depends on the sample size only through the ratios, $n_k/n_{0,T}$, $k=1,...,K$, and depends on these ratios continuously. For each $\boldsymbol{r} \in \mathbf{R}_{++}^K$, let $V_{P}(\boldsymbol{r})$ be the same as $V_{\boldsymbol{n},P}$ with $n_k/n_{0,T}$ replaced by $r_k$, for $k=1,...,K$, where $r_k$'s are positive constants in Assumption \ref{assump: nontrivial smple size}. Then, it is not hard to see that from Assumption \ref{assump: conditional average outcome bound},
\begin{align}
	\label{Omega P conv}
	\sup_{P \in \mathcal{P}} \left\| V_{\boldsymbol{n},P} - V_P(\boldsymbol{r})\right\| \rightarrow 0, \enspace (\boldsymbol{n} \rightarrow \infty)_{\boldsymbol{r}}.
\end{align}

We assume that $V_P(\boldsymbol{r})$ is positive definite uniformly over all $P \in \mathcal{P}$.

\begin{assumption}
  \label{assump: positive definite V_P}
  For each $\boldsymbol{r} \in \mathbf{R}_{++}^K$, there exists $\eta>0$ such that  
  \begin{align*}
	\inf_{P \in \mathcal{P}} \lambda_{\min}(V_P(\boldsymbol{r})) > \eta.
  \end{align*}
\end{assumption}

The following theorem shows that the estimators $\boldsymbol{\hat w}$ and $\hat \theta(\boldsymbol{\hat w})$ are $\sqrt{n_{0,T}}$-consistent for $\boldsymbol{w}_P$ and $\theta_P(\boldsymbol{w}_P)$ uniformly over $P \in \mathcal{P}$.

\begin{theorem}
	\label{thm: root n consist}
	Suppose that Assumptions \ref{assump: nontrivial smple size}-\ref{assump: positive definite V_P} hold. Then, for any $\epsilon>0$ and any $\boldsymbol{r} \in \mathbf{R}_{++}^K$, there exists $M_{\epsilon,\boldsymbol{r}} > 0$ such that 
	\begin{align*}
		\limsup_{(\boldsymbol{n} \rightarrow \infty)_{\boldsymbol{r}}} \sup_{P \in \mathcal{P}} P\left\{ \sqrt{n_{0,T}}\left\| \boldsymbol{\hat w} - \boldsymbol{w}_P  \right\| > M_{\epsilon,\boldsymbol{r}} \right\} < \epsilon.
	\end{align*}
\end{theorem}	

However, as noted earlier, depending on the sequence of probabilities in $\mathcal{P}$, $\sqrt{n_{0,T}}(\boldsymbol{\hat w} - \boldsymbol{w}_P)$ can be asymptotically non-normal, and so can $\sqrt{n_{0,T}}(\hat \theta(\boldsymbol{\hat w}) - \theta_P(\boldsymbol{w}_P))$ as a consequence. Nevertheless, the confidence interval $C_{1-\alpha}$ we propose in the main text turns out to be uniformly asymptotically valid as the following theorem shows.

\begin{theorem}
	\label{thm: validity}
	Suppose that Assumptions \ref{assump: nontrivial smple size}-\ref{assump: positive definite V_P} hold. Then, for each $\alpha \in (0,1)$ and $\boldsymbol{r} \in \mathbf{R}_{++}^K$,
	\begin{align*}
		\liminf_{(\boldsymbol{n} \rightarrow \infty)_{\boldsymbol{r}}} \inf_{P \in \mathcal{P}} P\left\{  \theta_P(\boldsymbol{w}_P)  \in C_{1-\alpha} \right\} \ge 1 - \alpha.
	\end{align*}
\end{theorem}

The proofs of these results are presented in the next section.

\subsection{Proofs}

Throughout the proofs below, we assume that Assumptions \ref{assump: nontrivial smple size}-\ref{assump: positive definite V_P} are satisfied.

\subsubsection{The Proof of Theorem \ref{thm: root n consist}}

Define
\begin{align*}
	\boldsymbol{\hat G}_P = \sqrt{n_{0,T}}( \hat H - H_P) \text{ and } \boldsymbol{\hat g}_P = \sqrt{n_{0,T}}(\boldsymbol{\hat h} - \boldsymbol{h}_P).
\end{align*}
The following lemma gives an asymptotic linear representation for $\boldsymbol{\hat G}_P$ and $\boldsymbol{\hat g}_P$.

\begin{lemma}
	\label{lemm: asym lin}
	For any $\boldsymbol{r} \in \mathbf{R}_{++}^K$,
	\begin{align*}
		\boldsymbol{\hat G}_P = \sum_{i \in N} \Psi_{i,P} + o_{\mathcal{P}}(1), \enspace (\boldsymbol{n} \rightarrow \infty)_{\boldsymbol{r}} \text{ and } 
		\boldsymbol{\hat g}_P = \sum_{i \in N} \boldsymbol{\psi}_{i,P} + o_{\mathcal{P}}(1), \enspace (\boldsymbol{n} \rightarrow \infty)_{\boldsymbol{r}},
	\end{align*}
	where $\Psi_{i,P}$ is the $K \times K$ matrix whose $(j,k)$-entry is given by
	\begin{align*}
		\psi_{i,P, jk} &= \frac{1}{\sqrt{n_j}} \psi_{j,0,P}(W_i;q_{k,0,P}) 1\{i \in N_j\} + \frac{1}{\sqrt{n_k}} \psi_{k,0,P}(W_i;q_{j,0,P}) 1\{i \in N_k\}\\
		&\quad + \frac{1}{\sqrt{n_{0,T}}}\left\{q_{j,0,P}(X_{i,T})q_{k,0,P}(X_{i,T}) - \mathbf{E}_P\left[q_{j,0,P}(X_{i,T})q_{k,0,P}(X_{i,T})\right]\right\} 1\{i \in N_{0,T}\},
	\end{align*}
	and $\boldsymbol{\psi}_{i,P}$ is the $K \times 1$ vector whose $k$-th entry is given by
	\begin{align*}
		\psi_{i,P, k} &= \frac{1}{\sqrt{n_k}} \psi_{k,0,P}(W_i;q_{0,0,P}) 1\{i \in N_k\} + \frac{1}{\sqrt{n_0}} \psi_{0,0,P}(W_i;q_{k,0,P}) 1\{i \in N_{0}\} \\
		&\quad \quad + \frac{1}{\sqrt{n_{0,T}}}\left\{q_{k,0,P}(X_{i,T})q_{0,0,P}(X_{i,T}) - \mathbf{E}_P\left[q_{k,0,P}(X_{i,T})q_{0,0,P}(X_{i,T})\right]\right\}1\{i \in N_{0,T}\}.
	\end{align*}
\end{lemma}

\noindent \textbf{Proof: } For $j,k=1,...,K$, let $\hat H_{jk}$ be the $(j,k)$-th entry of $\hat H$ and $H_{P,jk}$ the $(j,k)$-th entry of $H_P$. As for the first statement, for each $j,k=1,...,K$, we write
\begin{align*}
	\sqrt{n_{0,T}}(\hat H_{jk} - H_{P,jk}) &= \frac{1}{\sqrt{n_{0,T}}}\sum_{i \in N_{0,T}} (\hat q_{j,0}(X_{i,T}) - q_{j,0,P}(X_{i,T})) \hat q_{k,0}(X_{i,T})\\
	&\quad \quad + \frac{1}{\sqrt{n_{0,T}}}\sum_{i \in N_{0,T}} (\hat q_{k,0}(X_{i,T}) - q_{k,0,P}(X_{i,T})) q_{j,0,P}(X_{i,T})\\
	&\quad \quad + \frac{1}{\sqrt{n_{0,T}}}\sum_{i \in N_{0,T}}\left\{q_{j,0,P}(X_{i,T})q_{k,0,P}(X_{i,T}) - \mathbf{E}_P\left[q_{j,0,P}(X_{i,T})q_{k,0,P}(X_{i,T})\right]\right\}.
\end{align*}
By Assumption \ref{assump: asym lin}, we find that
\begin{align*}
	\sqrt{n_{0,T}}(\hat H_{jk} - H_{P,jk}) &= \frac{1}{\sqrt{n_j}}\sum_{i \in N_j} \psi_{j,0,P}(W_i;q_{k,0,P}) + \frac{1}{\sqrt{n_k}} \sum_{i \in N_k} \psi_{k,0,P}(W_i;q_{j,0,P}) \\
	&\quad + \frac{1}{\sqrt{n_{0,T}}}\sum_{i \in N_{0,T}}\left\{q_{j,0,P}(X_{i,T})q_{k,0,P}(X_{i,T}) - \mathbf{E}_P\left[q_{j,0,P}(X_{i,T})q_{k,0,P}(X_{i,T})\right]\right\}\\
	&\quad + o_{\mathcal{P}}(1), \enspace (\boldsymbol{n} \rightarrow \infty)_{\boldsymbol{r}}.
\end{align*}
The proof for the second statement is similar and is omitted. $\blacksquare$

\begin{lemma}
		\label{lemm: consistency H and h}
		For any $\boldsymbol{r} \in \mathbf{R}_{++}^K$, $\hat H = H_P + o_{\mathcal{P}}(1)$, $(\boldsymbol{n} \rightarrow \infty)_{\boldsymbol{r}}$ \text{ and } $\boldsymbol{\hat h} = \boldsymbol{h}_P + o_{\mathcal{P}}(1)$, $(\boldsymbol{n} \rightarrow \infty)_{\boldsymbol{r}}$.
\end{lemma}

\noindent \textbf{Proof: } Since $\sup_{P \in \mathcal{P}} \mathbf{E}_P\left[\left\| \Psi_{i,P}\right\|^2\right] < \infty$ and $\sup_{P \in \mathcal{P}} \mathbf{E}_P\left[\left\| \boldsymbol{\psi}_{i,P}\right\|^2\right] < \infty$, the result is immediate from Lemma \ref{lemm: asym lin}. $\blacksquare$\medskip

For each $\boldsymbol{w} \in \Delta_{K-1}$, we define
\begin{align*}
	\mathcal{\hat M}(\boldsymbol{w}) = \boldsymbol{w}^{\prime} \hat H \boldsymbol{w} - 2 \boldsymbol{w}'\boldsymbol{\hat h}, \text{ and } \mathcal{M}_P(\boldsymbol{w}) = \boldsymbol{w}^{\prime} H_P \boldsymbol{w} - 2 \boldsymbol{w}'\boldsymbol{h}_P.	
\end{align*}

\begin{lemma}
	\label{lemm: consistency}
	For any $\boldsymbol{r} \in \mathbf{R}_{++}^K$, $\boldsymbol{\hat w} = \boldsymbol{w}_P + o_{\mathcal{P}}(1)$, $(\boldsymbol{n} \rightarrow \infty)_{\boldsymbol{r}}$.
\end{lemma}	

\noindent \textbf{Proof}: First, we prove the following two claims.

(i) For each $\epsilon >0$,
\begin{align*}
	\sup_{P \in \mathcal{P}}P\left\{\sup_{\boldsymbol{w} \in \Delta_{K-1}} |\mathcal{\hat M}(\boldsymbol{w}) - \mathcal{M}_P(\boldsymbol{w}) | > \epsilon \right\} \rightarrow 0, \enspace (\boldsymbol{n} \rightarrow \infty)_{\boldsymbol{r}}.
\end{align*}

(ii) For each $\epsilon >0$,
\begin{align*}
	\inf_{P \in \mathcal{P}} \inf_{\boldsymbol{w} \in \Delta_{K-1} \setminus B(\boldsymbol{w}_P: \epsilon)} \left\{ \mathcal{M}_P(\boldsymbol{w}) -  \mathcal{M}_P(\boldsymbol{w}_P) \right\} > 0,
\end{align*}
where $B(\boldsymbol{w}_P; \epsilon) = \{\boldsymbol{w} \in \Delta_{K-1}: \|\boldsymbol{w} - \boldsymbol{w}_P\| < \epsilon\}$. 

Let us prove (i) first. For each $\boldsymbol{w} \in \Delta_{K-1}$, we write
\begin{align*}
	\mathcal{\hat M}(\boldsymbol{w}) - \mathcal{M}_P(\boldsymbol{w}) = \boldsymbol{w}^{\prime} (\hat H - H_P)\boldsymbol{w} - 2 (\boldsymbol{\hat h} - \boldsymbol{h}_P)^{\prime} \boldsymbol{w}.
\end{align*}
The desired result of (i) follows by Lemma \ref{lemm: consistency H and h}.

Let us turn to (ii). Note that
\begin{align}
	\label{M diff}
	\mathcal{M}_P(\boldsymbol{w}) - \mathcal{M}_P(\boldsymbol{w}_P) &= (\boldsymbol{w} - \boldsymbol{w}_P)^{\prime} H_P(\boldsymbol{w} - \boldsymbol{w}_P) + 2(\boldsymbol{w} - \boldsymbol{w}_P)^{\prime} (H_P \boldsymbol{w}_P - \boldsymbol{h}_P)\\ \notag
	&\ge \inf_{P \in \mathcal{P}} \lambda_{\min}(H_P) \| \boldsymbol{w} - \boldsymbol{w}_P\|^2,
\end{align}
because $(\boldsymbol{w} - \boldsymbol{w}_P)^{\prime} (H_P \boldsymbol{w}_P - \boldsymbol{h}_P) \ge 0$ for all $\boldsymbol{w} \in \Delta_{K-1}$ by the definition of $\boldsymbol{w}_P$. (See, e.g., Propositions 2.1.5 and 2.3.2 of \cite{Clarke:90:Optimization}.) The desired result follows from Assumption \ref{assump: lamda min}. 

Now that we have (i) and (ii), we follow the arguments in the proof of Theorem 2.1 of \cite{Newey/McFadden:94:Handbook} to complete the proof. More specifically, we invoke (ii) and take $\epsilon>0$, $\eta_\epsilon>0$ and $n _\epsilon$ such that for all $n_0,n_1,...,n_K \ge n_\epsilon$,
\begin{align*}
	\inf_{P \in \mathcal{P}} \inf_{\boldsymbol{w} \in \Delta_{K-1} \setminus B(\boldsymbol{w}_P: \epsilon)} \left\{ \mathcal{M}_P(\boldsymbol{w}) -  \mathcal{M}_P(\boldsymbol{w}_P) \right\} > \eta_\epsilon.
\end{align*}
The event of $\| \boldsymbol{\hat w} - \boldsymbol{w}_P \| > \epsilon$ implies $\mathcal{M}_P(\boldsymbol{\hat w}) - \mathcal{M}_P(\boldsymbol{w}_P) > \eta_\epsilon$, or 
\begin{align*}
	\mathcal{\hat M}(\boldsymbol{w}_P) - \mathcal{M}_P(\boldsymbol{w}_P) \ge \mathcal{\hat M}(\boldsymbol{\hat w}) - \mathcal{M}_P(\boldsymbol{\hat w}) + \eta_{\epsilon},
\end{align*}
where we use that $\mathcal{\hat M}(\boldsymbol{\hat w}) \le \mathcal{\hat M}(\boldsymbol{w}_P)$. The probability of this event is bounded by
\begin{align*}
	\sup_{P \in \mathcal{P}} P\left\{ \sup_{\boldsymbol{w} \in \Delta_{K-1}} |\mathcal{\hat M}(\boldsymbol{w}) - \mathcal{M}_P(\boldsymbol{w}) | \ge \frac{\eta_\epsilon}{2}  \right\} \rightarrow 0,\enspace (\boldsymbol{n} \rightarrow \infty)_{\boldsymbol{r}},
\end{align*}
by (i). Hence, we obtain the desired result of the lemma. $\blacksquare$

\begin{lemma}
	\label{lemm: uniform converg}
	(i) For any $\epsilon>0$ and $\boldsymbol{r} \in \mathbf{R}_{++}^K$, there exists $M>0$ such that
	\begin{align*}
		\limsup_{(\boldsymbol{n} \rightarrow \infty)_{\boldsymbol{r}}} \sup_{P \in \mathcal{P}} P\left\{\sup_{\boldsymbol{w} \in \Delta_{K-1}} \left| \mathcal{\hat M}(\boldsymbol{w}) - \mathcal{M}_P(\boldsymbol{w}) \right| > M n_{0,T}^{-1/2} \right\} < \epsilon.
	\end{align*}
	
	(ii) For any $\epsilon>0$ and $\boldsymbol{r} \in \mathbf{R}_{++}^K$, there exists $M>0$ such that for any $\delta_{\boldsymbol{n}} \rightarrow 0$, $(\boldsymbol{n} \rightarrow \infty)_{\boldsymbol{r}}$,
	\begin{align*}
		\limsup_{(\boldsymbol{n} \rightarrow \infty)_{\boldsymbol{r}}} \sup_{P \in \mathcal{P}} P\left\{ \sup_{\boldsymbol{w} \in \Delta_{K-1}: \left\| \boldsymbol{w}  - \boldsymbol{w}_P \right\| \le \delta_{\boldsymbol{n}}}\left| \mathcal{\hat M}^\Delta(\boldsymbol{w}) - \mathcal{\hat M}^\Delta(\boldsymbol{w}_P) \right| > M \delta_{\boldsymbol{n}} n_{0,T}^{-1/2} \right\} < \epsilon,
	\end{align*}
	where $\mathcal{\hat M}^\Delta(\boldsymbol{w}) = \mathcal{\hat M}(\boldsymbol{w}) - \mathcal{M}_P(\boldsymbol{w}).$
\end{lemma}

\noindent \textbf{Proof: } (i) First, we write
\begin{align*}
	\sqrt{n_{0,T}} (\mathcal{\hat M}(\boldsymbol{w}) - \mathcal{M}_P(\boldsymbol{w})) &= \boldsymbol{w}^{\prime} \boldsymbol{\hat G}_P \boldsymbol{w} - 2 \boldsymbol{w}^{\prime} \boldsymbol{\hat g}_P\\
	&= \boldsymbol{w}^{\prime} \left(\sum_{i \in N}  \Psi_{i,P}\right) \boldsymbol{w} - 2 \boldsymbol{w}^{\prime}\left(\sum_{i \in N}  \boldsymbol{\psi}_{i,P}\right) + o_{\mathcal{P}}(1),\enspace (\boldsymbol{n} \rightarrow \infty)_{\boldsymbol{r}},
\end{align*}
by Lemma \ref{lemm: asym lin}. The desired result follows, because $\boldsymbol{w} \in \Delta_{K-1}$, and from Assumption \ref{assump: nontrivial smple size} and the condition (\ref{psi bound}), 
\begin{align}
	\label{psi bound 2}
	\sup_{P \in \mathcal{P}} \mathbf{E}_P\left\| \sum_{i \in N}  \Psi_{i,P} \right\|^2 < \infty \text{ and } \sup_{P \in \mathcal{P}}\mathbf{E}_P\left\| \sum_{i \in N}  \boldsymbol{\psi}_{i,P} \right\|^2 < \infty.
\end{align}

(ii) From (\ref{M diff}), we write
\begin{align*}
	\mathcal{\hat M}^\Delta(\boldsymbol{w}) - \mathcal{\hat M}^\Delta(\boldsymbol{w}_P) &= (\boldsymbol{w} - \boldsymbol{w}_P)^{\prime} (\hat H - H_P) (\boldsymbol{w}- \boldsymbol{w}_P) \\
	&\quad + 2 (\boldsymbol{w} - \boldsymbol{w}_P)'((\hat H - H_P)\boldsymbol{w}_P - (\boldsymbol{\hat h} - \boldsymbol{h}_P)).
\end{align*}
Hence, again, by Lemma \ref{lemm: asym lin}, 
\begin{align*}
	&\sqrt{n_{0,T}}(\mathcal{\hat M}^\Delta(\boldsymbol{w}) - \mathcal{\hat M}^\Delta(\boldsymbol{w}_P)) \\
	&\quad = (\boldsymbol{w} - \boldsymbol{w}_P)^{\prime} \left(\sum_{i \in N}  \Psi_{i,P}\right) (\boldsymbol{w} - \boldsymbol{w}_P) - 2 (\boldsymbol{w} - \boldsymbol{w}_P)^{\prime}\left(\sum_{i \in N}  \boldsymbol{\psi}_{i,P}\right) + o_{\mathcal{P}}(1),\enspace (\boldsymbol{n} \rightarrow \infty)_{\boldsymbol{r}}.
\end{align*} 
The desired result follows from (\ref{psi bound 2}). $\blacksquare$

\begin{lemma}
	\label{lemm: aux}
	Suppose that for some positive numbers $\delta_{\boldsymbol{n},1}$ such that $\lim_{(\boldsymbol{n} \rightarrow \infty)_{\boldsymbol{r}}} \delta_{\boldsymbol{n},1} = 0$, we have 
	\begin{align*}
		\lim_{M \uparrow \infty} \limsup_{(\boldsymbol{n} \rightarrow \infty)_{\boldsymbol{r}}} \sup_{P \in \mathcal{P}} P\left\{\|	\boldsymbol{\hat w} - \boldsymbol{w}_P \| > M \delta_{\boldsymbol{n},1} \right\} = 0.
	\end{align*}
	Then, 
	\begin{align*}
		\lim_{M \uparrow \infty} \limsup_{(\boldsymbol{n} \rightarrow \infty)_{\boldsymbol{r}}} \sup_{P \in \mathcal{P}} P\left\{\|	\boldsymbol{\hat w} - \boldsymbol{w}_P \|^2 > M n_{0,T}^{-1/2} \delta_{\boldsymbol{n},1} \right\} = 0.
	\end{align*}
\end{lemma}

\noindent \textbf{Proof: } We take arbitrary $\epsilon>0$ and large $\overline M_\epsilon>0$ such that
\begin{align}
	\label{ineq32}
	\limsup_{(\boldsymbol{n} \rightarrow \infty)_{\boldsymbol{r}}} \sup_{P \in \mathcal{P}} P\left\{\|\boldsymbol{\hat w} -\boldsymbol{w}_P \| > \overline M_\epsilon \delta_{\boldsymbol{n},1}  \right\} \le \epsilon.
\end{align}
Recall the definition $\mathcal{\hat M}^\Delta(\boldsymbol{w}) = 	\mathcal{\hat M}(\boldsymbol{w}) - \mathcal{M}_P(\boldsymbol{w}).$ Since $\mathcal{\hat M}(\boldsymbol{w}_P) \ge \mathcal{\hat M}(\boldsymbol{\hat w})$, we have
\begin{align}
	\label{eqs}
	\mathcal{\hat M}^\Delta(\boldsymbol{w}_P) - \mathcal{\hat M}^\Delta(\boldsymbol{\hat w}) &\ge \mathcal{M}_P(\boldsymbol{\hat w}) - \mathcal{M}_P(\boldsymbol{w}_P) \\ \notag
	&\ge \inf_{P \in \mathcal{P}} \lambda_{\min}(H_P) \| \boldsymbol{\hat w} - \boldsymbol{w}_P\|^2 \ge \eta \| \boldsymbol{\hat w} - \boldsymbol{w}_P\|^2,
\end{align}
from (\ref{M diff}), where $\eta>0$ is the constant in Assumption \ref{assump: lamda min}. Define the event 
\begin{align*}
    E_n(\epsilon) = \left\{\|\boldsymbol{\hat w} -\boldsymbol{w}_P \| > \overline M_\epsilon \delta_{\boldsymbol{n},1}  \right\}.
\end{align*}
By Lemma \ref{lemm: uniform converg}(ii), for any $\epsilon_1>0$, there exists $M_{\epsilon_1} >0$ such that $M_{\epsilon_1} \rightarrow \infty$ as $\epsilon_1 \rightarrow 0$, and
\begin{align*}
	\limsup_{(\boldsymbol{n} \rightarrow \infty)_{\boldsymbol{r}}} \sup_{P \in \mathcal{P}} P\left(\left\{|\mathcal{\hat M}^\Delta(\boldsymbol{w}_P) - \mathcal{\hat M}^\Delta(\boldsymbol{\hat w})| > M_{\epsilon_1} n_{0,T}^{-1/2} \overline M_\epsilon \delta_{\boldsymbol{n},1} \right\} \cap E_n^c(\epsilon) \right) \le \epsilon_1.
\end{align*}
Therefore, from (\ref{eqs}),
\begin{align*}
	\liminf_{(\boldsymbol{n} \rightarrow \infty)_{\boldsymbol{r}}} \inf_{P \in \mathcal{P}} P\left\{ \eta \| \boldsymbol{\hat w} - \boldsymbol{w}_P\|^2 \le M_{\epsilon_1} n_{0,T}^{-1/2} \overline M_\epsilon \delta_{\boldsymbol{n},1} \right\} \ge 1 - \epsilon_1 - \epsilon.
\end{align*}
By sending $\epsilon_1,\epsilon \rightarrow 0$, we obtain the desired result. $\blacksquare$\medskip

\noindent \textbf{Proof of Theorem \ref{thm: root n consist} : } By Lemma \ref{lemm: consistency}, there exists $\delta_{\boldsymbol{n},1} \rightarrow 0$, $\enspace (\boldsymbol{n} \rightarrow \infty)_{\boldsymbol{r}}$, such that
\begin{align}
	\label{eq: consistency4}
	\lim_{M \uparrow \infty} \limsup_{(\boldsymbol{n} \rightarrow \infty)_{\boldsymbol{r}}} \sup_{P \in \mathcal{P}} P\left\{\|	\boldsymbol{\hat w} - \boldsymbol{w}_P \| > M \delta_{\boldsymbol{n},1} \right\} = 0.
\end{align}
By Lemma \ref{lemm: aux}, we find that the above result holds for $\delta_{\boldsymbol{n},1} = n_{0,T}^{-1/4}$. Now, we use mathematical induction. Suppose that (\ref{eq: consistency4}) holds with $\delta_{\boldsymbol{n},1}$ such that
\begin{align*}
	\log(\delta_{\boldsymbol{n},1}) = \log (n_{0,T}) \left( - \frac{1}{4} - \frac{1}{8} - ... - \frac{1}{2^{m}} \right),
\end{align*}
for some $m \ge 2$. Then, with this choice of $\delta_{\boldsymbol{n},1}$, we apply Lemma \ref{lemm: aux} again to find that (\ref{eq: consistency4}) holds with $\delta_{\boldsymbol{n},1}$ such that 
\begin{align*}
	\log(\delta_{\boldsymbol{n},1}) = \log (n_{0,T}) \left( - \frac{1}{4} - \frac{1}{8} - ... - \frac{1}{2^{m+1}} \right).
\end{align*}
Hence, we find that (\ref{eq: consistency4}) holds with $\delta_{\boldsymbol{n},1}$ such that 
\begin{align*}
	\log(\delta_{\boldsymbol{n},1}) = \log (n_{0,T}) \left( - \sum_{m=2}^\infty \frac{1}{2^m}\right) = - \frac{1}{2} \log (n_{0,T}).
\end{align*}
This gives the desired result. $\blacksquare$

\subsubsection{Proof of Theorem \ref{thm: validity}}

Define the bootstrap version of $\boldsymbol{\hat G}_P$ and $\boldsymbol{\hat g}_P$ as follows:
\begin{align*}
	\boldsymbol{\hat G}^* = \sqrt{n_{0,T}}( \hat H^* - \hat H) \text{ and } \boldsymbol{\hat g}^* = \sqrt{n_{0,T}}(\boldsymbol{\hat h}^* - \boldsymbol{\hat h}).
\end{align*}
The lemma below presents the bootstrap analogue of Lemma \ref{lemm: asym lin}.

\begin{lemma}
	\label{lemm: asym lin boot}
	For any $\boldsymbol{r} \in \mathbf{R}_{++}^K$, 
	\begin{align*}
		\boldsymbol{\hat G}^* = \sum_{i \in N} \Psi_{i,P}^* + o_{\mathcal{P}}(1), \enspace (\boldsymbol{n} \rightarrow \infty)_{\boldsymbol{r}}, \text{ and } 
		\boldsymbol{\hat g}_P^* = \sum_{i \in N} \boldsymbol{\psi}_{i,P}^* + o_{\mathcal{P}}(1), \enspace (\boldsymbol{n} \rightarrow \infty)_{\boldsymbol{r}},
	\end{align*}
	where $\Psi_{i,P}^*$ is the $K \times K$ matrix whose $(j,k)$-entry is given by
	\begin{align*}
		\psi_{i,P, jk}^* &= \frac{1}{\sqrt{n_j}} \hat \psi_{j,0,P}(W_i^*; q_{k,0,P})1\{i \in N_j\} + \frac{1}{\sqrt{n_k}} \hat \psi_{k,0,P}(W_i^*; q_{j,0,P})1\{i \in N_k\}\\
		&\quad + \frac{1}{\sqrt{n_{0,T}}}\left\{q_{j,0,P}(X_{i,T}^*)q_{k,0,P}(X_{i,T}^*) - \frac{1}{n_{0,T}} \sum_{i \in N_{0,T}} q_{j,0,P}(X_{i,T})q_{k,0,P}(X_{i,T}) \right\} 1\{i \in N_{0,T}\},
	\end{align*}
	and $\boldsymbol{\psi}_{i,P}^*$ is the $K \times 1$ vector whose $k$-th entry is given by
	\begin{align*}
		\psi_{i,P, k}^* &= \frac{1}{\sqrt{n_k}} \hat \psi_{k,0,P}(W_i^*;q_{0,0,P}) 1\{i \in N_k\} + \frac{1}{\sqrt{n_0}}\hat \psi_{0,0,P}(W_i^*;q_{k,0,P}) 1\{i \in N_{0}\} \\
		&\quad \quad + \frac{1}{\sqrt{n_{0,T}}}\left\{q_{k,0,P}(X_{i,T}^*)q_{0,0,P}(X_{i,T}^*)  - \frac{1}{n_{0,T}} \sum_{i \in N_{0,T}} q_{k,0,P}(X_{i,T})q_{0,0,P}(X_{i,T}) \right\}1\{i \in N_{0,T}\}.
	\end{align*}
\end{lemma}

\noindent \textbf{Proof: } The proof is similar to that of Lemma \ref{lemm: asym lin}. Since the arguments are standard, we provide a sketch of the proof of the first statement only for brevity. Let $\hat H_{jk}^*$ be the $(j,k)$-th entry of $\hat H^*$. We write 
\begin{align*}
	\sqrt{n_{0,T}}( \hat H_{jk}^* - \hat H_{jk}) = A_{\boldsymbol{n},1} + A_{\boldsymbol{n},2},
\end{align*}
where 
\begin{align*}
	A_{\boldsymbol{n},1} &= \frac{1}{\sqrt{n_{0,T}}} \sum_{i \in N_{0,T}} (\hat q_{j,0}^*(X_{i,T}^*) - \hat q_{j,0}(X_{i,T}^*)) \hat q_{k,0}^*(X_{i,T}^*) \\
	&\quad \quad + \frac{1}{\sqrt{n_{0,T}}} \sum_{i \in N_{0,T}} (\hat q_{k,0}^*(X_{i,T}^*) - \hat q_{k,0}(X_{i,T}^*)) \hat q_{j,0}(X_{i,T}^*), \text{ and }\\
    A_{\boldsymbol{n},2} &= \frac{1}{\sqrt{n_{0,T}}} \sum_{i \in N_{0,T}} \left\{\hat q_{j,0}(X_{i,T}^*)\hat q_{k,0}(X_{i,T}^*) - \frac{1}{n_{0,T}} \sum_{i \in N_{0,T}} \hat q_{j,0}(X_{i,T})\hat q_{k,0}(X_{i,T}) \right\}.
\end{align*}
From Assumptions \ref{assump: asym lin}-\ref{assump: asym lin boot}, we can show that 
\begin{align*}
	A_{\boldsymbol{n},1} = \frac{1}{\sqrt{n_j}} \sum_{i \in N_j} \hat \psi_{j,0,P}(W_i^*;q_{k,0,P}) 
	 + \frac{1}{\sqrt{n_k}} \sum_{i \in N_k} \hat \psi_{k,0,P}(W_i^*;q_{j,0,P}) + o_{\mathcal{P}}(1),\enspace (\boldsymbol{n} \rightarrow \infty)_{\boldsymbol{r}}.
\end{align*}
By Assumption \ref{assump: asym lin boot}(ii),
\begin{align*}
	A_{\boldsymbol{n},2} = \frac{1}{\sqrt{n_{0,T}}} \sum_{i \in N_{0,T}} \left\{q_{j,0,P}(X_{i,T}^*)q_{k,0,P}(X_{i,T}^*) - \frac{1}{n_{0,T}} \sum_{i \in N_{0,T}} q_{j,0,P}(X_{i,T})q_{k,0,P}(X_{i,T}) \right\} + o_{\mathcal{P}}(1),\enspace (\boldsymbol{n} \rightarrow \infty)_{\boldsymbol{r}}.
\end{align*}
Thus, we obtain the desired result. $\blacksquare$\medskip

Recall the definition of $V_{\boldsymbol{n},P}$ in (\ref{Omega_{n,P}}). We construct its bootstrap version. Define 
\begin{align*}
	\boldsymbol{\tilde \psi}_{i,P}^* = \Psi_{i,P}^* \boldsymbol{w}_P - \boldsymbol{\psi}_{i,P}^*,
\end{align*}
where $\Psi_{i,P}^*$ and $\boldsymbol{\psi}_{i,P}^*$ are defined in Lemma \ref{lemm: asym lin boot}. We let 
\begin{align*}
	\tilde V_{\boldsymbol{n},P} = \sum_{i \in N} \mathbf{E}\left[\boldsymbol{\tilde \psi}_{i,P}^* \boldsymbol{\tilde \psi}_{i,P}^{* \prime} \mid \mathcal{F}_{\boldsymbol{n}} \right],
\end{align*}
where $\mathcal{F}_{\boldsymbol{n}}$ denotes the $\sigma$-field generated by $(\tilde Y_i, \tilde X_i)_{i \in N}$ and $(X_{i,T})_{i \in N_{0,T}}$. Hence, $\mathbf{E}\left[\cdot \mid \mathcal{F}_{\boldsymbol{n}} \right]$ denotes the expectation under the bootstrap distribution.

\begin{lemma}
	\label{lemm: conv distribution}

	For any $\boldsymbol{n}$-indexed probabilities $P_{\boldsymbol{n}} \in \mathcal{P}$ and any $\boldsymbol{r} \in \mathbf{R}_{++}^K$, the following statements hold.
    
	(i)
	\begin{align*}
		\sup_{t \in \mathbf{R}^K}\left| P_{\boldsymbol{n}}\left\{ V_{\boldsymbol{n},P_{\boldsymbol{n}}}^{-1/2}\frac{1}{\sqrt{n_{0}}}\sum_{i \in N} \boldsymbol{\tilde \psi}_{i,P_{\boldsymbol{n}}} \le t \right\} - \Phi(t) \right| \rightarrow 0, \enspace (\boldsymbol{n} \rightarrow \infty)_{\boldsymbol{r}},
	\end{align*}
	where $\Phi$ is the CDF of $N(0,I_K)$.
	
	(ii) For any $\epsilon>0$,
	\begin{align*}
		P_{\boldsymbol{n}}\left\{ \sup_{t \in \mathbf{R}^K} \left| P_{\boldsymbol{n}} \left\{ \tilde V_{\boldsymbol{n},P_{\boldsymbol{n}}}^{-1/2} \frac{1}{\sqrt{n_{0}}} \sum_{i \in N} \boldsymbol{\tilde \psi}_{i,P_{\boldsymbol{n}}}^* \le t \mid \mathcal{F}_{\boldsymbol{n}} \right\} - \Phi(t) \right| > \epsilon \right\} \rightarrow 0, \enspace (\boldsymbol{n} \rightarrow \infty)_{\boldsymbol{r}}.
	\end{align*}
\end{lemma}

\noindent \textbf{Proof: } Both results follow from standard arguments involving the Central Limit Theorem and its bootstrap version for a sum of independent random variables. (See Chapter 3 of \cite{Shao/Tu:95:Jacknife}.) Details are omitted. $\blacksquare$

\begin{lemma}
	\label{lemm: variance est0}
	For any $\boldsymbol{r} \in \mathbf{R}_{++}^K$, $V_{\boldsymbol{n},P} = \tilde V_{\boldsymbol{n},P} + o_{\mathcal{P}}(1)$, $(\boldsymbol{n} \rightarrow \infty)_{\boldsymbol{r}}$.
\end{lemma}

\noindent \textbf{Proof: } Note that 
\begin{align*}
	\tilde V_{\boldsymbol{n},P} - V_{\boldsymbol{n},P} &=  \sum_{i \in N} \left( \mathbf{E}\left[ \Psi_{i,P}^* \boldsymbol{w}_P \boldsymbol{w}_P' \Psi_{i,P}^{*\prime} \mid \mathcal{F}_{\boldsymbol{n}} \right] - \mathbf{E}_P\left[ \Psi_{i,P}\boldsymbol{w}_P \boldsymbol{w}_P'\Psi_{i,P}^{\prime} \right] \right) \\
	& \quad - \sum_{i \in N} \left( \mathbf{E}\left[ \boldsymbol{\psi}_{i,P}^* \boldsymbol{w}_P' \Psi_{i,P}^{*\prime} \mid \mathcal{F}_{\boldsymbol{n}} \right] - \mathbf{E}_P\left[ \boldsymbol{\psi}_{i,P} \boldsymbol{w}_P' \Psi_{i,P}^{\prime} \right] \right) \\
	& \quad - \sum_{i \in N} \left( \mathbf{E}\left[ \Psi_{i,P}^* \boldsymbol{w}_P \boldsymbol{\psi}_{i,P}^{*'} \mid \mathcal{F}_{\boldsymbol{n}} \right] - \mathbf{E}_P\left[ \Psi_{i,P}\boldsymbol{w}_P \boldsymbol{\psi}_{i,P}^{'} \right] \right) \\
	& \quad + \sum_{i \in N} \left( \mathbf{E}\left[ \boldsymbol{\psi}_{i,P}^* \boldsymbol{\psi}_{i,P}^{*\prime} \mid \mathcal{F}_{\boldsymbol{n}} \right] - \mathbf{E}_P\left[ \boldsymbol{\psi}_{i,P} \boldsymbol{\psi}_{i,P}^{\prime} \right]\right) + o_{\mathcal{P}}(1), \enspace (\boldsymbol{n} \rightarrow \infty)_{\boldsymbol{r}}.
\end{align*} 
We can show that each sum on the right hand side is $o_{\mathcal{P}}(1)$. For brevity, we show this for the last term. The $(k,\ell)$-th entry of the last term is given by 
\begin{align*}
	\sum_{i \in N} \left( \mathbf{E}\left[ \psi_{i,P,k}^* \psi_{i,P,\ell}^* \mid \mathcal{F}_{\boldsymbol{n}} \right] - \mathbf{E}_P\left[ \psi_{i,P,k} \psi_{i,P,\ell} \right]\right) =\sum_{i \in N} \left( \psi_{i,P,k} \psi_{i,P,\ell} - \mathbf{E}_P\left[ \psi_{i,P,k} \psi_{i,P,\ell} \right]\right).
\end{align*} 
Again, for simplicity, we focus on the case where $k = \ell$, and show that the last sum is $o_{\mathcal{P}}(1)$. Note that 
\begin{align*}
	\sum_{i \in N} \mathbf{E}\left[(\psi_{i,P,k}^*)^2 \mid \mathcal{F}_{\boldsymbol{n}}\right] &= \frac{1}{n_k}\sum_{i \in N_k} \hat \psi_{k,0,P}^2(W_i;q_{0,0,P}) + \frac{1}{n_0}\sum_{i \in N_0} \hat \psi_{0,0,P}^2(W_i;q_{k,0,P})\\
	&\quad + \frac{1}{n_{0,T}} \sum_{i \in N_{0,T}} \left(q_{j,0,P}(X_{i,T})q_{k,0,P}(X_{i,T}) - \frac{1}{n_{0,T}} \sum_{i \in N_{0,T}} q_{j,0,P}(X_{i,T})q_{k,0,P}(X_{i,T}) \right)^2,
\end{align*} 
because $N_k$, $N_0$ and $N_{0,T}$ are disjoint. The three terms on the right hand side are the sample variances of the i.i.d.\ random variables, and likewise, $\sum_{i \in N} \mathbf{E}_P[\psi_{i,P,k}^2]$ is the sum of their population variances. Hence, using standard arguments, we can show that 
\begin{align*}
	\sum_{i \in N} \mathbf{E}\left[(\psi_{i,P,k}^*)^2 \mid \mathcal{F}_{\boldsymbol{n}}\right] = \sum_{i \in N} \mathbf{E}_P[\psi_{i,P,k}^2] + o_{\mathcal{P}}(1), \enspace (\boldsymbol{n} \rightarrow \infty)_{\boldsymbol{r}}.
\end{align*} 
$\blacksquare$\medskip

Recall the definition of $\boldsymbol{\hat \gamma}^* = \sqrt{n_{0,T}} \left(\hat H^* - \hat H \right) \boldsymbol{\hat w} - \sqrt{n_{0,T}} \left( \boldsymbol{\hat h}^* - \boldsymbol{\hat h}\right)$ in (\ref{hat gamma star}) in the main text. Note that 
\begin{align*}
	\boldsymbol{\hat \gamma}^* &= \sqrt{n_{0,T}} \left(\hat H^* - \hat H \right) \boldsymbol{w}_P - \sqrt{n_{0,T}} \left( \boldsymbol{\hat h}^* - \boldsymbol{\hat h}\right) + o_{\mathcal{P}}(1), \enspace (\boldsymbol{n} \rightarrow \infty)_{\boldsymbol{r}}\\
	&= \sum_{i \in N} \boldsymbol{\tilde \psi}_{i,P}^* + o_{\mathcal{P}}(1), \enspace (\boldsymbol{n} \rightarrow \infty)_{\boldsymbol{r}},
\end{align*}
by Lemma \ref{lemm: asym lin boot}. Let
\begin{align*}
	\boldsymbol{\hat \gamma}_P = \sqrt{n_{0,T}} \left(\hat H - H_P \right) \boldsymbol{w}_P - \sqrt{n_{0,T}} \left( \boldsymbol{\hat h} - \boldsymbol{h}_P\right).
\end{align*}
Recall the definition $\hat V = \text{Var}\left(\boldsymbol{\hat \gamma}^* \mid \mathcal{F}_n \right)$.

\begin{lemma}
	\label{lemm: conv distribution2}
	For any $\boldsymbol{n}$-indexed probabilities $P_{\boldsymbol{n}} \in \mathcal{P}$ and any $\boldsymbol{r} \in \mathbf{R}_{++}^K$, and for any $\epsilon>0$,
	\begin{align*}
		P_{\boldsymbol{n}}\left\{ \sup_{t \in \mathbf{R}^K} \left| P_{\boldsymbol{n}} \left\{ V_{\boldsymbol{n},P_{\boldsymbol{n}}}^{-1/2} \boldsymbol{\hat \gamma}^* \le t \mid \mathcal{F}_{\boldsymbol{n}} \right\} - \Phi(t) \right| > \epsilon \right\} \rightarrow 0, \enspace (\boldsymbol{n} \rightarrow \infty)_{\boldsymbol{r}}.
	\end{align*}
\end{lemma}

\noindent \textbf{Proof: } Note that by Lemma \ref{lemm: asym lin boot},
\begin{align*}
	\boldsymbol{\hat \gamma}^* - \sum_{i \in N} \boldsymbol{\tilde \psi}_{i,P_{\boldsymbol{n}}}^* = \left(\sum_{i \in N} \Psi_{i,P}^*\right) (\boldsymbol{\hat w} - \boldsymbol{w}_P) + o_{\mathcal{P}}(1), \enspace (\boldsymbol{n} \rightarrow \infty)_{\boldsymbol{r}},
\end{align*}
because $\boldsymbol{\hat w}, \boldsymbol{w} \in \Delta_{K-1}$. It is not hard to see that 
\begin{align*}
	\mathbf{E}\left[\left\|\sum_{i \in N} \Psi_{i,P}^*\right\|^2 \mid \mathcal{F}_{\boldsymbol{n}}\right] = O_{\mathcal{P}}(1), \enspace (\boldsymbol{n} \rightarrow \infty)_{\boldsymbol{r}}.
\end{align*}
Hence, by Lemma \ref{lemm: consistency}, we find that 
\begin{align*}
	\boldsymbol{\hat \gamma}^* - \sum_{i \in N} \boldsymbol{\tilde \psi}_{i,P}^* = o_{\mathcal{P}}(1), \enspace (\boldsymbol{n} \rightarrow \infty)_{\boldsymbol{r}}.
\end{align*}
Then, the desired result follows from Lemmas \ref{lemm: conv distribution}(ii) and \ref{lemm: variance est0}, (\ref{Omega P conv}) and Assumption \ref{assump: positive definite V_P}. $\blacksquare$

\begin{lemma}
	\label{lemm: asymp val}
	For any $\boldsymbol{n}$-indexed probabilities $P_{\boldsymbol{n}} \in \mathcal{P}$, any $\boldsymbol{r} \in \mathbf{R}_{++}^K$, and $t \in \mathbf{R}$, 
	\begin{align}
		\label{asymp val}
		\liminf_{(\boldsymbol{n} \rightarrow \infty)_{\boldsymbol{r}}} P_{\boldsymbol{n}}\left\{ \boldsymbol{\hat \gamma}_{P_{\boldsymbol{n}}}' \hat V^{-1} \boldsymbol{\hat \gamma}_{P_{\boldsymbol{n}}} \le t \right\} \ge P\{Z'Z \le t\},
	\end{align}
	where $Z \sim N(0,I_K)$.
\end{lemma}

\noindent \textbf{Proof: } First, for any subsequence of $\{\boldsymbol{n}_\ell: \ell = 1,2,...,\} \in \mathcal{N}_{K+1}(\boldsymbol{r})$, there exists a further subsequence of $\{\boldsymbol{n}_s: s = 1,2,...,\} \subset \{\boldsymbol{n}_\ell: \ell = 1,2,...,\}$, along which 
\begin{align*}
	\boldsymbol{w}_{P_{\boldsymbol{n}_s}} \rightarrow \boldsymbol{w}_0 \text{ and } V_{\boldsymbol{n}_s,P_{\boldsymbol{n}_s}} \rightarrow V_0, \enspace \text{ as } s \rightarrow \infty,
\end{align*}
for some $\boldsymbol{w}_0 \in \Delta_{K-1}$ and positive definite matrix $V_0$. Thus, for any $\epsilon>0$,
	\begin{align}
		\label{conv333}
		&\sup_{t \in \mathbf{R}^K}\left| P_{\boldsymbol{n}_s}\left\{ V_0^{-1/2}\boldsymbol{\hat \gamma}_{P_{\boldsymbol{n}_s}} \le t \right\} - \Phi(t) \right| \rightarrow 0, \text{ and }\\ \notag
		&P_{\boldsymbol{n}_s}\left\{ \sup_{t \in \mathbf{R}^K} \left| P_{\boldsymbol{n}_s} \left\{ V_0^{-1/2} \boldsymbol{\hat \gamma}^* \le t \mid \mathcal{F}_{\boldsymbol{n}_s} \right\} - \Phi(t) \right| > \epsilon \right\} \rightarrow 0,
	\end{align}
as $s \rightarrow \infty$ by Lemmas \ref{lemm: asym lin}, \ref{lemm: conv distribution}(i) and \ref{lemm: conv distribution2}. To complete the proof, it suffices to show (\ref{asymp val}) along this $\boldsymbol{n}_s$, $s=1,2,...$. Following the arguments in the proof of Theorem 2 of \cite{Hahn/Liao:21:Ecma}, we find that for any $\epsilon>0$ and $t \in \mathbf{R}$,
\begin{align*}
	\limsup_{s \rightarrow \infty} P_{\boldsymbol{n}_s}\left\{ \boldsymbol{\hat \gamma}_{P_{\boldsymbol{n}_s}}' \hat V^{-1} \boldsymbol{\hat \gamma}_{P_{\boldsymbol{n}_s}} > t \right\}
	\le  P\{Z'Z \le t\} + \limsup_{s \rightarrow \infty} R_s(\epsilon),
\end{align*} 
where, with $S_K$ denoting the unit sphere in $\mathbf{R}^K$,
\begin{align*}
	R_s(\epsilon) = P_{\boldsymbol{n}_s}\left\{ \inf_{\alpha \in S_K} \alpha' (V_0^{-1/2} \hat V V_0^{-1/2}) \alpha \ge 1 - \frac{\epsilon}{2}\right\}.
\end{align*}
Hence, the desired result follows once we obtain the following: for each $\epsilon>0$, 
\begin{align}
	\label{conv33}
	\limsup_{s \rightarrow \infty} R_s(\epsilon) = 0.
\end{align}
This result can be proved as in the proof of Theorem 3 of \cite{Hahn/Liao:21:Ecma}. More specifically, first define 
\begin{align*}
	W_s = \sup_{t \in \mathbf{R}^K} \left| P_{\boldsymbol{n}_s} \left\{ V_0^{-1/2} \boldsymbol{\hat \gamma}^* \le t \mid \mathcal{F}_{\boldsymbol{n}} \right\} - \Phi(t) \right|. 
\end{align*} 
Then, for any subsequence of $\{s\}$, there exists a further subsequence $\{s'\}$ along which 
\begin{align*}
	P\left\{ \lim_{s' \rightarrow \infty} W_{s'} = 0 \right\} = 1.
\end{align*} 
(See, e.g., Theorem 20.5 of (\cite{Billingsley:95:ProbMeasure}).) It suffices to focus on this subsequence $\{s'\}$. Of course, along this subsequence, we have 
\begin{align*}
	\limsup_{s'\rightarrow \infty} \sup_{t \in \mathbf{R}^K} \left|P_{\boldsymbol{n}_{s'}} \left\{ V_0^{-1/2} \boldsymbol{\hat \gamma}_{P_{\boldsymbol{n}_{s'}}} \le t \right\} - \Phi(t) \right| = 0,
\end{align*} 
by (\ref{conv333}). Using these results, and following the proof of Theorem 3 of \cite{Hahn/Liao:21:Ecma}, we obtain (\ref{conv33}). $\blacksquare$

\begin{lemma}
	\label{lemm: asym val w_0}
	For any $\kappa \in (0,1)$ and $\boldsymbol{r} \in \mathbf{R}_{++}^K$, we have
	\begin{align*}
		\liminf_{(\boldsymbol{n} \rightarrow \infty)_{\boldsymbol{r}}} \inf_{P \in \mathcal{P}}P\left\{ \boldsymbol{w}_P \in \tilde C_{1- \kappa}\right\} \ge 1 - \kappa.
	\end{align*}
\end{lemma}

\noindent \textbf{Proof}: First, we define the following, infeasible confidence set:
\begin{align}
	\label{tilde C22}
	\overline C_{1-\alpha} = \left\{ \boldsymbol{w} \in \Delta_{K-1}: \overline T_P(\boldsymbol{w}) \le \hat c_{1-\alpha}(\boldsymbol{w}) \right\},
\end{align}
where 
\begin{align}
	\label{bar T(u)}
	\overline T_P(\boldsymbol{w}) = n_{0,T} \left( \hat \varphi(\boldsymbol{w}) - \hat\lambda(\boldsymbol{w}) \right)'B_2 (\overline \Omega_{\boldsymbol{n},P})^{-1}B_2' \left( \hat \varphi(\boldsymbol{w}) - \hat\lambda(\boldsymbol{w}) \right).
\end{align}
and \begin{align*}
	\overline \Omega_{\boldsymbol{n},P} = B_2' V_{\boldsymbol{n},P} B_2.
\end{align*}
Then, we first show that 
\begin{align}
	\label{overline C}
	\liminf_{(\boldsymbol{n} \rightarrow \infty)_{\boldsymbol{r}}} \inf_{P \in \mathcal{P}}P\left\{ \boldsymbol{w}_P \in \overline C_{1- \kappa}\right\} \ge 1 - \kappa.
\end{align}

We write $\overline C_{1- \kappa}$, $\hat V$, and $\hat \varphi$ as $\overline C_{\boldsymbol{n},1- \kappa}$, $\hat V_{\boldsymbol{n}}$ and $\hat \varphi_{\boldsymbol{n}}$ making their dependence on $\boldsymbol{n}$ explicit. Take any arbitrary sequence $P_{\ell} \in \mathcal{P}$ and $\boldsymbol{w}_{P_\ell}$, $\ell = 1,2,...$. It suffices to show that
\begin{align*}
	\liminf_{\ell \rightarrow \infty} P_{\ell} \left\{ \boldsymbol{w}_{P_{\ell}} \in \overline C_{\boldsymbol{n}_{\ell},1- \kappa}\right\} \ge 1 - \kappa,
\end{align*}
for any subset $\{\boldsymbol{n}_{\ell}: \ell = 1,2,...\} \in \mathcal{N}_{K+1}(\boldsymbol{r})$. We apply Lemma 3.1 of \cite{Canen/Song:25:arXiv}. For this, we let
\begin{align*}
	Y_\ell = \overline \Omega_{\boldsymbol{n}_{\ell},P_\ell}^{1/2} Z_\ell + \mu_\ell \text{ and } Z_\ell = \sqrt{n_{0,T}^{(\ell)}} (\overline \Omega_{\boldsymbol{n}_{\ell},P_\ell})^{-1/2} B_2' (\hat \varphi_{\boldsymbol{n}_{\ell}}(\boldsymbol{w}_{P_\ell}) - \varphi_{P_\ell}(\boldsymbol{w}_{P_\ell})),
\end{align*}
where 
\begin{align*}
	\mu_{\ell} = \sqrt{n_{0,T}^{(\ell)}} B_2' \varphi_{P_\ell}(\boldsymbol{w}_{P_\ell}) \text{ and } \overline \Omega_{\boldsymbol{n}_{\ell},P_\ell} = B_2' V_{\boldsymbol{n}_{\ell},P_\ell} B_2,
\end{align*}
and $n_{0,T}^{(\ell)}$ is the $n_{0,T}$-component of $\boldsymbol{n}_{\ell}$, where 
\begin{align*}
	\varphi_{P_\ell} = H_{P_\ell} \boldsymbol{w}_{P_\ell} - \boldsymbol{h}_{P_\ell}.
\end{align*}
For Lemma 3.1, we need to check Assumption 3.1 of \cite{Canen/Song:25:arXiv}. This assumption is satisfied once the following statements hold.\medskip

(i) $Z_\ell \rightarrow_d N(0,I_{K-1})$, as $\ell \rightarrow \infty$.

(ii) $\overline \Omega_{\boldsymbol{n}_{\ell},P_{\ell}} - \Omega_{P_{\ell}} \rightarrow 0$, as $\ell \rightarrow \infty$, where $\Omega_{P_{\ell}} = B_2' V_{P_{\ell}}(\boldsymbol{r}) B_2$.

(iii) There exist $\overline B, \epsilon >0$ such that $\|\Omega_{P_\ell}\| \le \overline B$ and $\lambda_{\min}(\Omega_{P_\ell}) > \epsilon$ for all $\ell \ge 1$.\medskip

Note that 
\begin{align*}
	Z_\ell = (\overline \Omega_{\boldsymbol{n}_{\ell},P_\ell})^{-1/2} B_2' (\boldsymbol{\hat G}_{P_\ell} \boldsymbol{w}_{P_\ell} - \boldsymbol{\hat g}) = (B_2' V_{\boldsymbol{n}_{\ell},P_\ell}^{-1/2} B_2)^{-1/2} B_2' \boldsymbol{\hat \gamma}_{P_\ell}.
\end{align*} 
Condition (i) follows from Lemmas \ref{lemm: asym lin} and \ref{lemm: conv distribution}. Condition (ii) follows from (\ref{Omega P conv}). Condition (iii) follows from the following: there exist $\overline B, \epsilon >0$ such that for all $\boldsymbol{n} \in \mathbb{N}^{K+1}$,
\begin{align}
	\label{bounds}
	\sup_{P \in \mathcal{P}} \| V_{\boldsymbol{n},P} \| \le \overline B \text{ and } \inf_{P \in \mathcal{P}} \lambda_{\min}(V_{\boldsymbol{n},P}) > \epsilon.
\end{align}
To see how these bounds are obtained, first note that 
\begin{align*}
	\|V_{\boldsymbol{n},P}\| \le 2 K^2 \mathbf{E}\left[ \| \Psi_{i,P}\|^2 \right] + \mathbf{E}\left[ \|\boldsymbol{\psi}_{i,P}\|^2 \right].
\end{align*}
The two expectations on the right hand side are bounded uniformly over $P \in \mathcal{P}$ and $\boldsymbol{n} \in \mathbb{N}^{K+1}$ by Assumption \ref{assump: conditional average outcome bound} and (\ref{psi bound}). The second bound in (\ref{bounds}) is due to Assumption \ref{assump: positive definite V_P}. Thus, we obtain (\ref{overline C}).

Now, let us prove the main result. Now, observe that as in the proof of Theorem 5 of \cite{Hahn/Liao:21:Ecma}, we can use (\ref{conv33}) to show that
\begin{align*}
	\liminf_{(\boldsymbol{n} \rightarrow \infty)_{\boldsymbol{r}}} P_{\boldsymbol{n}}\left\{\inf_{a \in S_K} a' (\hat V_{\boldsymbol{n}} - V_{\boldsymbol{n}, P_{\boldsymbol{n}}}) a \ge - \epsilon \right\} = 1,
\end{align*}
for any $\epsilon>0$. This means that
\begin{align*}
	\liminf_{(\boldsymbol{n} \rightarrow \infty)_{\boldsymbol{r}}} P_{\boldsymbol{n}}\left\{T(\boldsymbol{w}_{P_{\boldsymbol{n}}}) \le \overline T(\boldsymbol{w}_{P_{\boldsymbol{n}}})\right\} = 1.
\end{align*}
Therefore, 
\begin{align*}
	\liminf_{(\boldsymbol{n} \rightarrow \infty)_{\boldsymbol{r}}} \inf_{P \in \mathcal{P}}P\left\{ \boldsymbol{w}_P \in \tilde C_{1- \kappa}\right\} \ge \liminf_{(\boldsymbol{n} \rightarrow \infty)_{\boldsymbol{r}}} \inf_{P \in \mathcal{P}}P\left\{ \boldsymbol{w}_P \in \overline C_{1- \kappa}\right\} \ge 1- \kappa.
\end{align*}
$\blacksquare$

\begin{lemma}
	\label{lemm: asym repre}
	For each $\boldsymbol{r} \in \mathbf{R}_{++}^K$, we have
	\begin{align*}
		\sup_{\boldsymbol{w} \in \Delta_{K-1}} \left|\sqrt{n_{0,T}}\left(\hat \theta(\boldsymbol{w}) - \theta_P(\boldsymbol{w})\right) - \left(A_{0, \boldsymbol{n}} + \sum_{k=1}^K w_k A_{k, \boldsymbol{n}} \right)\right| = o_{\mathcal{P}}(1), \enspace(\boldsymbol{n} \rightarrow \infty)_{\boldsymbol{r}},
	\end{align*}
	where
	\begin{align*}
		A_{0, \boldsymbol{n}} &= \frac{1}{\sqrt{n_0}} \sum_{i \in N_0} \psi_{0,0,P}(W_i;1) + \frac{1}{\sqrt{n_{0,T}}} \sum_{i \in N_{0,T}}\left(q_{0,0,P}(X_{i,T}) - \mathbf{E}_P\left[q_{0,0,P}(X_{i,T})\right]\right), \text{ and } \\
		A_{k, \boldsymbol{n}} &= \frac{1}{\sqrt{n_k}} \sum_{i \in N_k} \psi_{k,1,P}(W_i;1) + \frac{1}{\sqrt{n_{0,T}}} \sum_{i \in N_{0,T}}\left(q_{k,1,P}(X_{i,T}) - \mathbf{E}_P\left[q_{k,1,P}(X_{i,T})\right]\right).
	\end{align*}
\end{lemma}

\noindent \textbf{Proof: }  We write
\begin{align*}
	\sqrt{n_{0,T}}\left(\hat \theta(\boldsymbol{w}) - \theta_P(\boldsymbol{w})\right) &= \frac{1}{\sqrt{n_{0,T}}} \sum_{i \in N_{0,T}} \left(\hat q_{0,0}(X_{i,T}) - q_{0,0,P}(X_{i,T})\right) \\
	& \quad + \frac{1}{\sqrt{n_{0,T}}} \sum_{i \in N_{0,T}} \left(q_{0,0,P}(X_{i,T}) - \mathbf{E}_P\left[q_{0,0,P}(X_{i,T})\right]\right)\\
	&\quad +  \frac{1}{\sqrt{n_{0,T}}} \sum_{k=1}^K w_k \sum_{i \in N_{0,T}} \left\{ \hat q_{k,1}(X_{i,T}) - q_{k,1,P}(X_{i,T}) \right\}\\
	&\quad +  \frac{1}{\sqrt{n_{0,T}}} \sum_{k=1}^K w_k \sum_{i \in N_{0,T}} \left\{ q_{k,1,P}(X_{i,T}) - \mathbf{E}_P\left[q_{k,1,P}(X_{i,T})\right]\right\}.
\end{align*}
By Assumption \ref{assump: asym lin} and by the fact that $\sum_{k=1}^K w_k = 1$, we find 
\begin{align*}
	\sqrt{n_{0,T}}\left(\hat \theta(\boldsymbol{w}) - \theta_P(\boldsymbol{w})\right) &= \frac{1}{\sqrt{n_0}} \sum_{i \in N_{0}} \psi_{0,0,P}(W_i;1) + \sum_{k=1}^K w_k \sum_{i \in N_k} \frac{1}{\sqrt{n_k}} \psi_{k,1,P}(W_i;1)\\
	&\quad + \frac{1}{\sqrt{n_{0,T}}} \sum_{i \in N_{0,T}} \left(q_{0,0,P}(X_{i,T}) - \mathbf{E}_P\left[q_{0,0,P}(X_{i,T})\right]\right)\\
	&\quad + \sum_{k=1}^K w_k \frac{1}{\sqrt{n_{0,T}}} \sum_{i \in N_{0,T}} \left(q_{k,1,P}(X_{i,T}) - \mathbf{E}_P\left[q_{k,1,P}(X_{i,T})\right]\right)  + o_{\mathcal{P}}(1), \enspace(\boldsymbol{n} \rightarrow \infty)_{\boldsymbol{r}}.
\end{align*}
We obtain the desired result. $\blacksquare$\medskip

Recall the definitions of $A_{0, \boldsymbol{n}}$ and $A_{k, \boldsymbol{n}}$ in Lemma \ref{lemm: asym repre}. Define 
\begin{align*}
	\sigma_{\boldsymbol{n},P}^2(\boldsymbol{w}) = \mathbf{E}_P\left[ \left(A_{0, \boldsymbol{n}} + \sum_{k=1}^K w_k A_{k, \boldsymbol{n}} \right)^2 \right].
\end{align*}

\begin{lemma}
	\label{lemm: weak conv}
	For each $\boldsymbol{r} \in \mathbf{R}_{++}^K$, we have
	\begin{align*}
		\sup_{P \in \mathcal{P}} \sup_{t \in \mathbf{R}}\left|P\left\{\frac{\sqrt{n_{0,T}}\left(\hat \theta(\boldsymbol{w}_P) - \theta_P(\boldsymbol{w}_P) \right)}{\sigma_{\boldsymbol{n},P}(\boldsymbol{w}_P)} \le t \right\}  - \Phi(t)\right| \rightarrow 0, \enspace(\boldsymbol{n} \rightarrow \infty)_{\boldsymbol{r}},
	\end{align*}
	where $\Phi$ is the CDF of $N(0,1)$.
\end{lemma}

\noindent \textbf{Proof: } The result follows from Lemma \ref{lemm: asym repre} and the Central Limit Theorem for independent random variables. $\blacksquare$\medskip

Recall the definition of $\hat \sigma$ in (\ref{hat sigma}) in the main text.

\begin{lemma}
	\label{lemm: consistency2}
	For each $\boldsymbol{r} \in \mathbf{R}_{++}^K$, $\hat \sigma = \sigma_{\boldsymbol{n},P}(\boldsymbol{w}_P) + o_{\mathcal{P}}(1)$, $(\boldsymbol{n} \rightarrow \infty)_{\boldsymbol{r}}.$
\end{lemma}

\noindent \textbf{Proof: } For $t \in \mathbf{R}$, define 
\begin{align*}
	\hat G_{\boldsymbol{n}}(t) = P\left\{\frac{\sqrt{n_{0,T}}\left(\hat \theta^*(\boldsymbol{\hat w}) - \hat \theta(\boldsymbol{\hat w})\right)}{\sigma_{\boldsymbol{n},P}(\boldsymbol{w}_P)} \le t \mid \mathcal{F}_{\boldsymbol{n}} \right\}.
\end{align*}
We first show that for all $\alpha \in (0,1)$,
\begin{align}
	\label{conv34}
	\hat G_{\boldsymbol{n}}^{-1}(1-\alpha) = \Phi^{-1}(1-\alpha) + o_{\mathcal{P}}(1), \enspace (\boldsymbol{n} \rightarrow \infty)_{\boldsymbol{r}}.
\end{align} 
For this, we follow the arguments in the proof of Lemma 1.2.1 of \cite{Politis/Romano/Wolf:99:Subsampling}. 
Similarly as in the proof of Lemma \ref{lemm: conv distribution2}, we find that for each $\epsilon>0$,
\begin{align}
	\label{conv35}
	\sup_{P \in \mathcal{P}} P\left\{\sup_{t \in \mathbf{R}} \left|\hat G_{\boldsymbol{n}}(t) - \Phi(t) \right| > \epsilon\right\} \rightarrow 0, \enspace (\boldsymbol{n} \rightarrow \infty)_{\boldsymbol{r}}.
\end{align} 
We fix $\delta>0$ and let $\epsilon \in (0,\delta]$. Let $y = \Phi^{-1}(1-\alpha)$. Then, from (\ref{conv35}), we find that 
\begin{align*}
	\hat G_{\boldsymbol{n}}(y - \epsilon) &= \Phi(y - \epsilon) + o_{\mathcal{P}}(1) < 1-\alpha + o_{\mathcal{P}}(1)  \text{ and } \\
	\hat G_{\boldsymbol{n}}(y + \epsilon) &= \Phi(y + \epsilon)  + o_{\mathcal{P}}(1) > 1-\alpha + o_{\mathcal{P}}(1), \enspace (\boldsymbol{n} \rightarrow \infty)_{\boldsymbol{r}}.
\end{align*}
This implies that for all sufficiently large $\boldsymbol{n}$, 
\begin{align*}
	y - \epsilon \le \hat G_{\boldsymbol{n}}^{-1}(1 - \alpha) \le y + \epsilon,
\end{align*}
uniformly over $P \in \mathcal{P}$. Since the choice of $\epsilon>0$ was arbitrary, we obtain (\ref{conv34}). 

Now, define 
\begin{align*}
	\hat F_{\boldsymbol{n}}(t) = P\left\{\sqrt{n_{0,T}}\left(\hat \theta^*(\boldsymbol{\hat w}) - \hat \theta(\boldsymbol{\hat w})\right) \le t \mid \mathcal{F}_{\boldsymbol{n}} \right\}.
\end{align*}
Then, (\ref{conv34}) implies that 
\begin{align*}
	\frac{\hat F_{\boldsymbol{n}}^{-1}(0.75) - \hat F_{\boldsymbol{n}}^{-1}(0.25)}{\sigma_{\boldsymbol{n},P}(\boldsymbol{w}_P)} = \Phi^{-1}(0.75) - \Phi^{-1}(0.25) + o_{\mathcal{P}}(1), \enspace (\boldsymbol{n} \rightarrow \infty)_{\boldsymbol{r}},
\end{align*}
Or, 
\begin{align*}
	\hat \sigma = \frac{\hat F_{\boldsymbol{n}}^{-1}(0.75) - \hat F_{\boldsymbol{n}}^{-1}(0.25)}{\Phi^{-1}(0.75) - \Phi^{-1}(0.25)} = \sigma_{\boldsymbol{n},P}(\boldsymbol{w}_P) + o_{\mathcal{P}}(1), \enspace (\boldsymbol{n} \rightarrow \infty)_{\boldsymbol{r}},
\end{align*}
which delivers the desired result. $\blacksquare$\medskip

\noindent \textbf{Proof of Theorem \ref{thm: validity} : } Note that
\begin{align*}
	P\left\{ \theta_P(\boldsymbol{w}_P) \notin C_{1- \alpha}\right\} &= P\left\{ \inf_{\boldsymbol{w} \in \tilde C_{1 - \kappa}} \left( \frac{\sqrt{n_{0,T}}(\hat \theta(\boldsymbol{w}) - \theta_P(\boldsymbol{w}_P))}{\hat \sigma} \right)^2 > c_{1 - \alpha + \kappa}(1) \right\}\\
	&\le P\left\{ \left( \frac{\sqrt{n_{0,T}}(\hat \theta(\boldsymbol{w}_P) - \theta_P(\boldsymbol{w}_P))}{\hat \sigma} \right)^2 > c_{1 - \alpha + \kappa}(1) \right\} + P\left\{\boldsymbol{w}_P \notin \tilde C_{1 - \kappa} \right\}.
\end{align*}
The desired result follows by Lemmas \ref{lemm: weak conv}, \ref{lemm: consistency2} and \ref{lemm: asym val w_0}. $\blacksquare$

\subsection{Without Requiring the Weights to be Identified}
\label{subsec: without identifiability}

Let us discuss the case where $H_P$ is not necessarily invertible. In this case, we show how we can still obtain uniformly valid confidence intervals for $\theta_0$. First, we provide a modification of the method to accommodate this setting, and then present the uniform validity result. 

We define 
\begin{align*}
     \mathbb{W}_P = \arg \min_{\boldsymbol{w} \in \Delta_{K-1}} \rho_P^2(\boldsymbol{w}),
\end{align*}
where
\begin{align*}
	\rho_P^2(\boldsymbol{w}) = \boldsymbol{w}^{\prime} H_P \boldsymbol{w} - 2 \boldsymbol{w}^{\prime} \boldsymbol{h}_P.
\end{align*}
Let us explain how we construct the confidence interval for $\theta_0(\boldsymbol{w}_0)$ for a fixed $\boldsymbol{w}_0 \in \Delta_{K-1}$. We first define
\begin{align}
	\label{hat theta2}
	\hat \theta(\boldsymbol{w}) &=  \frac{1}{n_{0,T}}\sum_{i \in N_{0,T}} \hat m_0\left(\hat \mu_0(X_{i,T},v_{0,T}^*),X_{i,T} \right) 1\{X_{i,T} \in \mathcal{\hat X}_{0,T}^\mathsf{M}\} \\ \notag
	&\quad \quad + \sum_{k=1}^K \frac{1}{n_{0,T}}\sum_{i \in N_{0,T}} m_k\left( \hat \mu_k(X_{i,T},v_{0,T}^*),X_{i,T}\right) w_{k} 1\{X_i \in \mathcal{\hat X}_{0,T} \setminus \mathcal{\hat X}_{0,T}^\mathsf{M}\}.
\end{align}
First, we define $\hat \Omega(\boldsymbol{w}) = B_2' \hat V(\boldsymbol{w}) B_2$, where $\hat V(\boldsymbol{w})$ is constructed as in (\ref{hat Omega}) in the main text, except with $\boldsymbol{w}$ replacing $\boldsymbol{\hat w}$. Let
\begin{align*}
	\hat \lambda(\boldsymbol{w}) = \argmin_{\lambda} \left( \hat \varphi(\boldsymbol{w}) - \lambda \right)'B_2 \hat \Omega^{-1}(\boldsymbol{w}) B_2' ( \hat \varphi(\boldsymbol{w}) - \lambda ),
\end{align*}
where the minimization over $\lambda$ is done under the constraints: $\boldsymbol{w}'\lambda = 0$ and $\lambda \ge 0$. Let $\hat d(\boldsymbol{w})$ be the number of zeros in the vector $B_2 \hat \Omega^{-1} (\boldsymbol{w}) B_2'( \hat \varphi(\boldsymbol{w}) - \hat \lambda(\boldsymbol{w}))$, and let $\hat c_{1-\alpha}(\boldsymbol{w})$ be the $1-\alpha$ percentile of the $\chi_{\hat k(\boldsymbol{w})}^2$ distribution, where
\begin{align*}
	\hat k(\boldsymbol{w}) = \max \left\{K-1 - \hat d(\boldsymbol{w}),1 \right\}.
\end{align*} 

As in (\ref{T(u)}) in the main text, we construct 
\begin{align}
	\label{T(u)2}
	T'(\boldsymbol{w}) = n_{0,T}\left( \hat \varphi(\boldsymbol{w}) - \lambda \right)'B_2 \hat \Omega^{-1}(\boldsymbol{w}) B_2' \left( \hat \varphi(\boldsymbol{w}) - \lambda \right).
\end{align}
Then, the confidence set for $\boldsymbol{w}_0$ is given by 
\begin{align}
	\label{tilde C2}
	\tilde C_{1-\kappa}' = \left\{\boldsymbol{w} \in \Delta_{K-1}: T'(\boldsymbol{w}) \le \hat c_{1 - \kappa}(\boldsymbol{w})\right\}.
\end{align}
Let $\hat \theta^*(\boldsymbol{w})$ be as defined in (\ref{hat theta star}) in the main text. Define
\begin{align*}
	\hat \tau^*(\boldsymbol{w}) = \sqrt{n_{0,T}}\left(\hat \theta^*(\boldsymbol{w}) - \hat \theta(\boldsymbol{w})\right).
\end{align*}
We read the $0.75$ quantile and $0.25$ quantile of the bootstrap distribution of $\hat \tau^*(\boldsymbol{w})$, and denote them to be $\hat q_{0.75}(\boldsymbol{w})$ and $\hat q_{0.25}(\boldsymbol{w})$, respectively. Define
\begin{align*}
	\hat \sigma(\boldsymbol{w}) = \frac{\hat q_{0.75}(\boldsymbol{w}) - \hat q_{0.25}(\boldsymbol{w})}{z_{0.75} - z_{0.25}},
\end{align*}
where $z_{0.75}$ and $z_{0.25}$ are the $0.75$- and $0.25$-quantiles of $N(0,1)$. 

Define
\begin{align*}
	\hat \tau'(\boldsymbol{w}, \theta) = \frac{\sqrt{n_{0,T}} (\hat \theta(\boldsymbol{w}) - \theta)}{\hat \sigma(\boldsymbol{w})}.
\end{align*}
We construct the $(1-\alpha)$-level confidence interval using the Bonferroni approach as follows:
\begin{align}
	\label{conf interval2}
	C_{1- \alpha}' = \left\{ \theta \in \Theta: \inf_{\boldsymbol{w} \in \tilde C_{1- \kappa}} \left(\hat T'(\boldsymbol{w}, \theta)\right)^2 \le c_{1 - \alpha + \kappa}(1) \right\},
\end{align}
where $\kappa>0$ is a small constant, such as $\kappa = 0.005$, $c_{1 - \alpha + \kappa}(1)$ denotes the $(1 - \alpha + \kappa)$-quantile of the $\chi_1^2$ distribution. By modifying the arguments in the proof of Theorem \ref{thm: validity}, we can show that
	\begin{align*}
		\liminf_{n \rightarrow \infty} \inf_{P \in \mathcal{P}} \inf_{\boldsymbol{w} \in \mathbb{W}_P} P\left\{  \theta_P(\boldsymbol{w})  \in C_{1-\alpha} \right\} \ge 1 - \alpha.
	\end{align*}
We can show this using similar arguments as before. The proof is simpler because we do not need to deal with the estimation error of $\boldsymbol{\hat w}$ in $\hat V$. We omit the details.

\section{Further Details on Empirical Applications}
\setcounter{equation}{0}
\renewcommand\theequation{B.\arabic{equation}}

\subsection{Details on Empirical Application: Estimation of Conditional Average Outcomes}

In this section, we explain the estimation of $m_k(\mu_k(X_{i,t}, \underline W_0^*))$ used in our empirical application. First, we estimate $\gamma_k$ using the pairwise difference estimation method of \cite{Honore/Powell:94:JOE}. (Note that $\delta_k$ is not identified in this semiparametric setting.) More specifically, we first define
\begin{align*}
	s(y_1,y_2,z) = \left\{\begin{array}{ll}
		y_1^2 - (y_2 + z) y_1,& \text{ if } z \le - y_2\\
		(y_1 - y_2 - z)^2,& \text{ if } - y_2 < z < y_1\\
		(-y_2)^2 - (z - y_1)(-y_2),& \text{ if } z \ge y_1.
		\end{array}
		\right.
\end{align*}
For each $k=1,...,K$, we let $\hat \gamma_k$ be the estimator obtained as a solution to the following optimization problem:
\begin{align*}
	\min_\gamma \sum_{t=1}^{T} \sum_{i,j \in N_{k,t}: j > i} s\left(\log W_{i,t} - \log \underline W_{k,t}, \log W_{j,t} - \log \underline W_{k,t}, (X_{i,t} - X_{j,t})^{\prime}\gamma\right)
\end{align*}
and let $\hat \gamma_0$ be the estimator obtained as a solution to the following optimization problem: 
\begin{align*}
	\min_\gamma \sum_{t=1}^{T-1} \sum_{i,j \in N_{0,t}: j > i} s\left(\log W_{i,t} - \log \underline W_{0,t}, \log W_{j,t} - \log \underline W_{0,t}, (X_{i,t} - X_{j,t})^{\prime}\gamma\right).
\end{align*}
From this, we obtain $\hat \gamma_k$.

\begin{align*}
	\hat \mu_k(X_{i,t}, \underline W_{k,t}) = \tilde X_{i,t}'\hat \gamma_k - \log \underline W_{k,t} \text{ and } \hat \mu_k(X_{i,t}, \underline W_0^*) = X_{i,t}'\hat \gamma_k - \log \underline W_0^*,
\end{align*}
and construct
\begin{align*}
	\hat m_{0}(\overline \mu) &= \frac{\displaystyle \sum_{t=1}^{T-1}\sum_{\ell \in N_{0,t}} K_h\left(\overline \mu - \hat \mu_0(X_{\ell,t},\underline W_{k,t}) \right) Y_{i,t} }{\displaystyle \sum_{t=1}^{T-1}\sum_{\ell \in N_{0,t}} K_h\left(\overline \mu - \hat \mu_0(X_{\ell,t},\underline W_{k,t})\right)} \text{ and } \\
	\hat m_{k}(\overline \mu) &= \frac{\displaystyle \sum_{t=1}^{T}\sum_{\ell \in N_{k,t}} K_h\left(\overline \mu - \hat \mu_k(X_{\ell,t},\underline W_{k,t})\right) Y_{i,t} }{\displaystyle \sum_{t=1}^{T}\sum_{\ell \in N_{k,t}} K_h\left(\overline \mu - \hat \mu_k(X_{\ell,t},\underline W_{k,t})\right)},
\end{align*}
where $K_h(x) = K(x/h)/h$ and $K$ is a univariate kernel. In particular, we use a quartic kernel and choose $h$ by cross-validation. We obtain the estimator of $m_k(\mu_k(X_{i,t},\underline W_0^*))$ as follows:
\begin{align*}
	\hat m_{k}(\hat \mu_k(X_{i,t},\underline W_0^*)).
\end{align*} 
Once the conditional average outcomes are estimated, we can proceed to construct a synthetic prediction after the minimum wage changes as described in the main text.
 
\subsection{Support Conditions for the Source Populations}

In the empirical application, we require the support conditions for the source populations in the following form: for all $k=1,...,K$,
\begin{align}
	\label{support condition}
	\mathcal{S}_{k0} \subset \mathcal{S}_{kk},
\end{align}
where
\begin{align*}
	\mathcal{S}_{k0} &= \left\{\mu_k(x,\underline W_0^*): x \in \mathcal{X}_{0,T} \right\} \text{ and }\\
	\mathcal{S}_{kk} &= \left\{\mu_k(x,\underline W_0^*): x \in \mathcal{X}_{k,t} \text{ for some } t=1,...,T \right\}.
\end{align*}
(Recall Assumption \ref{assump: support source2}(ii).) Using the estimated conditional average outcomes $\mu_k(x,\underline W_0^*)$ and the data points in $\mathcal{X}_{0,T}$ and $\mathcal{X}_{k,t}$, we can gauge whether this support condition is satisfied or not. We present the results in Table \ref{support_b}.

Table \ref{support_b} reports the estimated sets of $\mathcal{S}_{k0}$ and $\mathcal{S}_{kk}$, and the fraction $\pi_k$ of the samples in the set $\mathcal{S}_{k0}$ but not in the set $\mathcal{S}_{kk}$. The results suggest strongly that the support condition appears to be satisfied by all the source regions in our empirical application. 
\FloatBarrier
	
		\begin{table}[!ht]
		\caption{\small Evidence of Support Condition for the Source Populations}
		
		\label{support_b}
		
		\small 
		
		\begin{centering}
			\small
			\begin{tabular}{c|cccc}
				\hline 
				\hline
				\\ 
				& \multicolumn{4}{c}{Specifications from Table \ref{emp} in the Main Text}\\
				\\
	& (1) & (2) & (3) & (4)\\
				\\
				\hline 
				\\
	
	$\mathcal{S}_{k0}$			 & $\begin{pmatrix} [ -0.117, 0.694] \\ [-0.217, 0.741]\\ - \\ - \\ [0.106, 1.438]\end{pmatrix}$ & $\begin{pmatrix} [ -0.167, 0.619] \\ [-0.251, 0.583] \\ - \\ - \\ [0.045, 1.295]\end{pmatrix}$ & $\begin{pmatrix}  [ -0.117, 0.694] \\ [-0.217, 0.741]\\ [-0.655, 0.052] \\ [1.969, 3.649] \\ [0.106, 1.438]\end{pmatrix}$ & $\begin{pmatrix} [ -0.167, 0.619] \\ [-0.251, 0.583] \\ [-0.691, -0.048] \\ [1.462, 2.958] \\ [0.045, 1.295]\end{pmatrix}$\\
				 
				\\
	$\mathcal{S}_{kk}$									  & $\begin{pmatrix} 
				 [ -0.117, 0.694] \\ [-0.217, 0.741]\\ - \\ - \\ [0.183, 1.438]\end{pmatrix}$ & $\begin{pmatrix} 
				[ -0.213, 0.732] \\ [-0.256, 0.590] \\ - \\ - \\ [0.020, 1.368]\end{pmatrix}$ &
				 $\begin{pmatrix}[ -0.117, 0.694] \\ [-0.217, 0.741]\\ [-0.655, 0.082] \\ [1.969, 3.649] \\ [0.183, 1.438]\end{pmatrix}$ & $\begin{pmatrix} [ -0.213, 0.732] \\ [-0.256, 0.590] \\ [-0.702, -0.023] \\ [1.439, 2.971] \\ [0.020, 1.368]\end{pmatrix}$\\
				 \\
			$\pi_k$ & $\begin{pmatrix} 0 \\ 0  \\ - \\ - \\  0.002 \end{pmatrix}$ & $\begin{pmatrix} 0 \\ 0  \\ - \\ - \\ 0\end{pmatrix}$ & $\begin{pmatrix} 0 \\ 0 \\ 0 \\ 0 \\ 0.002 \end{pmatrix}$ & $\begin{pmatrix} 0 \\ 0 \\ 0 \\ 0 \\ 0 \end{pmatrix}$\\
				\hline
				\multicolumn{1}{c}{} &  &    \tabularnewline
			\end{tabular}
			\par\end{centering}
		\parbox{6.4in}{\footnotesize
			Notes: The table presents the estimated supports of $\mathcal{S}_{k0} = \{\mu_k(x,\underline W_0^*): x \in \mathcal{X}_{0,T} \}$ (first row), $\mathcal{S}_{kk} = \{\mu_k(x,\underline W_0^*): x \in \mathcal{X}_{k,t} \text{ for some } t=1,...,T \}$ (second row), across the four specifications of Table 2. Each column represents the respective specification in Table \ref{emp}. Each entry in each row represents the respective result for a given state, where states are sorted by $(\text{CA},\text{CT},\text{FL},\text{LA},\text{WA})'$.	The third row shows the fraction of the observations in $\mathcal{S}_{k0}$ that are not in $\mathcal{S}_{kk}$.}
	\end{table}

	\FloatBarrier

	\newpage
\putbib[SyntheticDecomp]
\end{bibunit} 
\end{document}